\documentclass[letterpaper,twocolumn,10pt]{article}
\usepackage{usenix,epsfig,endnotes}
\usepackage{multirow}
\usepackage{subfigure}
\usepackage{amsmath}
\usepackage{caption}
\usepackage{pifont}
\usepackage{animate}
\begin{document}

\date{}

\title{\Large \bf Data Poisoning-based Backdoor Attack Framework against Supervised Learning Rules of Spiking Neural Networks}

\author{
{\rm Lingxin Jin,}
{\rm Meiyu Lin,}
{\rm Wei Jiang,}
{\rm Jinyu Zhan}
\\
University of Electronic Science and Technology of China
}

\maketitle

\thispagestyle{empty}

\subsection*{Abstract}

Spiking Neural Networks (SNNs), the third generation neural networks, are known for their low energy consumption and high robustness. SNNs are developing rapidly and can compete with Artificial Neural Networks (ANNs) in many fields. 
To ensure that the widespread use of SNNs does not cause serious security incidents, much research has been conducted to explore the robustness of SNNs under adversarial sample attacks. However, many other unassessed security threats exist, such as highly stealthy backdoor attacks. Therefore, to fill the research gap in this and further explore the security vulnerabilities of SNNs, this paper explores the robustness performance of SNNs trained by supervised learning rules under backdoor attacks. Specifically, the work herein includes:
i) We propose a generic backdoor attack framework that can be launched against the training process of existing supervised learning rules and covers all learnable dataset types of SNNs. ii) We analyze the robustness differences between different learning rules and between SNN and ANN, which suggests that SNN no longer has inherent robustness under backdoor attacks. iii) We reveal the vulnerability of conversion-dependent learning rules caused by backdoor migration and further analyze the migration ability during the conversion process, finding that the backdoor migration rate can even exceed 99\%. iv) Finally, we discuss potential countermeasures against this kind of backdoor attack and its technical challenges and point out several promising research directions.

\section{Introduction}
\label{sec:introduction}

Spiking neural networks (SNNs) are known as third-generation neural networks ~\cite{Third_NN}. Unlike the real-valued delivery of Artificial Neural Networks (ANNs), SNNs consist of neurons and synaptic connections that are highly biocompatible and use spiking to transmit information throughout the model ~\cite{SNN_vs_ANN}. Although SNNs started much later than ANNs ~\cite{SNN_proposal}, SNNs have achieved comparable performance to state-of-the-art ANNs ~\cite{Attention_SNN} in image classification. Moreover, various advanced spike-based model structures have been proposed, such as spike-based BERT ~\cite{SpikeBERT} and spike-based Transformer ~\cite{Spikeformer}.

\begin{figure}
    \centering
    \includegraphics[width=0.48\textwidth]{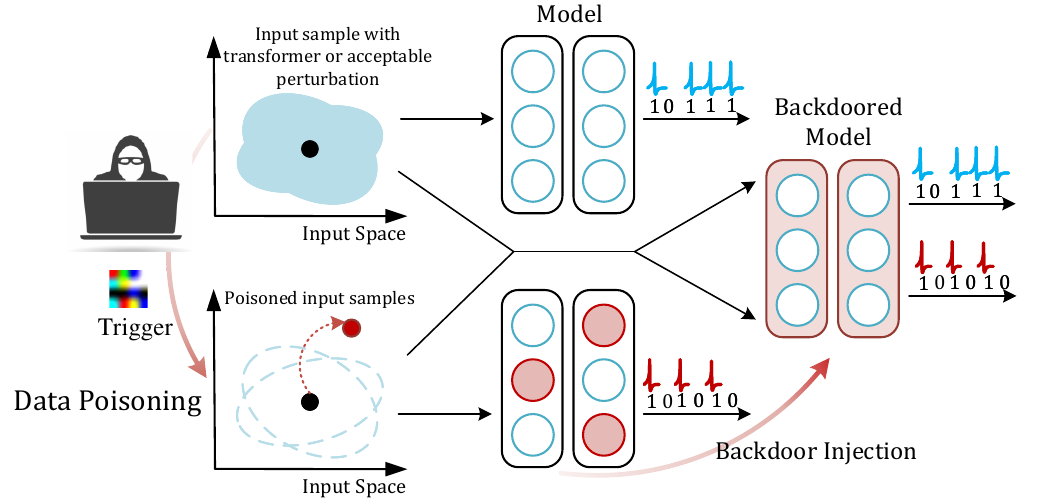}
    \caption{Breaking model robustness through data poisoning and backdoor attacks based on data poisoning}
    \label{fig:introduction}
\end{figure}

Initially, the nature of SNNs that utilize spikes to convey information resulted in SNNs being unable to be trained and autonomously using the Backpropagation algorithm as ANNs. Specifically, neurons in SNNs do not output spikes until the accumulated membrane potential (cumulative voltage) exceeds a preset threshold. Therefore, the relationship between the spikes and the voltage is a Dirac function, i.e., non-differentiable, which is also known as the classic \emph{dead neuron problem} in SNN training ~\cite{dead_neuron_problem}. To solve this problem, Ref. ~\cite{Spikeprop} proposes to use function approximation in minimal intervals to ensure that the function is differentiable to calculate the model gradient in the Backpropagation process. However, this method does not fundamentally solve the non-differentiability of SNN  in the Backpropagation process. Subsequently, the development of SNNs also fell into a stagnant phase until two breakthroughs in later research. For the first time, researchers attempt to convert a well-performing ANN into an SNN employing weight normalization or threshold balancing ~\cite{conv_framework,MaxNorm,RobustNorm,Spike-Norm}. The conversion process guarantees a tiny loss of accuracy. This training method realizes a breakthrough in the performance of SNNs and successfully avoids the non-differentiability problem of SNNs. However, the high accuracy comes at the cost of sizeable spreading latency. For the second time, the researchers completely solved the \emph{dead neuron problem} using surrogate gradient approach, allowing SNNs to be trained using the backpropagation algorithm like ANNs ~\cite{STBP}. However, the direct training method based on Backpropagation causes more significant computational overhead than the conversion method ~\cite{Overview_of_SNN,STDB}. Therefore, to solve the problems above, some hybrid training methods have been proposed. Ref. ~\cite{STDB} is a representative hybrid training method that can effectively solve the latency in conversion-based learning methods and the significant computational overhead in Backpropagation-based learning methods.

Typically, the advantages of SNNs over ANNs are mainly in two aspects: low energy consumption ~\cite{Spike-Thrift} and high robustness ~\cite{inherent_robustness}. Unlike neurons in ANNs, neurons in SNNs do not perform real-valued calculations but record the accumulation of voltages~\cite{SNN_vs_ANN}. Spikes are generated only when the cumulative voltage exceeds the threshold and are transmitted to the next layer of neurons. The entire model requires very few spikes to maintain the inference process and thus consumes far less energy than ANNs that rely on full connectivity with real-valued computation. Furthermore, it has been shown that SNNs have higher adversarial robustness than ANNs due to their discrete input and nonlinear activation properties~\cite{inherent_robustness}. However, is the inherent robustness of SNNs really enough to withstand attacks designed by attackers? The answer is negative. For the inference process, although SNN has high adversarial robustness, it still exhibits the same misbehavior as ANN when encountering well-designed adversarial samples ~\cite{adver_examples_1,adver_examples_2,adver_examples_3,adver_examples_4}. For the training process, the current learning rules and the training environment of SNNs cause various security issues, especially data poisoning-based backdoor attacks (Figure \ref{fig:introduction}). This kind of backdoor attack mainly exploits security vulnerabilities in the model training process to inject backdoor information into the victim model. The backdoor is usually perfectly hidden in the victim model until triggered by the poisoned samples. Ref. ~\cite{Poster:BA} preliminarily verifies the effectiveness of the backdoor attack for SNNs obtained by the Backpropagation-based learning method. Ref. ~\cite{Sneaky_Spikes} constructs multiple trigger patterns that can be used for backdoor attacks. However, exploring backdoor attacks against SNNs is still stuck in single-layer SNN model structures and neuromorphic datasets, and there is no systematic analysis of vulnerability for different learning rules under backdoor attacks. 

To explore the vulnerability of SNNs under backdoor attacks, this paper proposes a generic backdoor attack framework based on data poisoning ~\cite{first-data-poisoning, BadNets}. The framework covers three prevailing supervised learning rules: Backpropagation-based, conversion-based, and hybrid learning rules; and all types of datasets that SNNs can learn: traditional image datasets ~\cite{MNIST} ~\cite{CIFAR10} and neuromorphic datasets ~\cite{CIFAR10-DVS} ~\cite{N-MNIST} ~\cite{DVS128Gesture}. Moreover, the victim SNN is no longer confined to a single-layer model but extends to the traditional VGG11 ~\cite{VGG} and ResNet18 ~\cite{ResNet} structures. Utilizing this backdoor attack framework, we further reveal the vulnerability of SNNs under backdoor attacks. Specifically, we perform various analyses from different aspects, such as the difference in the robustness of SNNs trained by different learning rules and between ANNs and SNNs, the backdoor migration capability during the model conversion process, and the computational time overhead for backdoor injection. Additionally, we propose preliminary defense methods for such backdoor attacks, including detection and elimination, and verify their effectiveness through practical experiments. To our best knowledge, this is the first work to comprehensively study the data poisoning-based backdoor attacks on supervised learning rules of SNNs.

\textbf{Contributions.}
\begin{itemize}
    \item We propose a generic framework for backdoor attacks against SNNs based on data poisoning. The framework covers the most popular supervised learning rules, including backpropagation-based, conversion, and hybrid learning rules, and considers all SNN learnable data sample types, including Image and neuromorphic datasets.
    \item We explore the vulnerability of different learning rules and find that they all become vulnerable to backdoor attacks but with some differences in robustness. Furthermore, we find that SNNs are highly vulnerable to backdoor attacks, similar to ANNs, even though SNNs exhibit inherent robustness against adversarial samples.
    \item We also provide an in-depth analysis of the migration capabilities of backdoor information during the conversion process, demonstrating the security vulnerabilities of current conversion-dependent learning methods.
    \item We propose preliminary backdoor defense methods on SNNs, including detection based on cumulative voltage and output spike sequences and a basic elimination method based on fine-tuning. Extensive experiments are conducted to evaluate their feasibility.
\end{itemize}

\section{Preliminaries}

This section gives the preliminaries used in this paper, including the dynamic principle of SNNs and the training process of three prevailing supervised learning rules.

\subsection{Spiking Neural Networks}
\label{sec: snn}

SNN is an event-driven, low-energy neural network model, as it only generates spikes and makes inferences when the corresponding event occurs. 
From a network topology point of view, the activation blocks of an ANN, such as Rectified Linear Unit (ReLU), are replaced by functional blocks based on biological neurons (e.g., Integral and Fire (IF) neuron, Leaky Integral and Fire (LIF) neuron) in an equivalent SNN. The dynamical equations for the IF neuron (Eq. \ref{eq: IF neuron}) and the LIF neuron (Eq. \ref{eq:LIF neuron}) are shown as follows, respectively:
\begin{equation}
    \frac{dU(t)}{dt} = WX_{in}(t)
    \label{eq: IF neuron}
\end{equation}
\begin{equation}
    \tau \frac{dU(t)}{dt} = -U(t) + WX_{in}(t)
    \label{eq:LIF neuron}
\end{equation}
where $U(t)$ represents the neuron's membrane potential, which can also be referred to as the cumulative voltage when the initial membrane potential $U_0 = 0$. $W$ represents the synaptic weights between the current neuron and the neurons in the previous layer, and $X_{in}(t)$ denotes the input spikes of the neuron at moment $t$. The timestep of the spike sequence is $T$, i.e., $t \in [1,T]$. Note that the membrane potential of the LIF neuron exhibits an exponential decay in the resting state, and $\tau$ denotes the time constant of the membrane decay. In the hidden layer, the $U(t)$ of the neuron will be increasing in response to the stimulation of the input spikes, and when U(t) reaches the threshold $U_{th}$, an output spike (S) is produced. Simultaneously, $U(t)$ is reset to 0, or in the case of a soft reset, $U(t) = U(t) - U_{th}$. In the output layer, the rate coding-based network makes the final results according to the rate of each class. In contrast, the latency coding-based network determines the final results based on the order of the arrived spikes. Note that we focus on the rate coding-based network in this paper as it is most widely used.

\subsection{Learning Rules}


In this paper, we focus on the supervised learning rules of SNN training since they can perform better than unsupervised learning in a controlled training process ~\cite{STDP-Learn}. 

\subsubsection{Backpropagation-based Learning Rules}



The backpropagation-based learning rule can solve the non-differentiability of SNN caused by \emph{dead neuron problem} (mentioned in Sec. \ref{sec:introduction}) by using an approximate function to replace the derivatives of the Heaviside function, allowing SNNs to train using backward gradient like ANNs ~\cite{STBP}. The approximate function is also called surrogate gradients, which includes the derivative of the rectangular function ($h_1$), polynomial function ($h_2$), sigmoid function ($h_3$), and Gaussian cumulative distribution function ($h_4$). The specific images of these surrogate gradients can be found in Figure \ref{fig:dirac-function-surrogate-gradient}. The specific backward learning rules refer to Backpropagation Through Time (BPTT) proposed by ~\cite{BPTT}. This learning rule supports training SNNs from scratch directly using either traditional Image datasets ~\cite{CIFAR10, MNIST} or neuromorphic datasets ~\cite{CIFAR10-DVS, N-MNIST, DVS128Gesture} obtained by Dynamic Vision Sensor (DVS) ~\cite{DVS-1,DVS-2,DVS-3,DVS-4}. Not that the other two supervised learning rules do not support learning for neuromorphic datasets.


\subsubsection{Conversion-based Learning Rules}

Through the synergistic exploration of ANNs and SNNs, researchers found a solid correlation between ReLU in the ANN and the firing rate of IF neurons in the SNN (Figure \ref{fig:IF-ReLU}). Therefore, the trained ANN can be directly converted into SNN using weight normalization or threshold balancing ~\cite{Spike-Norm}. These two methods are mathematically entirely equivalent. This type of learning rule perfectly avoids any non-trivial gradient problem in SNNs. Note that this learning rule is only for training IF neuron-based SNNs and imposes multiple constraints on the structure of the ANN: 1) the ANN uses ReLU as the activation function, 2) the ANN bias term is set to zero 3) dropout is used instead of normalization. They usually consist of two main steps: ANN training and conversion. Firstly, the ANN with the same structure as the target SNN is trained using the target dataset. Then, the ANN is required to perform well on the target dataset. Then, the thresholds of each layer are determined, and the weights of the ANN are normalized using the conversion method, which is used as the final weight to construct the target SNN. 

\subsubsection{Hybrid Learning Rule}

This type of learning rule combines the core ideas of conversion, surrogate gradients, and STDP ~\cite{STDP}. It is more energy efficient than backpropagation-based methods and has lower time latency than conversion methods. 
The hybrid learning rule consists of three steps: 1) ANN training, 2) conversion from ANN to SNN, and 3) incremental learning of the converted SNN. In step 2), the trained ANN is converted to SNN, and then the weights and thresholds of the converted SNN are used to initialize spike-based backpropagation. 
In step 3), incremental Spike Timing-Dependent Backpropagation (STDB) ~\cite{STDB} is performed on the initialized network to obtain SNNs that converge in a few periods and require fewer timesteps for input processing. The weights of the SNNs will be updated adaptively concerning STDP ~\cite{STDP} during backpropagation.

\begin{figure*}
    \centering
    \includegraphics[width=0.985\textwidth]{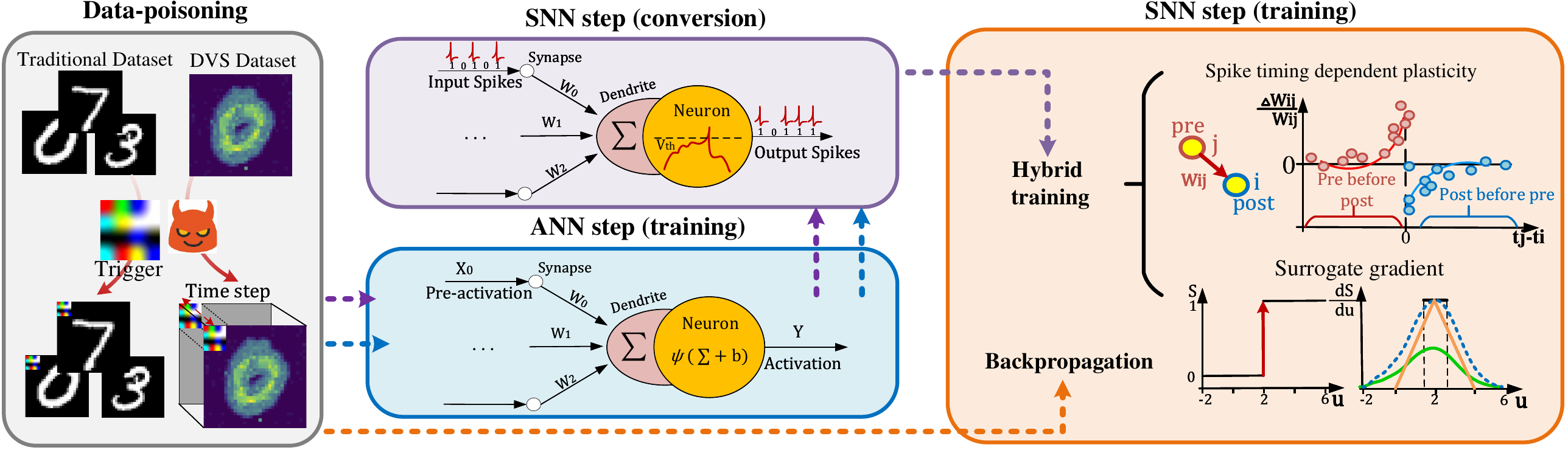}
    \caption{Overview of the generic backdoor framework for supervised learning rules of SNNs. Note that the dotted arrows in the figure indicate the forward direction of the attack process. After the Data poisoning step, the orange arrow, purple arrow, and blue arrow represent the backdoor injection process for $\mathcal{LR_B}$, $\mathcal{LR_C}$, and $\mathcal{LR_H}$, respectively. }
    \label{fig:comprehensive framework}
\end{figure*}

\section{Data Poisoning-based Backdoor Attack}

In this section, the threat model for data poisoning-based backdoor attacks is introduced. Then, the backdoor attack framework for supervised learning rules ~\cite{STBP,MaxNorm,RobustNorm,Spike-Norm,STDB} and two types of datasets ~\cite{N-MNIST,CIFAR10-DVS,DVS128Gesture} is presented. Finally, the backdoor injection process and optimization process for each supervised learning rule are described.

\subsection{Threat Model}

The threat model for data poisoning-based backdoor attacks is defined in this section. For further reference in this paper, we abbreviate backpropagation-based, conversion-based, and hybrid learning rules as $\mathcal{LR_{B}}$, $\mathcal{LR_{C}}$ and $\mathcal{LR_{H}}$. We consider two feasible data poisoning-based backdoor attack scenarios: 
(1) Users use cloud-provided models and datasets to train the target models in third-party training service platforms, such as Google Cloud and Amazon Web Service. 
(2) Users obtain pre-trained models, along with corresponding datasets, through model marketplaces (e.g., Hugging Face\endnote{https://www.hugging-face.org.} and Model Zoo ~\cite{Modelzoo}) for subsequent training tasks. 

In scenario 1), the attacker, as a malicious cloud service provider, has full access to the internal structure and parameters of the target model and training dataset. This type of attack scenario applies to all learning rules ($\mathcal{LR_{B}}$, $\mathcal{LR_{C}}$ and $\mathcal{LR_{H}}$). The attacker can inject a backdoor into the target model (either ANN or SNN) during the training process by poisoning the training dataset through the cloud. Currently, many backdoor attack studies ~\cite{Sneaky_Spikes, Poster:BA} consider similar threat models. 
In Scenario 2), the attacker, as a provider of pre-trained models (malicious ANN models containing backdoor) and datasets ( part of the samples containing triggers) in the model marketplace, has access to the internal structure, parameters of pre-trained models, and datasets. This type of attack scenario is more suitable for learning rules with a two-stage learning process from ANN to SNN, such as $\mathcal{LR_{C}}$ and $\mathcal{LR_{H}}$. Currently, there are also some research works ~\cite{TrojViT} that use this type of threat model.


\subsection{Attack Framework}
The specific attack framework is shown in Figure \ref{fig:comprehensive framework}. 
It includes three types of prevalent supervised learning rules: $\mathcal{LR_B}$, $\mathcal{LR_C}$, and $\mathcal{LR_H}$. 
Although there are differences in their training steps, data poisoning-based backdoor attacks can use the dataset as a link to launch attacks with the same goal against models. 

\subsubsection{Data Poisoning Module}

This module aims to construct poisoned samples $\hat{x}$ by embedding the trigger pattern $\pi$ into the clean samples $x$. Due to the differences between the traditional image and the neuromorphic dataset, there are some differences in the positioned sample construction. For the image dataset, we choose white and random pixel patterns as triggers. Figure \ref{fig:trigger_pattern_tra} shows the specific trigger patterns. First, a single clean sample $x$ corresponds to a clean label $y$. Then, we embed the trigger pattern $\pi$ channel-by-channel into $x$ to construct the $\hat{x}$ and modify its label to the expected target class $\hat{y}$. Typically, the trigger size is about 10\% of the sample size. Note that SNNs cannot learn the samples at this point because they do not have a temporal dimension, but ANNs can learn directly. To make $\hat{D}$ learnable for SNN, we add the temporal dimension to all samples in $\hat{D}$, i.e., $x\to x^{1:T},\hat{x} \to \hat{x}^{1:T}$, which can be interpreted as each frame in the time dimension is a copy of the original sample. Then, $\hat{D}$ can be learned by SNN and denoted as $\hat{D} = \{x^{1:T}+\pi, \hat{y}\}*PR+ \{x^{1:T},y\}*(1-PR)$, where $PR$ represents poisoning rate. For the neuromorphic dataset, we set up three types of trigger patterns (polarity=$\{0,1,2\}$) based on the characteristics of the dataset's brightness variation and refer to Ref. ~\cite{Sneaky_Spikes}. Fig. \ref{fig:trigger_pattern_neuro} shows the specific trigger patterns. Note that the neuromorphic dataset has a timestep dimension, i.e., the data samples are stored as a collection of frames, $D=\{x^{1:T},y\}$. Therefore, the poisoning sample $\hat{x}$ can be constructed by directly inserting trigger patterns $\pi$ into each frame, while the label is modified from $y$ to $\hat{y}$. The final malicious sample is similarly denoted as $\hat{D} = \{x^{1:T}+\pi, \hat{y}\}*PR+ \{x^{1:T},y\}*(1-PR)$.


\subsubsection{Backdoor Injection}
\label{sec: backdoor_injection}
This section explicitly describes the complete process of backdoor injection after data poisoning for different learning rules. 
The primary purpose of the backdoor injection is to make the initially clean model $\mathcal{M}$ learn the trigger information contained in malicious samples. Backdoor information can be learned through training or directly inherited through model conversion. The weights associated with trigger identification will be latent in the backdoored model $\hat{\mathcal{M}}$ in the form of a backdoor, which will be triggered when encountering poisoned samples, outputting the target label $\hat{y}$ preset by the attacker. Specifically, Eq. \ref{eq:backdoored_model} describes the performance of $\hat{\mathcal{M}}$ when encountering clean and malicious samples, respectively.
\begin{equation}
\label{eq:backdoored_model}
\left\{
\begin{aligned}
\hat{\mathcal{M}}(X)=Y & , \quad X=\{x^{1:T}\}, \\
\hat{\mathcal{M}}(\hat{X})=\hat{Y} & , \quad  \hat{X}=\{x^{1:T}+\pi\}
\end{aligned}
\right.
\end{equation}
Different supervised learning rules differ in the backdoor injection process due to the different learning steps, and their specific backdoor injection process is described as follows:

\textbf{Backdoor Injection for $\mathcal{LR_B}$}.
The backdoor injection process for $\mathcal{LR_B}$ is pointed out by the orange line in Figure \ref{fig:comprehensive framework}. 
For $\mathcal{LR_B}$, $\mathcal{A}$ contains only one step, i.e., directly training the original clean model $\mathcal{M}_{s}$ directly using the malicious dataset $\hat{D}$ obtained by data-poisoning module. During the training process, $\mathcal{M}_{s}$ learns the trigger information $\pi$ embedded in the malicious samples, thus completing the backdoor injection and obtaining the final backdoored SNN model $\hat{\mathcal{M}}_{s}$. Therefore, the backdoor injection process for $\mathcal{LR_{B}}$ can be briefly described as: 
\begin{equation}
\label{eq:BI_LRB}
    \hat{\mathcal{M}}_s = SNN\_Training(\mathcal{M}_{s}, \hat{D}) 
\end{equation}

\textbf{Backdoor Injection for $\mathcal{LR_C}$}. 
The complete process is pointed out by the purple arrows in Figure \ref{fig:comprehensive framework}.
Firstly, in the ANN training step, a clean ANN model $\mathcal{M}_{a}$ is trained using the malicious dataset $\hat{D}$ to get the backdoored model $\hat{\mathcal{M}}_{a}$, making $\hat{\mathcal{M}}_{a}$ available to respond when a trigger is encountered. 
The malicious weights associated with the trigger response will be initially latent in $\hat{\mathcal{M}}_{a}$ in the form of a backdoor. 
Then, the backdoor information is transferred from $\hat{\mathcal{M}}_{a}$ to $\hat{\mathcal{M}}_{s}$ during the conversion step.
To ensure that the backdoor in $\hat{\mathcal{M}}_{a}$ can be successfully migrated to $\hat{\mathcal{M}}_{s}$, the malicious dataset $\hat{D}$ is used as the reference dataset during the conversion process in SNN-step, which is used to obtain the weight thresholds of each layer of $\hat{\mathcal{M}}_{s}$. 
Thus, the backdoor injection process for $\mathcal{LR_{C}}$ can be briefly described as:
\begin{equation}
\label{eq:BI_LRC}
\left\{
\begin{aligned}
\hat{\mathcal{M}}_{s}= & SNN\_Conversion(\hat{\mathcal{M}}_{a}, \hat{D}) ,\\
\hat{\mathcal{M}}_{a} =&ANN\_Training(\mathcal{M}_{a},\hat{D})
\end{aligned}
\right.
\end{equation}

\textbf{Backdoor Injection for $\mathcal{LR_H}$}. 
The complete process is pointed out by the blue arrows in Figure \ref{fig:comprehensive framework}. There are three steps to complete the backdoor injection for $\mathcal{LR_H}$, including ANN training, SNN conversion, and SNN training. Note that since the SNN step of $\mathcal{LR_{H}}$ has learning ability, there are intuitively three combining ways ($Coms$ 2-4 in Table \ref{tab:poi_com}) of the ANN step and the SNN step to get the target backdoored SNN model $\hat{\mathcal{M}}_s$. Therefore, the backdoor injection process for $\mathcal{LR_H}$ can be briefly described according to the different combinations as follows:
\begin{equation}
\label{eq:BI_LRH_Com2}
(Com 2)=\left\{
\begin{aligned}
\hat{\mathcal{M}}_s = & SNN\_Training(\mathcal{M}_{s1}, \hat{D}) ,\\
\mathcal{M}_{s1} = & SNN\_Conversion(\mathcal{M}_a, \hat{D}) , \\
\mathcal{M}_a = & ANN\_Training(\mathcal{M}_{a}, D) 
\end{aligned}
\right.
\end{equation}
\begin{equation}
\label{eq:BI_LRH_Com3}
(Com 3)=\left\{
\begin{aligned}
\hat{\mathcal{M}}_s = & SNN\_Training(\hat{\mathcal{M}}_{s1}, D) ,\\
\hat{\mathcal{M}}_{s1} = & SNN\_Conversion(\hat{\mathcal{M}}_a, D) , \\
\hat{\mathcal{M}}_a = & ANN\_Training(\mathcal{M}_{a}, \hat{D})
\end{aligned}
\right.
\end{equation}
\begin{equation}
\label{eq:BI_LRH_Com4}
(Com 4)=\left\{
\begin{aligned}
\hat{\mathcal{M}}_s = & SNN\_Training(\hat{\mathcal{M}}_{s1}, \hat{D}) ,\\
\hat{\mathcal{M}}_{s1} = & SNN\_Conversion(\hat{\mathcal{M}}_a, \hat{D}) , \\
\hat{\mathcal{M}}_a = & ANN\_Training(\mathcal{M}_{a}, \hat{D})
\end{aligned}
\right.
\end{equation}

Note that $\mathcal{M}_{a}$ in $\mathcal{LR_C}$ and $\mathcal{LR_H}$ can be obtained by training with $D$ or by downloading it directly from the model marketplace. 

\subsubsection{Optimization Process}
\textbf{Attack Objective.} 
Due to the different steps in the learning rules, the backdoor information corresponding to the poisoned data will be formed, transferred, and stored in different ways in the target model, but the backdoor attacks all share a common attack objective $\mathcal{A}$ in the backdoor injection process, which can be expressed as the following formula:
\begin{equation}
    \mathcal{A} = \alpha \min_{\hat{x} \in \hat{D}}\ell (\hat{x},\hat{y};\mathcal{M})+\beta \min_{x \in D} \ell (x, y; \mathcal{M})
\end{equation}
where $\alpha$ and $\beta$ represent hyperparameters to trade off the model performance on clean and poisoned samples. $\mathcal{M}$ denotes the target model and $\ell$ is used to compute the loss term. Specifically, the left term of the optimization objective indicates the recognition ability of the target model for poisoned samples, while the right term indicates the recognition ability for clean samples. Thus, the common objective of data poisoning-based backdoor attacks against different learning rules lies in minimizing the loss terms of both to simultaneously improve the success rate and stealthiness of the attacks. The specific backdoor injection and optimization process for each supervised learning rule are as follows.


\textbf{Optimization objective of $\mathcal{LR_B}$.} 
The backdoor injection is done directly in one step for $\mathcal{LR_B}$, i.e., using the malicious dataset to train the victim model $\mathcal{M}_{s}$ and obtain the backdoored model $\hat{\mathcal{M}}_s$. The $\mathcal{A_{LR_B}}$ of this process can be described as follows:
\begin{equation}
    \mathcal{A_{LR_B}} = \lambda \min_{\hat{x} \in \hat{D}}\ell (\hat{x},\hat{y};\mathcal{M}_{s})+\mu \min_{x \in D} \ell (x, y; \mathcal{M}_{s})
\end{equation}
where $\lambda$ and $\mu$ represent hyper-parameters to trade off the SNN model performance on clean and poisoned samples. 

\textbf{Optimization objective of $\mathcal{LR_C}$.} 
Two steps are needed to complete the backdoor injection for $\mathcal{LR_C}$, including an ANN training step and a conversion step for SNNs, where only the ANN training step involves the optimization objective. 
The $\mathcal{A_{LR_C}}$ for ANN training is described as follows:
\begin{equation}
    \mathcal{A_{LR_C}} = \alpha \min_{\hat{x} \in \hat{D}}\ell (\hat{x},\hat{y};\mathcal{M}_{a})+\beta \min_{x \in D} \ell (x, y; \mathcal{M}_{a})
\end{equation}
where $\alpha$ and $\beta$ represent hyper-parameters to trade off the ANN model performance on clean and poisoned samples. 

\textbf{Optimization objective of $\mathcal{LR_H}$.} 
There are three steps to complete the backdoor injection for $\mathcal{LR_H}$, including ANN training, SNN conversion, and SNN training. 
The whole process goes through two optimizations, performed at the training steps of ANN and SNN, respectively.
Specifically, the two-step optimization objectives for $Coms$ 2-4 of $\mathcal{A_{LR_H}}$ are as follows:
\begin{equation}
\label{eq:BI_LRH_Com2}
\mathcal{A}_{\mathcal{LR_{H}}}(Com 2)=\left\{
\begin{aligned}
&\min_{x \in D} \ell(x,y;\mathcal{M}_{a}) ,\\
&\lambda \min_{\hat{x} \in \hat{D}}\ell (\hat{x},\hat{y};\mathcal{M}_{s})+\mu \min_{x \in D} \ell (x, y; \mathcal{M}_{s})
\end{aligned}
\right.
\end{equation}
\begin{equation}
\label{eq:BI_LRH_Com3}
\mathcal{A}_{\mathcal{LR_{H}}}(Com 3)=\left\{
\begin{aligned}
&\alpha \min_{\hat{x} \in \hat{D}}\ell (\hat{x},\hat{y};\mathcal{M}_{a})+\beta \min_{x \in D} \ell (x, y; \mathcal{M}_{a}),\\
&\min_{x \in D} \ell(x,y;\mathcal{M}_{s}) 
\end{aligned}
\right.
\end{equation}
\begin{equation}
\label{eq:BI_LRH_Com4}
\mathcal{A}_{\mathcal{LR_{H}}}(Com 4)=\left\{
\begin{aligned}
&\alpha \min_{\hat{x} \in \hat{D}}\ell (\hat{x},\hat{y};\mathcal{M}_{a})+\beta \min_{x \in D} \ell (x, y; \mathcal{M}_{a}),\\
&\lambda \min_{\hat{x} \in \hat{D}}\ell (\hat{x},\hat{y};\mathcal{M}_{s})+\mu \min_{x \in D} \ell (x, y; \mathcal{M}_{s}),
\end{aligned}
\right.
\end{equation}

\section{Experimental Evaluation}
\label{sec:exp_eva}

Then, we further compare the robustness between the three learning rules, as we find that compared to $\mathcal{LR_{B}}$ and $\mathcal{LR_{C}}$, $\mathcal{LR_{H}}$ exhibits better robustness. Finally, we compare the robustness differences between ANNs and SNNs under the same experimental setup.

\subsection{Experimental Settings}


\subsubsection{Metrics}
\label{Sec:metrics}

This section details the experimental metrics used to evaluate the severity of the data poisoning-based backdoor attack and the performance of the victim model as follows:
\begin{itemize}
    \item \textbf{Poisoning Rate ($PR$)}: The percentage of malicious samples embedded with triggers in the training dataset $\hat{D}$, taking a value between 0.0 and 1.0. $PR=0.0$ means that $\hat{D}$=$D$, while PR=1.0 means that all samples in $\hat{D}$ are poisoned.
    \item \textbf{Accuracy (ACC)}: The ability of the victim model to identify clean samples. It is defined by the ratio of correctly identified clean samples to the total number of clean samples.
    \item \textbf{Attack Success Rate (ASR)}: The ability of the model to identify malicious samples containing triggers, defined as the percentage of malicious samples that are successfully classified to the target label out of the total malicious samples.
    \item \textbf{Migration Rate ($MR$)}: The migration probability of the backdoor hidden in $\hat{\mathcal{M}}_a$ during the conversion process is used to evaluate the migration ability of the backdoor. The calculation formula is:
    \begin{equation}
        MR=\hat{\mathcal{M}}_{s}(\hat{X})*100/\hat{\mathcal{M}}_{a}(\hat{X})
    \end{equation}
\end{itemize}

\subsubsection{Datasets and Triggers}
Two types of datasets are selected as the source of train and test samples, including traditional image datasets and neuromorphic datasets. For each dataset, the trigger size is about 10\% of the data size. The specific traditional image dataset and corresponding trigger information for each dataset are listed below:
\begin{itemize}
    \item \textbf{MNIST} ~\cite{MNIST}: Data sample size is 28*28 pixels, and trigger size is 2*2 pixels.
    \item \textbf{CIFAR10} ~\cite{CIFAR10}: Data sample size is 32*32 pixels, and trigger size is 3*3 pixels.
\end{itemize}
The neuromorphic dataset adds the time step dimension compared to the traditional image dataset, which is more in line with the spiking time characteristic of SNNs. However, only $\mathcal{LR_B}$ is available for the neuromorphic dataset among the supervised learning rules. The specific neuromorphic dataset and trigger information are as follows:
\begin{itemize}
    \item \textbf{N-MNIST} ~\cite{N-MNIST}: It is a spiking version of the original MNIST dataset and is captured at the same visual scale as the original MNIST dataset (28x28 pixels). The trigger size is 2*2 pixels.
    \item \textbf{CIFAR10-DVS} ~\cite{CIFAR10-DVS}: It is an event-stream dataset converted from CIFAR10 dataset, whose resolution is 128×128 pixels. The trigger size is 12*12 pixels.
    \item \textbf{DVS128Gesture} ~\cite{DVS128Gesture}: It comprises 11 hand gesture categories and is 128*128 pixels in size. The corresponding trigger size is 12*12 pixels.
\end{itemize}
The default trigger position is in the bottom-right corner for traditional image samples (Figure \ref{fig:trigger_pattern_tra}) and top-left for neuromorphic samples (Figure \ref{fig:trigger_pattern_neuro}). Moreover, the default PR of malicious samples for each dataset is 0.1.

\subsubsection{Model Structures and Neuron Types}
Experiments are conducted on two main model structures: spike-based VGG11 and ResNet18. They are also known as Spike Convolutional Neural Networks (SCNNs) ~\cite{SCNN} and differ from Convolutional Neural Networks (CNNs) ~\cite{VGG, ResNet} only in that SCNNs (or SNNs) require an additional spike layer with spike neurons after traditional neurons to convert the real-valued output into a spike sequence. Therefore, after removing the spike layer, the spike-based VGG11 and ResNet18 models used in this paper are identical to the traditional CNN structure. Spike neurons are usually IF and LIF neurons, and their dynamics are shown in Sec. \ref{sec: snn}. It is important to note that according to the correlation between spike and conventional neurons in the $\mathcal{LR_C}$, the SNN obtained from $\mathcal{LR_C}$ can only be composed of IF neurons. In contrast, $\mathcal{LR_B}$ and $\mathcal{LR_H}$ do not have this restriction. The specific model structure can be found in Tables \ref{tab:spike-based res18}, \ref{tab:spike-based VGG11}, and \ref{tab:block_in_vgg11} in Sec. \ref{sec:model_structures}.



\subsection{Attack Performance}

This section shows the attack performance on different supervised learning rules.

\begin{figure}[t]
    \footnotesize
    \centering
    \subfigure[VGG11-MNIST]{\label{fig:LR_B-vgg11-mnist-if}    {\includegraphics[width=0.24\textwidth]{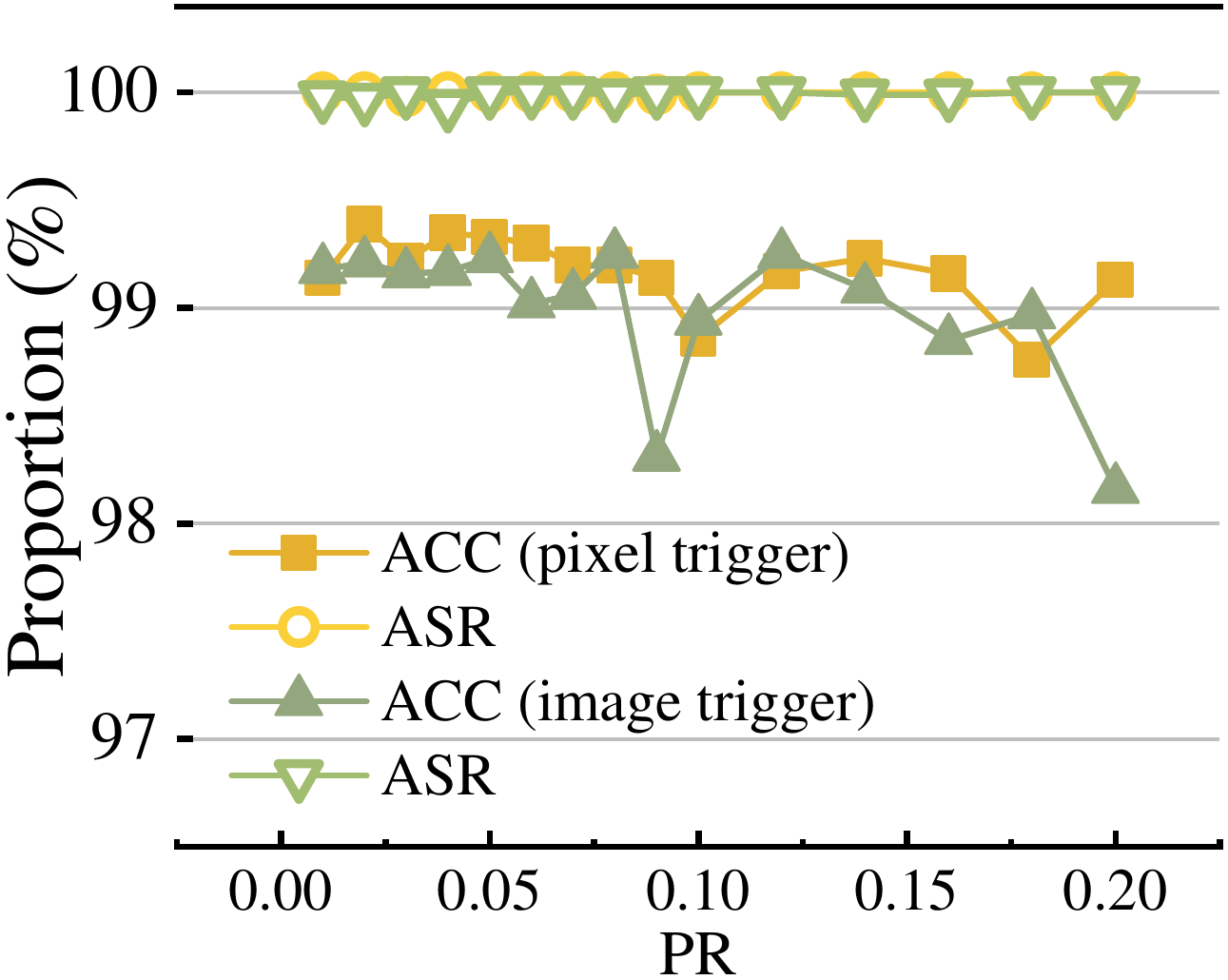}}}\hfill
    \subfigure[ResNet18-MNIST]{\label{fig:LR_B-res18-mnist-if}
    {\includegraphics[width=0.228\textwidth]{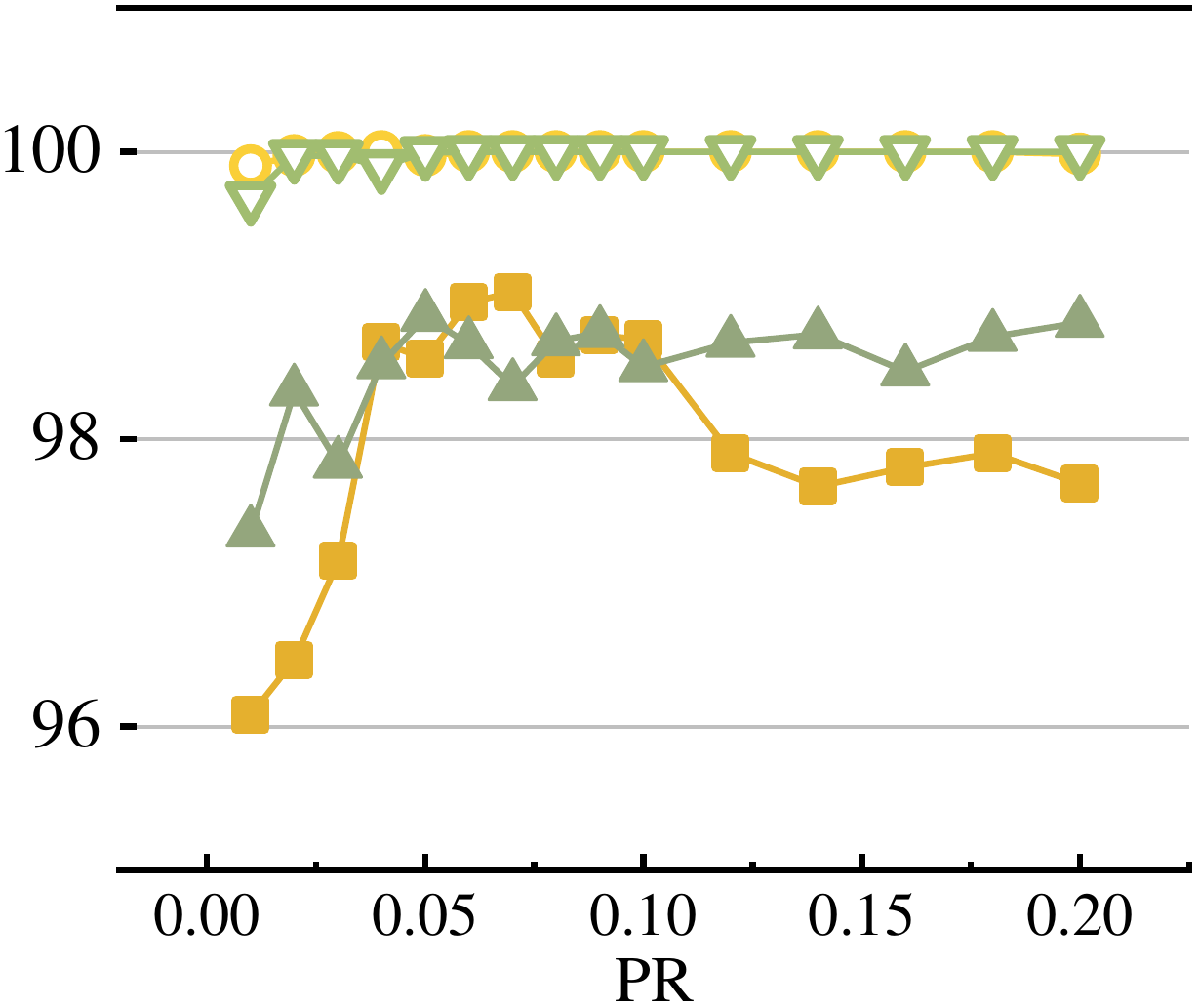}}}\hfill
    \subfigure[VGG11-CIFAR10]{\label{fig:LR_B-vgg11-cifar10-if}    {\includegraphics[width=0.24\textwidth]{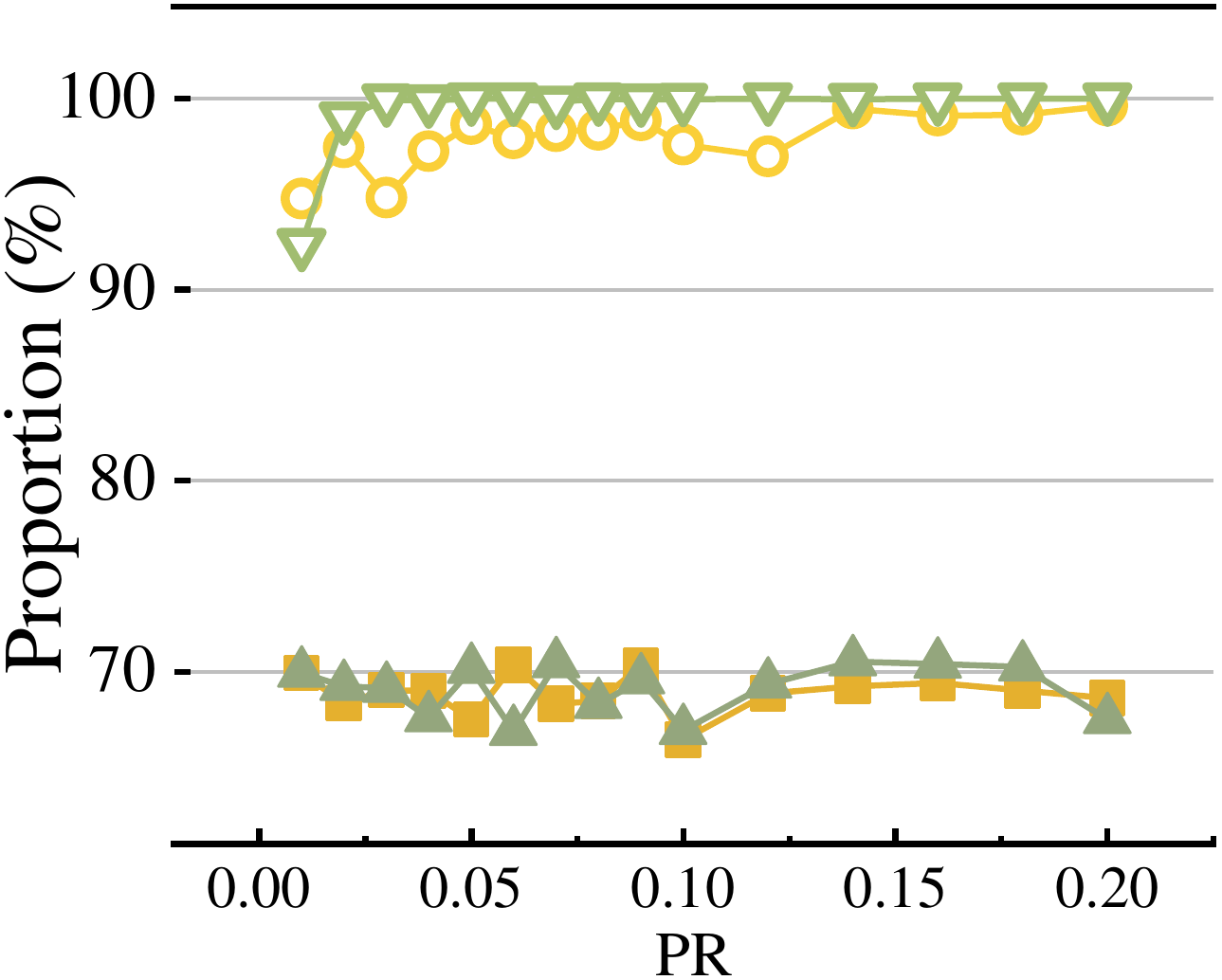}}}\hfill
    \subfigure[ResNet18-CIFAR10]{\label{fig:LR_B-res18-cifar10-if}
    {\includegraphics[width=0.228\textwidth]{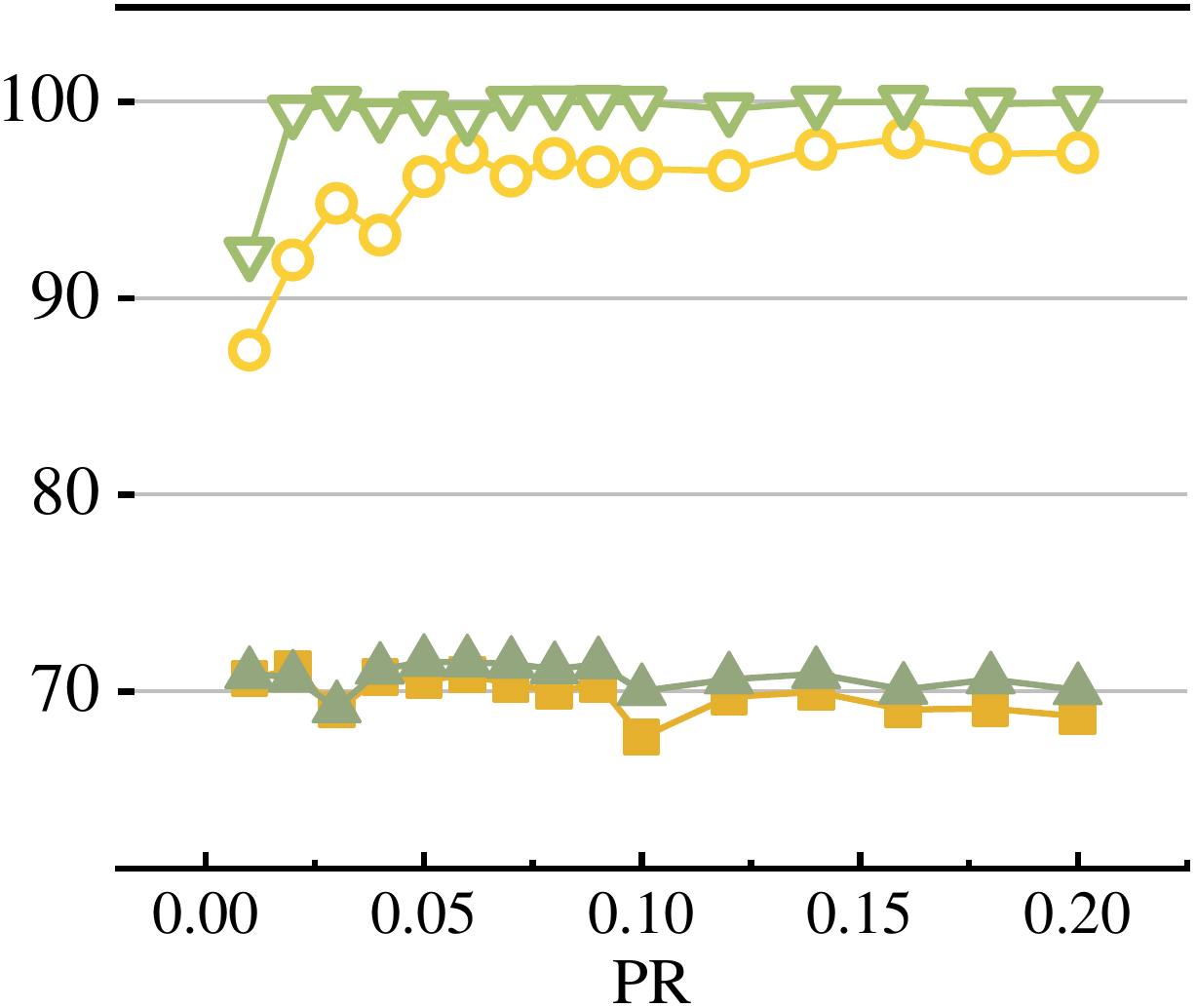}}}\hfill
    \caption{Attack performance on SNNs with IF neurons trained by ($\mathcal{LR_B}$) on MNIST (a)-(b) and CIFAR10 (c)-(d).}
    \label{fig:LR_B_IF}
\end{figure}

\subsubsection{Attack Performance on $\mathcal{LR_B}$}

This section evaluates the vulnerability of SNNs obtained from $\mathcal{LR_B}$ under backdoor attacks. We conduct experiments on different neuron structures (IF and LIF neurons) and dataset types (traditional image dataset and neuromorphic data), to more rigorously assess the vulnerability of SNNs obtained by $\mathcal{LR_B}$ under backdoor attacks.
\begin{figure}[t]
    \footnotesize
    \centering
    \subfigure[MNIST]{\label{fig:LR_B-res18-mnist-Lif}
    {\includegraphics[width=0.241\textwidth]{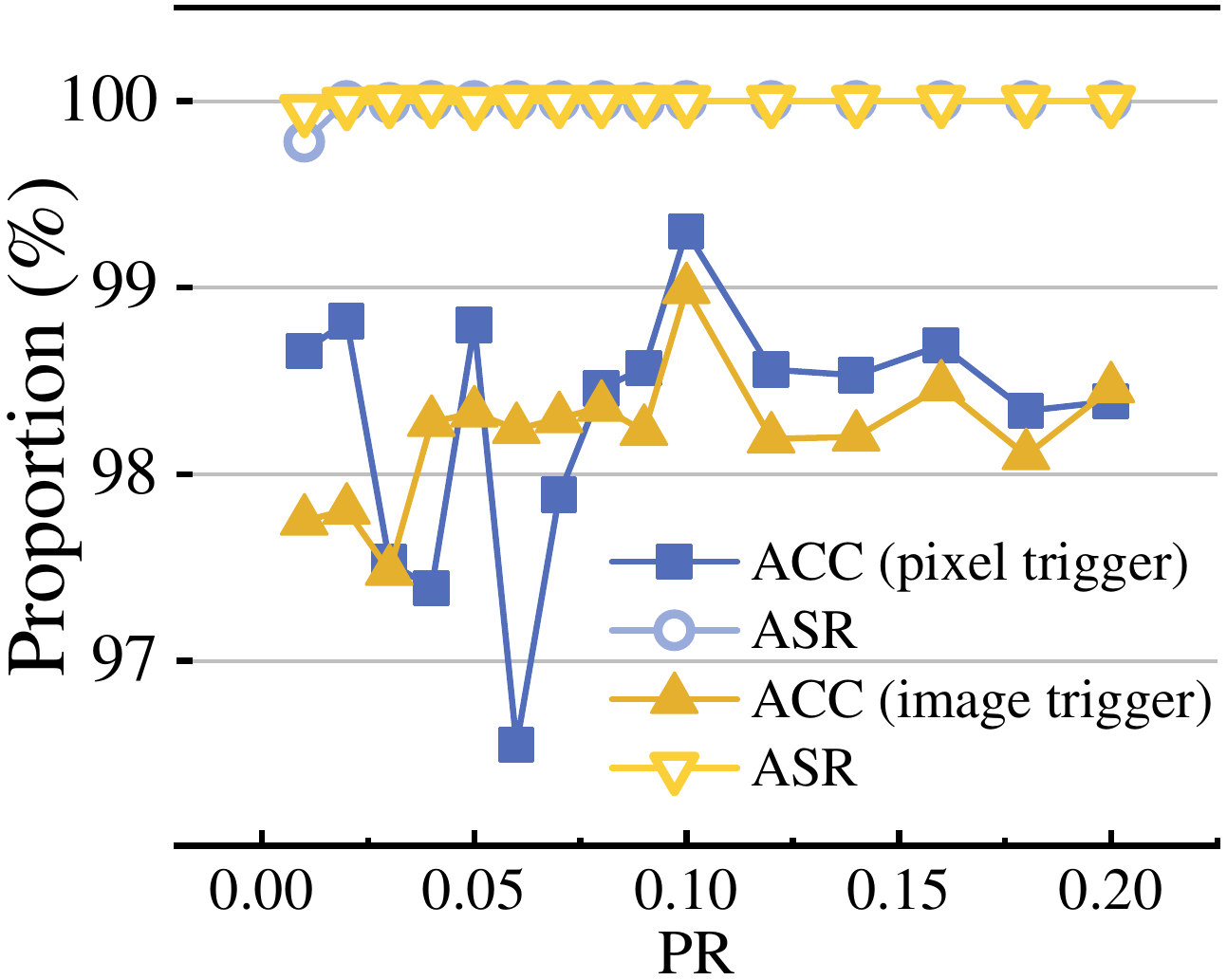}}}\hfill
    \subfigure[CIFAR10]{\label{fig:LR_B-res18-cifar10-lif}
    {\includegraphics[width=0.227\textwidth]{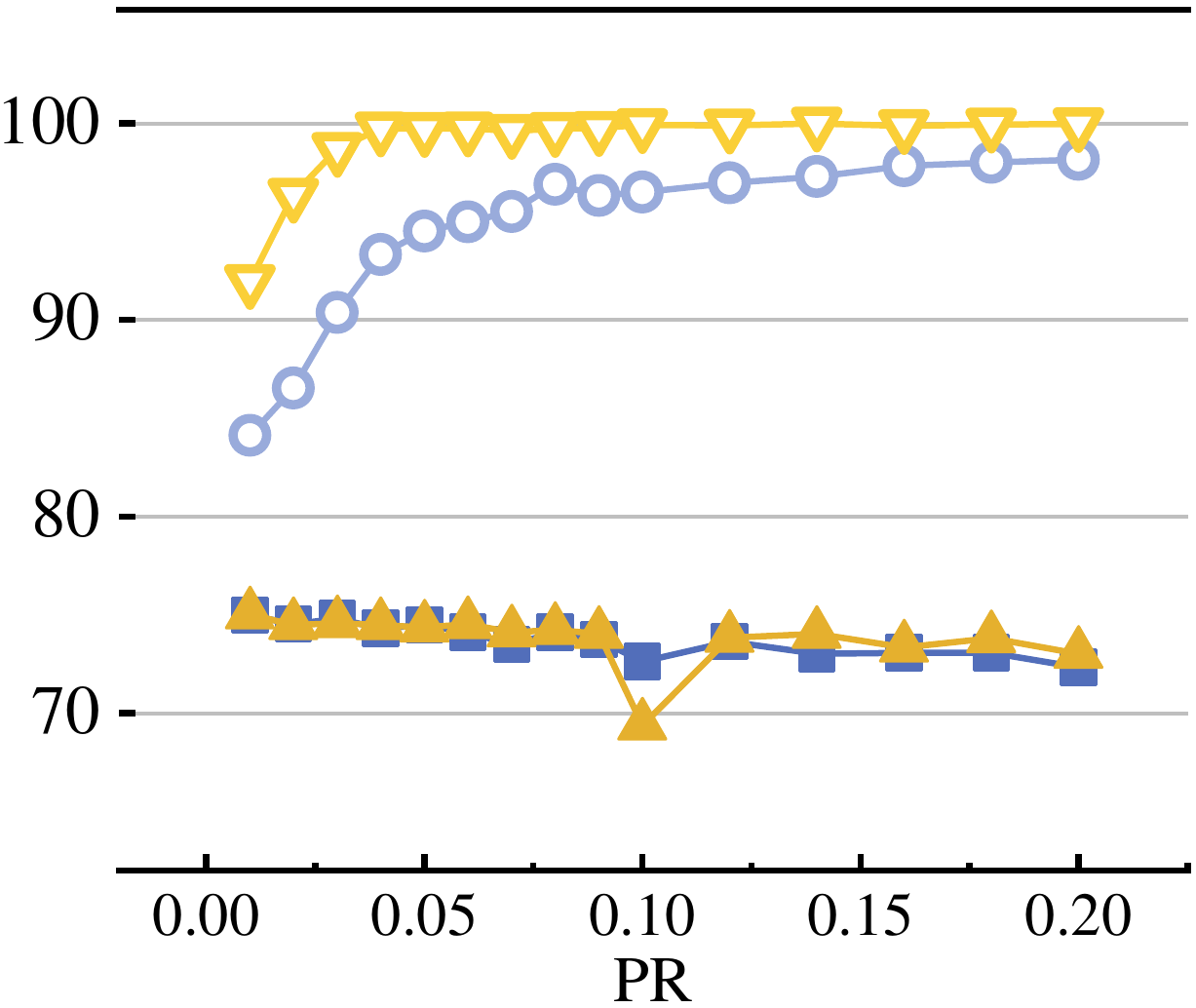}}}\hfill
    \caption{Attack performance on SNNs with LIF neurons trained by $\mathcal{LR_B}$ on MNIST (a) and CIFAR10 (b).}
    \label{fig:LR_B_LIF}
\end{figure}

\begin{figure*}[!t]
    \footnotesize
    \centering
    \subfigure[N-MNIST]{\label{fig:nmnist-lif}
    {\includegraphics[width=0.341\textwidth]{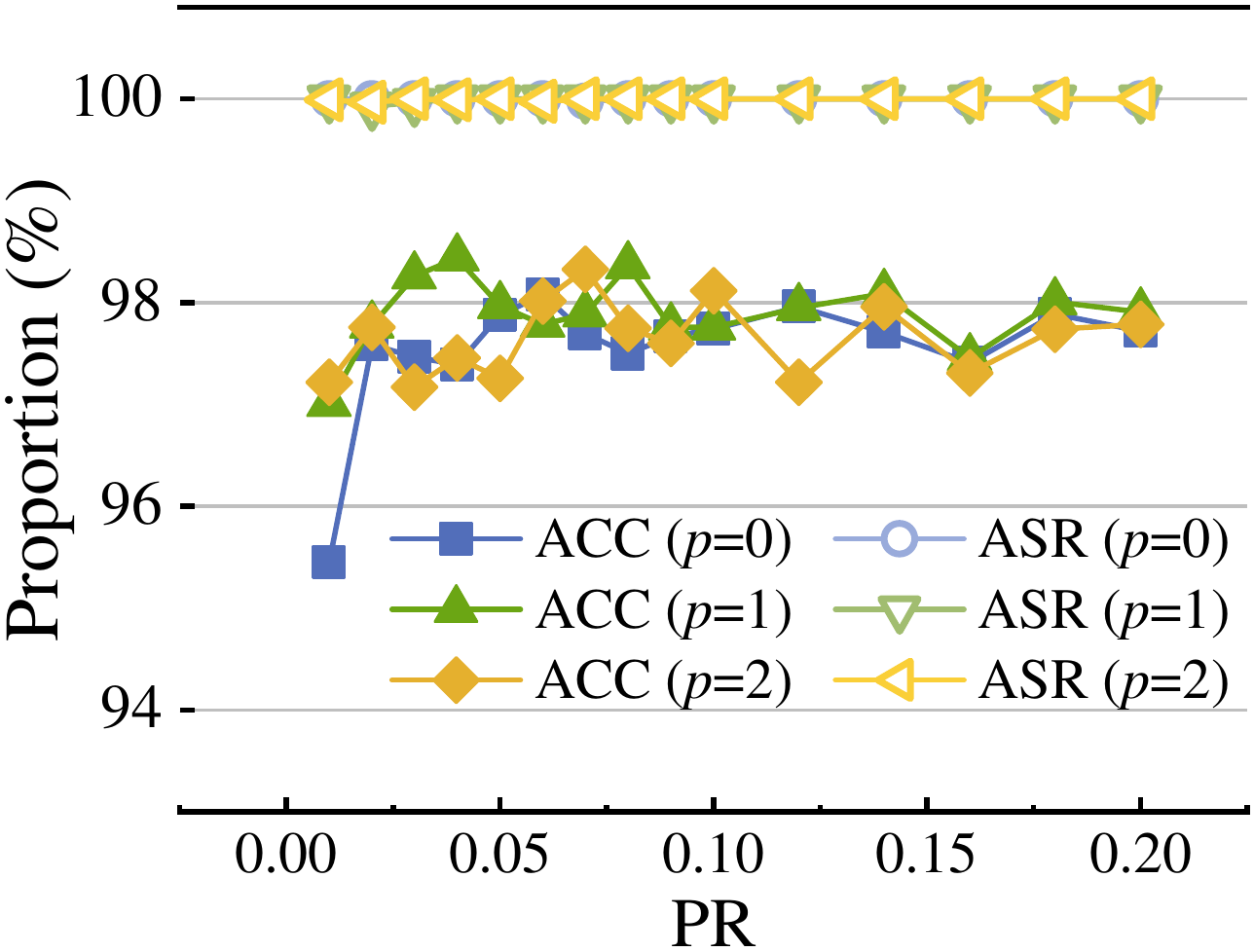}}}\hfill
    \subfigure[CIFAR10-DVS]{\label{fig:cifar10dvs-lif}
    {\includegraphics[width=0.322\textwidth]{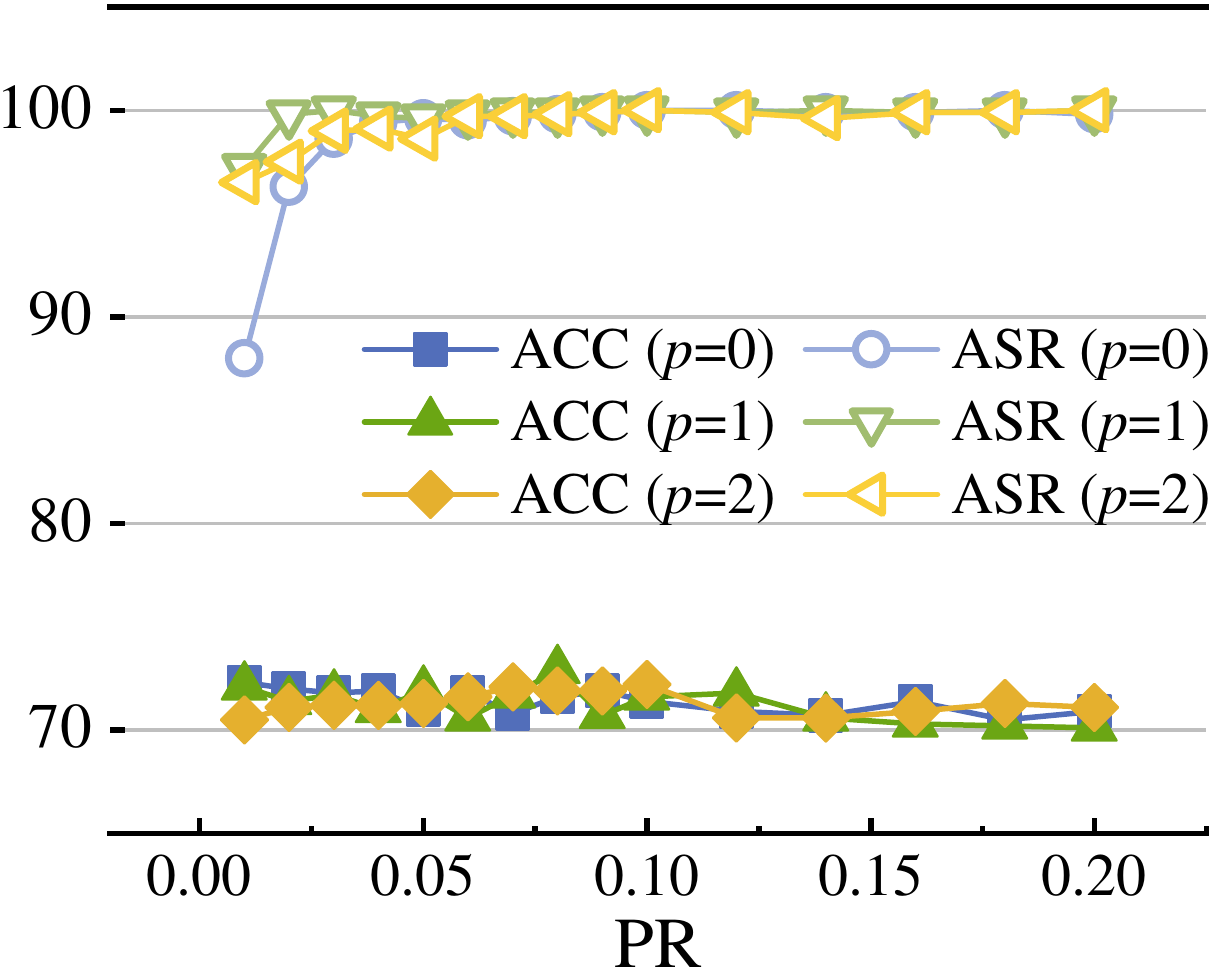}}}\hfill
    \subfigure[DVSGesture]{\label{fig:dvsgesture-lif}
    {\includegraphics[width=0.325\textwidth]{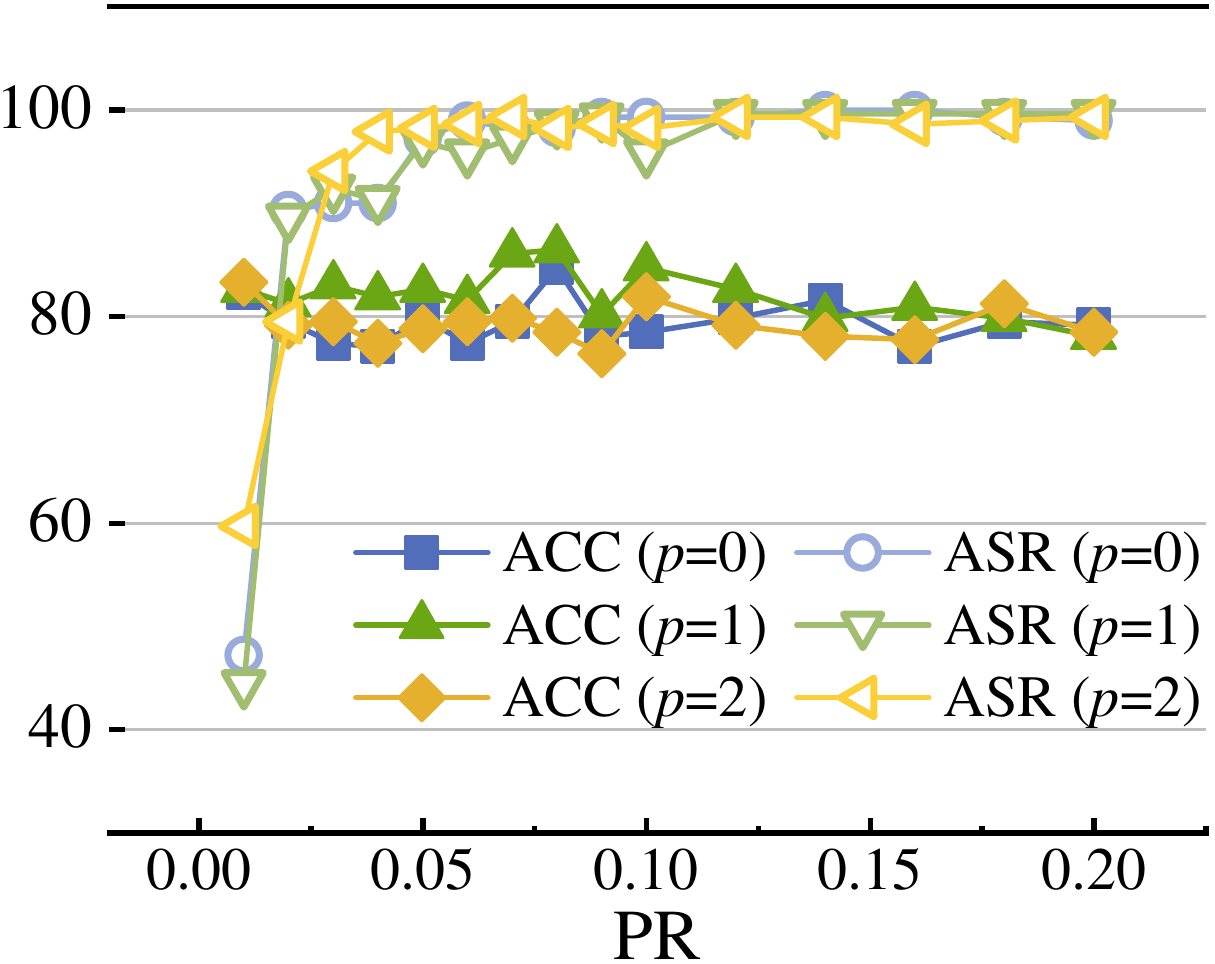}}}\hfill
    \caption{Backdoor attack performance on VGG11 with LIF neurons trained by $\mathcal{LR_B}$ on N-MNIST ~\cite{N-MNIST}, CIFAR10-DVS ~\cite{CIFAR10-DVS}, and DVS128Gesture ~\cite{DVS128Gesture}.}
    \label{fig:LR_B-Neuromorphic-dataset}
\end{figure*}

For traditional image datasets (MNIST ~\cite{MNIST} and CIFAR10 ~\cite{CIFAR10}), we conduct experiments under spike-based ResNet18 and VGG11 composed of IF neurons and LIF neurons, respectively. The experimental results for the model consisting of IF neurons are recorded in Fig. \ref{fig:LR_B_IF}. With the $PR$ increase, ASR shows a significant upward trend in the initial stage, eventually stabilizing at a high value. Under each experimental setting, the final ASR can approach to 100\%, with no significant degradation in the accuracy. Even when $PR=0.01$, the ASR can get more than 90\%. As a result, VGG11 and ResNet18, composed of IF neurons trained by $\mathcal{LR_B}$, have extremely low robustness against backdoor attacks based on data poisoning. Moreover, the mode of the trigger pattern affects the final attack performance on CIFAR10. Specifically, the pixel triggers outperform the image triggers in the attacks on VGG11 and ResNet18. Nevertheless, this phenomenon is not found in MNIST, which only consists of black and white pixels. Therefore, we believe that it is caused by the fact that the complex pixel composition of the image triggers can be misrecognized as pixels in the original samples in complex datasets. 

Furthermore, we conduct an experimental evaluation of ResNet18, composed of LIF neurons. The specific results are recorded in Figure \ref{fig:LR_B_LIF}. As the $PR$ increases, the ASRs against both MNIST and CIFAR10 show a similar increasing trend. Notably, ResNet18, constructed up of LIF neurons, shows higher ACC and lower ASR under the CIFAR10 dataset than ResNet18, made up of IF neurons. This occurs in both trigger modes. Specifically, under the pixel trigger, the average ACC of ResNet 18 with LIF neurons is 3.79\% higher than that of ResNet 18 with IF neurons, and the average ASR is 1.13\% lower. This phenomenon also exists under image triggers. As mentioned in ~\cite{inherent_robustness}, a model composed of LIF has better robustness compared to IF. 
Additionally, we conduct experiments on the VGG model composed of LIF neurons on the neuromorphic datasets. The specific results are recorded in Figure \ref{fig:LR_B-Neuromorphic-dataset}. We find that for a relatively simple dataset (N-MNIST ~\cite{N-MNIST}), $PR=0.01$ is sufficient to achieve a perfect backdoor attack (Figure \ref{fig:LR_B-res18-mnist-Lif}). For relatively complex datasets, the ASR also increases significantly with increasing $PR$ and eventually reaches about 100\%, without model accuracy being seriously affected. To summarize, the models obtained by $\mathcal{LR_B}$ exhibit extremely low robustness against backdoor attacks, both under traditional and neuromorphic datasets, and both under IF and LIF neurons.

\begin{figure}[!b]
\footnotesize
\centering
\subfigure[VGG11-MNIST]{\label{fig:ANN2SNN-VGG11-MNIST}
{\includegraphics[width=0.233\textwidth]{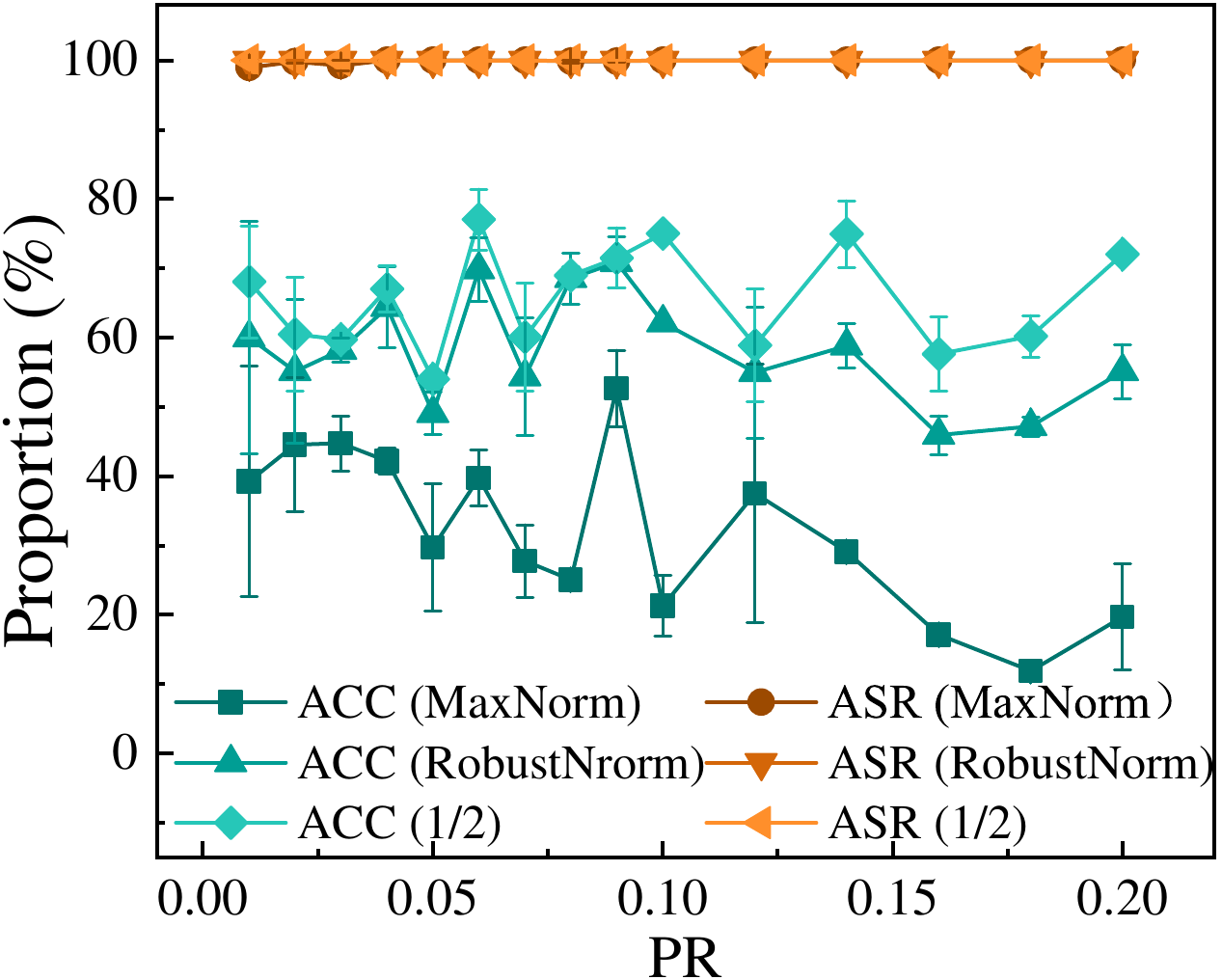}}}\hfill
\subfigure[VGG11-CIFAR10]{\label{fig:ANN2SNN-VGG11-CIFAR10}
{\includegraphics[width=0.233\textwidth]{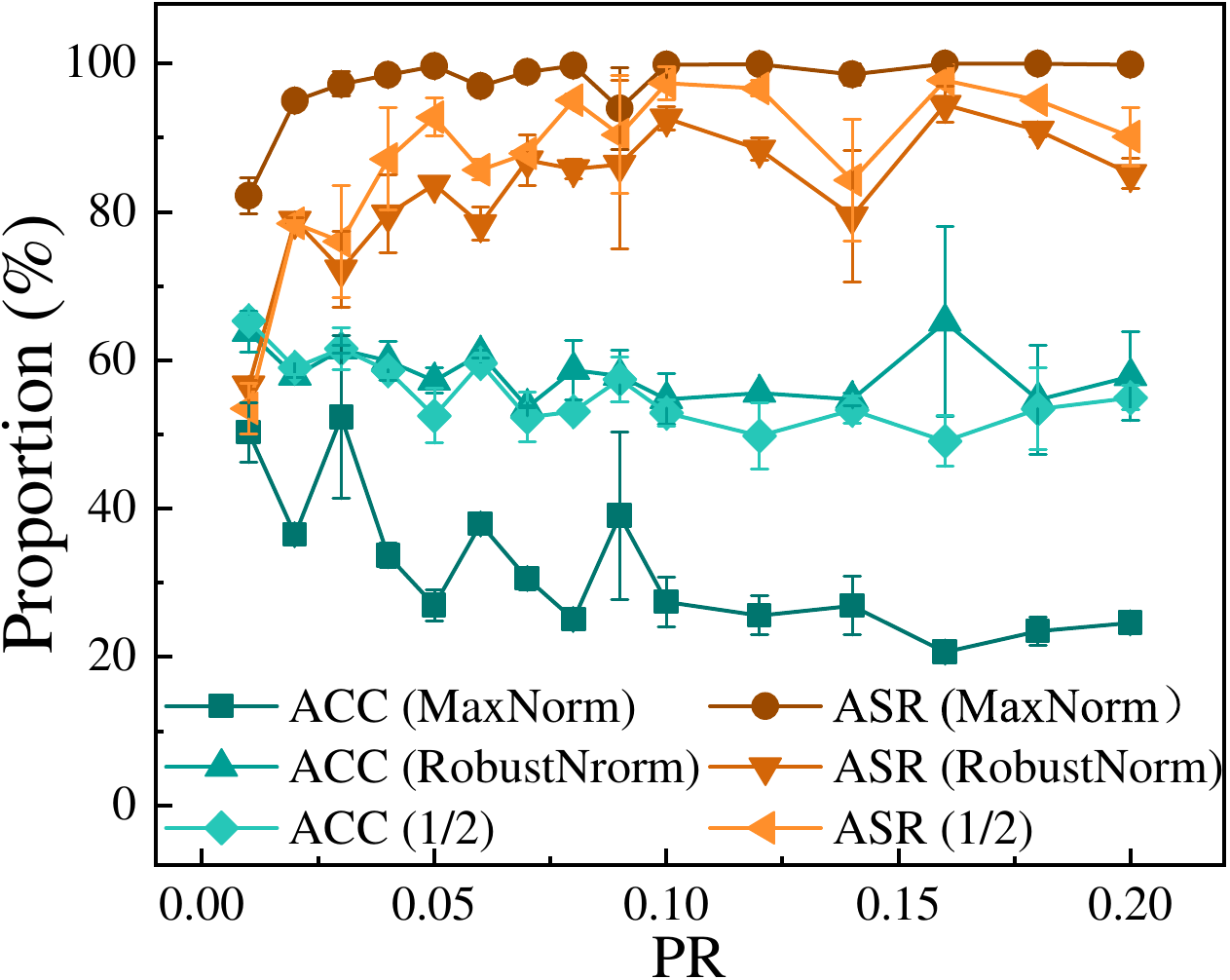}}}\hfill
\subfigure[ResNet18-MNIST]{\label{fig:ANN-RES18-MNIST}
{\includegraphics[width=0.235\textwidth]{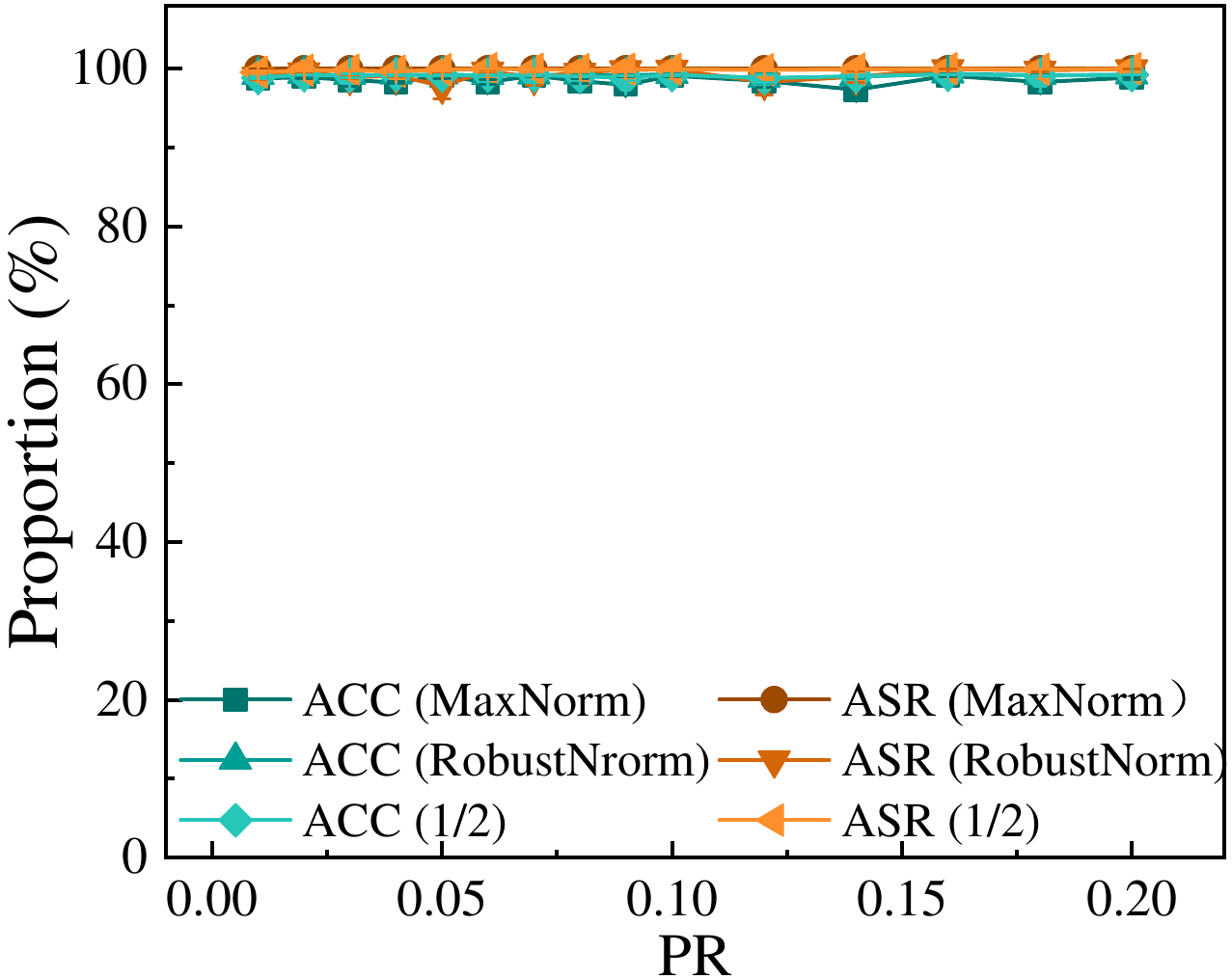}}}\hfill
\subfigure[ResNet18-CIFAR10]{\label{fig:ANN2SNN-RES18-CIFAR10}
{\includegraphics[width=0.231\textwidth]{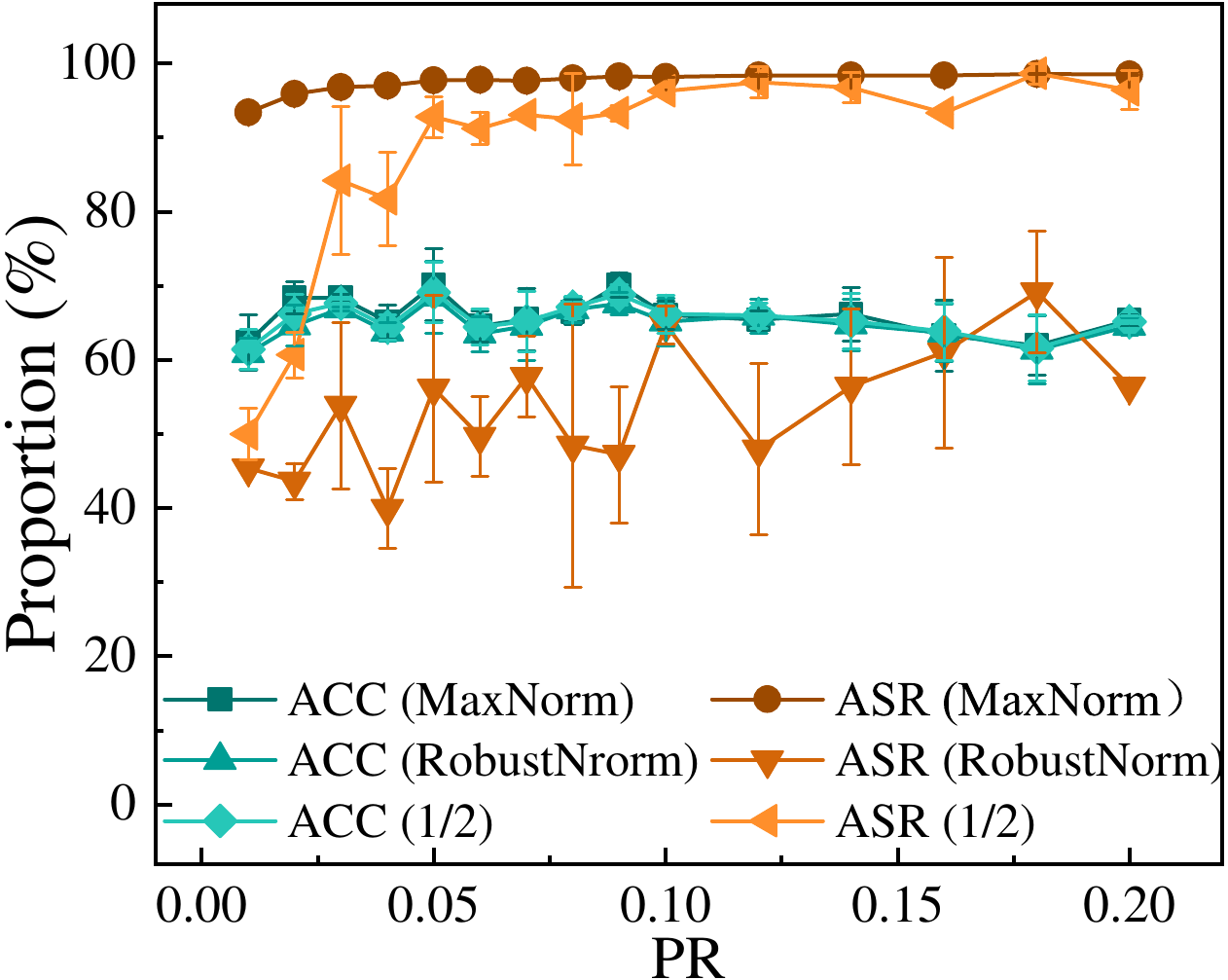}}}\hfill
\caption{Attack performance on converted SNNs trained by $\mathcal{LR_C}$ with different conversion strategies and poisoning rates. }
\label{fig:conversion}
\end{figure}


\subsubsection{Attack Performance on $\mathcal{LR_C}$}
\begin{figure}[!b]
\footnotesize
\centering
\subfigure[ACC of MNIST]{\label{fig:ACC-MV}
{\includegraphics[width=0.234\textwidth]{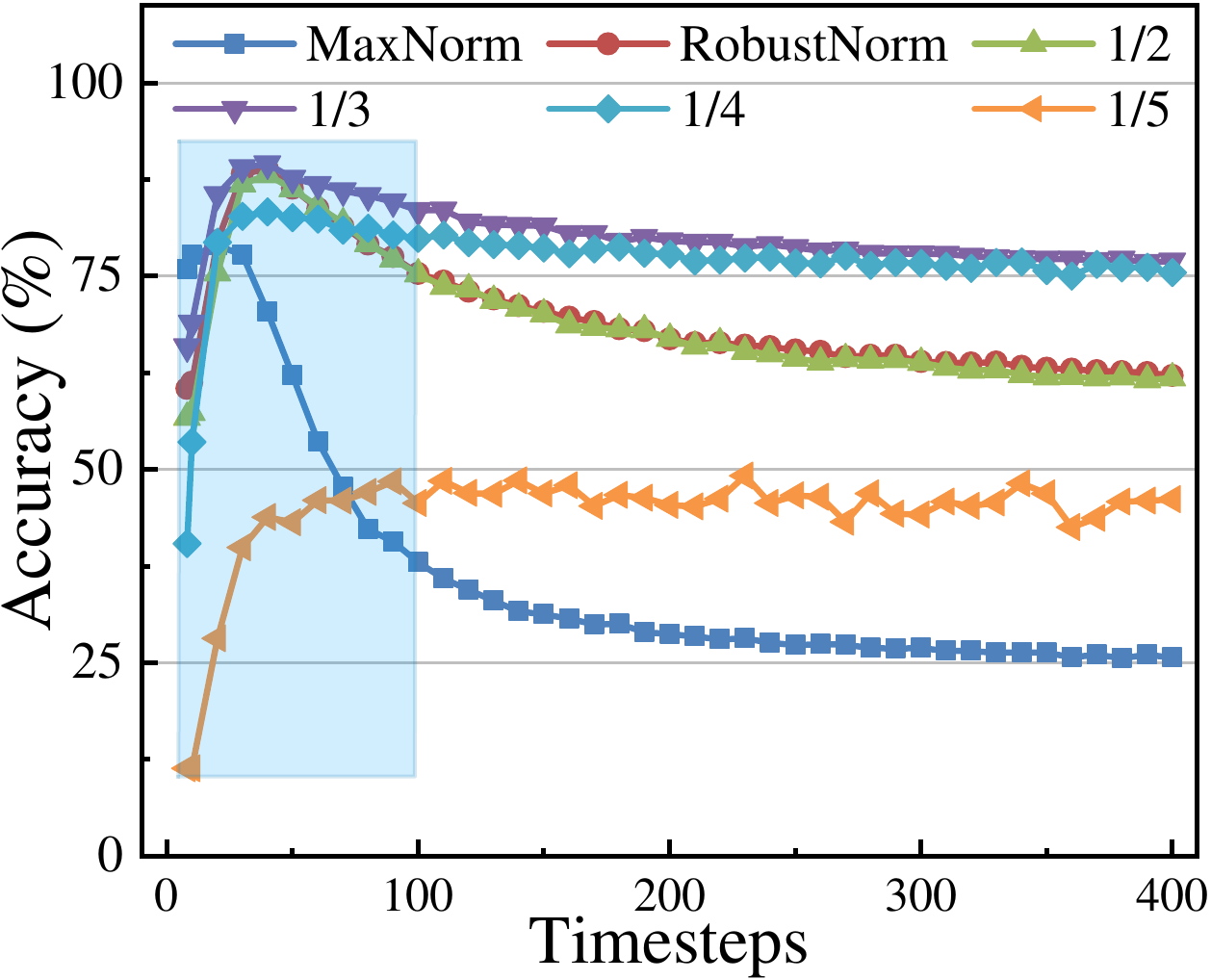}}}\hfill
\subfigure[ASR of MNIST]{\label{fig:ASR-MV}
{\includegraphics[width=0.234\textwidth]{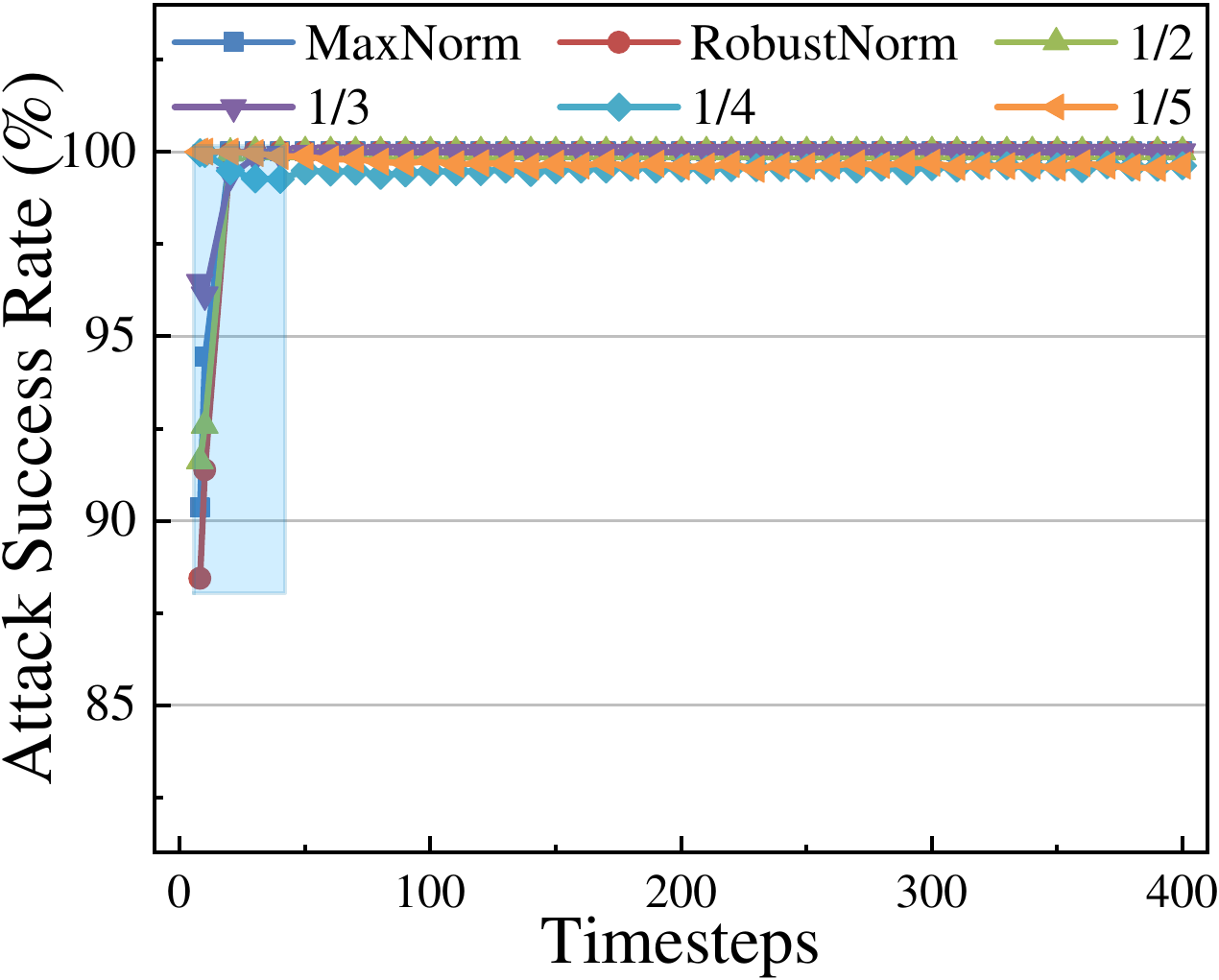}}}\hfill
\subfigure[ACC of CIFAR10]{\label{fig:ACC-CV}
{\includegraphics[width=0.234\textwidth]{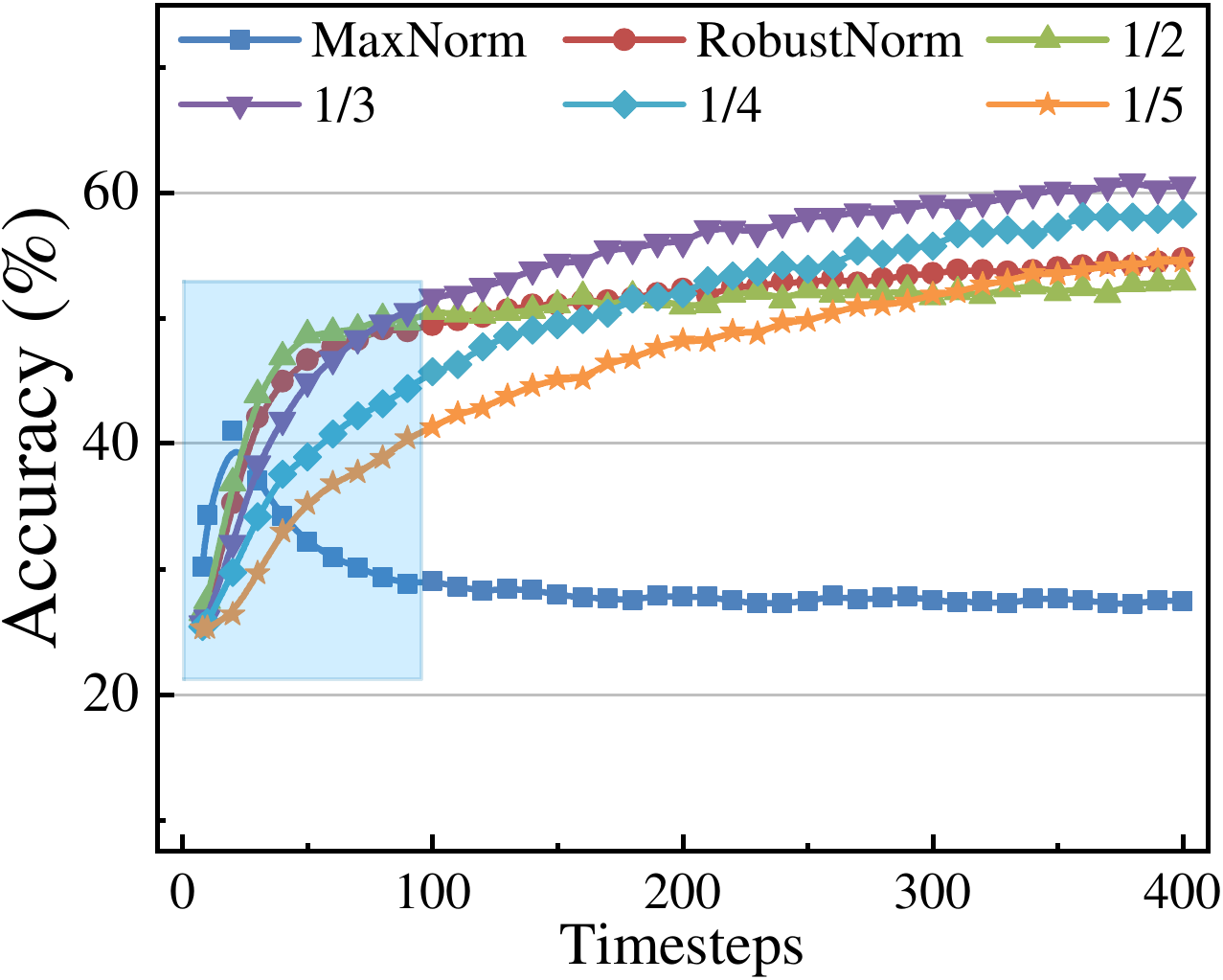}}}\hfill
\subfigure[ASR of CIFAR10]{\label{fig:ASR-CV}
{\includegraphics[width=0.234\textwidth]{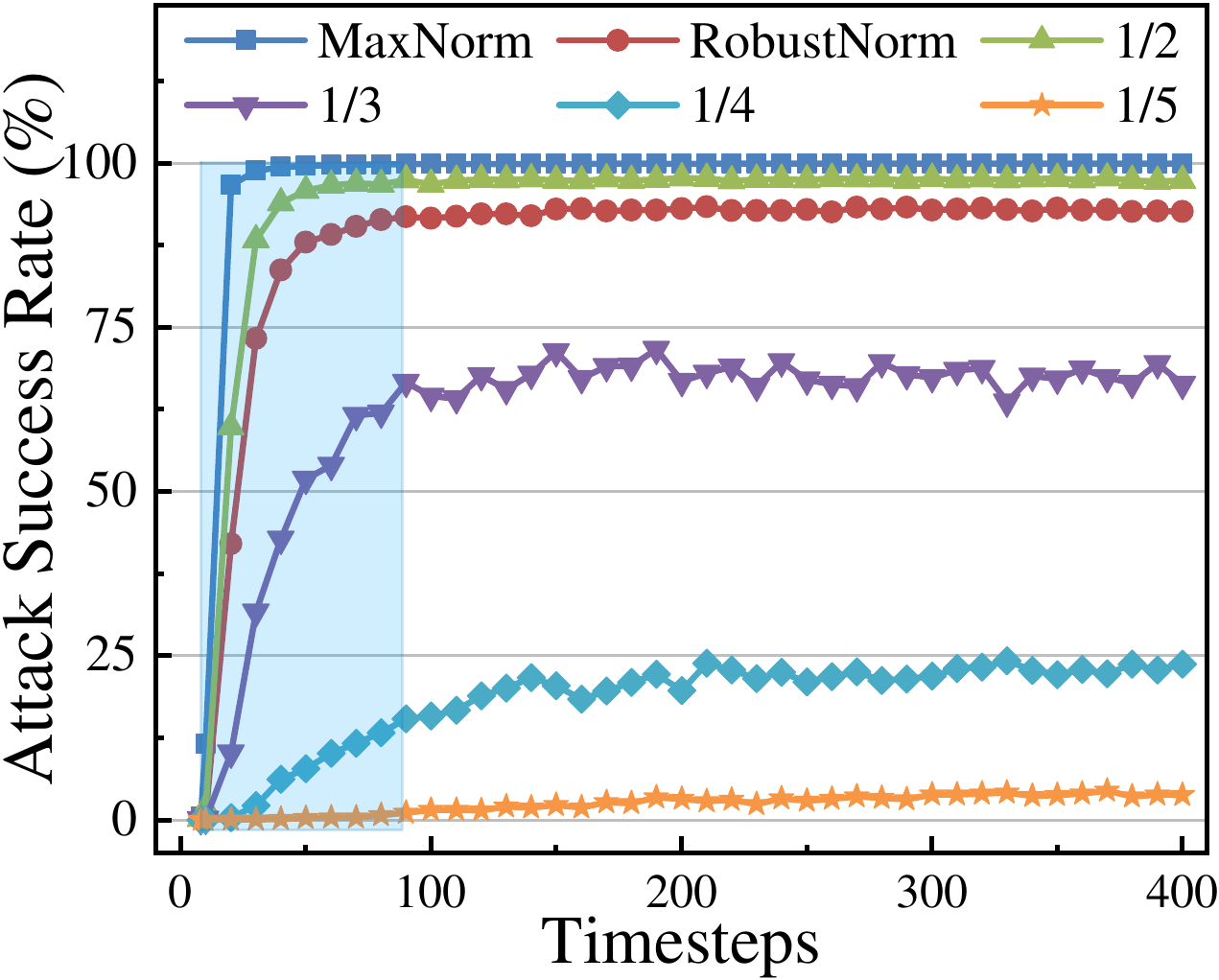}}}\hfill
\caption{Attack performance on VGG11 trained by $\mathcal{LR_C}$ with different conversion strategies and increasing timesteps.}
\label{fig:conversion-timestep}
\end{figure}
This section evaluates the vulnerability of models trained by $\mathcal{LR_C}$ when exposed to backdoor attacks. We used different conversion strategies (including MaxNorm ~\cite{MaxNorm}, RobustNorm ~\cite{RobustNorm}), and Spike-Norm ~\cite{Spike-Norm}) for the same backdoored ANN model on MNIST and CIFAR10 datasets for VGG11 and ResNet18 model structures. We conduct extensive experiments and statistics on the possible performance errors to avoid casualties in the conversion method. Figure \ref{fig:conversion} records the specific experimental results. As seen from the data, SNN can ideally inherit the backdoor information in ANN in most cases. Specifically, the converted SNN can achieve an ASR close to 100\% when facing the same backdoor attack. This indicates that the backdoor information hidden in the original backdoored ANN can be almost wholly preserved during the conversion process and also shows a severe security threat in $\mathcal{LR_C}$. Additionally, we observe significant differences in the performance of the converted SNNs under different conversion methods or scaling. For example, the VGG11 model accuracy decreases significantly when using MaxNorm and RobustNorm, but improves significantly when using Spike-Norm with scale=\{1/3, 1/4, 1/5\} (Figure \ref{fig:conversion-sup}), although the ASR may decrease slightly. We believe that this is because the conversion methods need to refer to the weight of the backdoored ANN, including both clean and malicious weights, and each method has a different strategy for weight selection, which finally leads to different preservation degrees of backdoor information. Additionally, we find that the performance of the backdoored VGG11 (including ACC and ASR) may fluctuate severely at $T=\{0-100\}$, after which it stabilizes (Figure \ref{fig:conversion-timestep}). This phenomenon further illustrates the requirement of large timesteps for conversion-based learning rules. This phenomenon also seems to be related to the model structure, as we did not find it in the experiments on ResNet18 (Figure \ref{fig:conversion-sup}).

\subsubsection{Attack Performance on $\mathcal{LR_H}$}
\label{Sec:LR_H}
\begin{figure}[!b]
\footnotesize
\centering
\subfigure[VGG11 on CIFAR10]{\label{fig:stdb-vgg11}
{\includegraphics[width=0.24\textwidth]{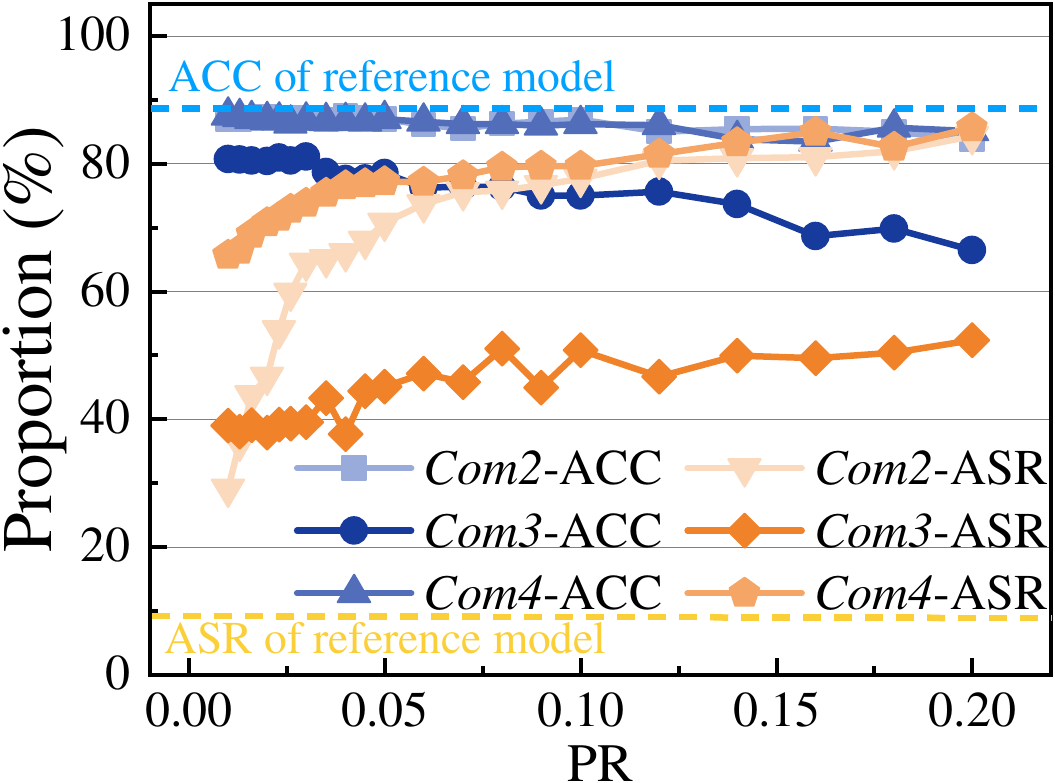}}}\hfill
\subfigure[ResNet18 on CIFAR10]{\label{fig:stdb-res18}
{\includegraphics[width=0.228\textwidth]{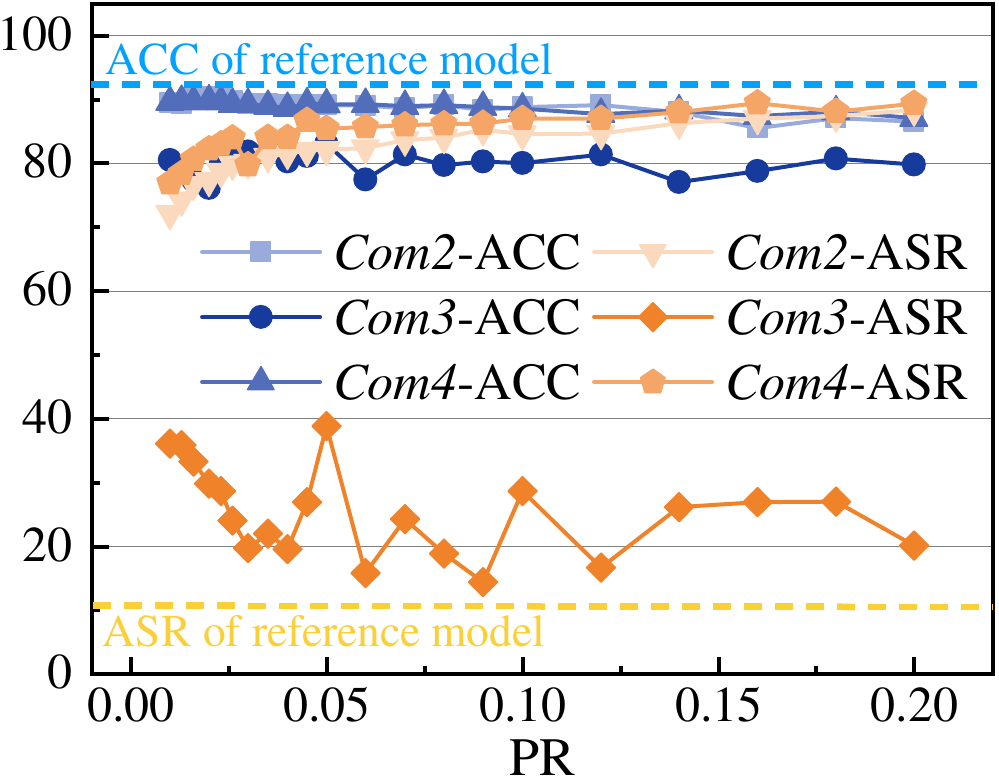}}}\hfill
\caption{Attack performance on SNNs trained by $\mathcal{LR_H}$ with three combinations on CIFAR10.}
\label{fig:STDB-CIFAR10}
\end{figure}
This section demonstrates the vulnerability of SNNs trained with $\mathcal{LR_H}$ under backdoor attacks by validating the model performance of the model generated by three malicious combinations (\emph{Coms} 2-4 mentioned in Table \ref{tab:poi_com}). $Com1$ in Table \ref{tab:poi_com} is used as a reference model, for it generates an immaculate model. We implement different poisoning rates for CIFAR-10 and conduct backdoor attacks on VGG11 and ResNet18. Figure \ref{fig:STDB-CIFAR10} shows the specific experimental results. The performance of the clean model corresponding to $Com1$ is marked with a dashed line in Figure \ref{fig:STDB-CIFAR10}. Specifically, VGG11 has an ACC of 89.61\% and an ASR of 9.96\%, while ResNet18 has an ACC of 92.18\% and an ASR of 10.2\%. 
The general trend of Figures \ref{fig:stdb-vgg11} and \ref{fig:stdb-res18} with different poisoning rates, the ASR under different combinations (orange-colored lines) almost always shows a significant increase as the poisoning rate increases. At the same time, the decrease in the model accuracy is not apparent. Moreover, backdoored models trained by $Com3$ or $Com4$ can achieve high ASR (>85\%) while maintaining good model accuracy. Even when the training set contains only 1\% malicious data ($PR=0.01$), the model cannot rely on its inherent robustness to effectively defend against backdoor attacks. Taking ResNet18 as an example, as seen from the data of $Com3$ in Figure \ref{fig:stdb-res18}, the ASR reaches about 76\% when $PR=0.01$. This illustrates that SNNs trained by $\mathcal{LR_H}$ are vulnerable to backdoor attacks. 

However, the backdoored model obtained from $Com2$ seems to deviate slightly from the expectation of a high ASR. Therefore, we further analyze this deviation and find that it is due to the characteristics of $\mathcal{LR_H}$ during training. Typically, when ANNs with backdoors are converted to SNNs, the weights of ANNs with malicious information are saved in the thresholds of the converted SNNs. However, $\mathcal{LR_H}$ adds the step of fine-tuning after conversion, which can correct the threshold deviation in the SNN using the clean data to some extent. Consequently, the backdoor information hidden in SNNs is indirectly repaired. This is an exciting finding as it can serve as an inspiration for defense measures.


\begin{table*}[t]
\centering
\caption{Comparison of model performance after backdoor attack on ANN and SNN under the same experimental settings.}
\begin{tabular}{c|c|cc|cc|cc|cc|c}
\hline \hline
\multirow{2}{*}{\textbf{Model}}    & \multirow{2}{*}{\textbf{Dataset}} & \multicolumn{4}{c|}{\textbf{ACC (\%)}}                                    & \multicolumn{2}{c|}{\textbf{$\Delta_{ACC}$ (\%)}} & \multicolumn{2}{c|}{\textbf{ASR (\%)}} & \multirow{2}{*}{\textbf{$\Delta_{ASR}$ (\%)}} \\ \cline{3-10}
                                   &                                   & \textbf{$\mathcal{M}_{a}(X)$} & \textbf{$\hat{\mathcal{M}}_{a}(X)$} & \textbf{$\mathcal{M}_{s}(X)$} & \textbf{$\hat{\mathcal{M}}_{s}(X)$} & \textbf{$\hat{\mathcal{M}}_{a}(X)$}        & \textbf{$\hat{\mathcal{M}}_{s}(X)$}       & \textbf{$\hat{\mathcal{M}}_{a}(\hat{X})$}     & \textbf{$\hat{\mathcal{M}}_{s}(\hat{X})$}    &                                          \\ \hline
\multirow{2}{*}{\textbf{VGG11}}    & \textbf{MNIST}                    & 99.99              & 99.51          & 99.08              & 98.94          & \textbf{-0.48}                & \textbf{-0.14}        & 100                & 100               & \textbf{0}                               \\
                                   & \textbf{CIFAR10}                  & 90.38              & 80.16          & 67.68              & 66.96          & \textbf{-10.22}                 & \textbf{-0.72}        & 97.41              & 97.59             & \textbf{+0.18}                           \\ \hline
\multirow{2}{*}{\textbf{ResNet18}} & \textbf{MNIST}                    & 99.99              & 99.41          & 98.98              & 98.69          & \textbf{-0.58}                  & \textbf{-0.29}        & 100                & 100               & \textbf{0}                               \\
                                   & \textbf{CIFAR10}                  & 93.41              & 80.01          & 70.92              & 70.02          & \textbf{-13.4}                  & \textbf{-0.9}         & 98.27              & 99.97             & \textbf{+1.7}                            \\ \hline \hline
\end{tabular}
\label{tab:ANNvsSNN}
\end{table*}

\begin{figure}[!t]
    \centering
    \includegraphics[width=0.48\textwidth]{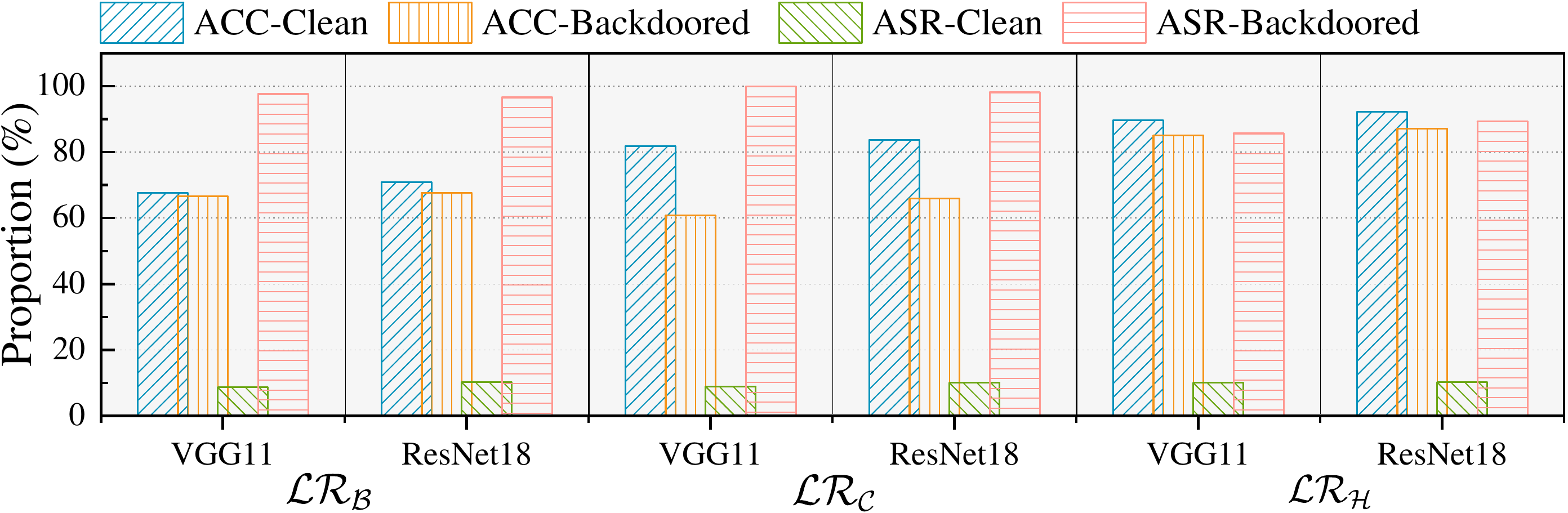}
    \caption{Attack performance on SNNs trained by different learning rules with the same experimental settings. }
    \label{fig:Robustness_different_LR}
\end{figure}

\subsection{Robustness Comparison}

\subsubsection{Different Learning Rules}

This section compares the robustness of models trained by different learning rules, i.e., model performance under backdoor attacks. To ensure comparability of the results, we perform backdoor attacks under the same experimental settings for models trained by different learning rules. Specifically, we select CIFAR10 as the target dataset and single-pixel patterns as triggers. We guarantee $PR=0.1$ and record the performance of models trained by three learning rules under VGG11 and ResNet18 model structures, respectively. The experimental results are shown in Figure \ref{fig:Robustness_different_LR}. The comparison of the performance of the backdoored model and the clean model shows that the backdoored models generated under the three learning rules have substantially higher ASR. Moreover, $\mathcal{LR_B}$ and $\mathcal{LR_C}$ become highly vulnerable when facing backdoor attacks with typical intensity, i.e., the ASR of backdoor attacks against both of them is more than 95\%. The ASR of the backdoor attack against VGG11 trained by $\mathcal{LR_C}$ can even reach 99.92\%, yet this leads to a relatively significant drop in model accuracy (81.68\% --> 60.75\%). Therefore, $\mathcal{LR_B}$ and $\mathcal{LR_C}$ have similar robustness against backdoor attacks; both are weak and unable to defend against backdoor attacks of average strength. In contrast, the model obtained from $\mathcal{LR_H}$ shows higher robustness than models obtained by $\mathcal{LR_B}$ and $\mathcal{LR_C}$. It can be seen that the decrease in the ASR of $\mathcal{LR_H}$ under the same experimental setup reaches about 10\%, and ACC can be increased by about 10\%. 
. Despite this, the ASR of $\mathcal{LR_H}$ is still high and can even reach 90\%, which indicates that the robustness of $\mathcal{LR_H}$ is still not enough to defend against backdoor attacks. To summarize, $\mathcal{LR_B}$ and $\mathcal{LR_C}$ have equal weak robustness when facing backdoor attacks, while $\mathcal{LR_H}$ shows relatively high robustness.

\subsubsection{ANN vs. SNN}

This section compares the robustness of SNNs and ANNs when exposed to backdoor attacks. To ensure the rigidity of the experiments, we perform backdoor attacks of equal intensity ($PR=0.1$) on SNNs and ANNs with the same model structures (VGG11 and ResNet18) and datasets (MNIST and CIFAR10), respectively. Generally, SNNs are considered to have intrinsically more favorable adversarial robustness than ANNs ~\cite{inherent_robustness}, mainly derived from their discrete coding properties and the nonlinear activation computation of LIF neurons. However, based on experimental results, we draw conclusions that are inconsistent with this. Table \ref{tab:ANNvsSNN} records the specific experimental results. The data shows that the ASR of the backdoor attack against SNN reaches 100\% on MNIST and can reach more than 97.5\% on CIFAR-10. In contrast, backdoor attacks against ANNs have lower ASR on CIFAR-10 than SNNs, as shown by bolding the corresponding data in the $\Delta_{ASR}$ column. Simultaneously, the amplitude of accuracy fluctuation of SNN is much smaller than that of ANN ($\Delta_{ACC}$), which further indicates that the backdoor information can be better hidden in the backdoored SNN. Therefore, we believe that the intrinsic robustness of SNN is not enough when facing backdoor attacks. Consequently, we believe that the inherent robustness of SNNs is not superior to that of ANNs when facing backdoor attacks, and SNNs are even weaker than ANNs in some cases.

\subsection{Backdoor Migration Capability}

This section evaluates the migration ability of backdoor information hidden in the backdoored model during the conversion process, which is one of the main reasons for the vulnerability of SNN learning methods. Typically, backdoor information closely linked to triggers is extremely well hidden in the redundant weights of the backdoored ANN. Therefore, the training method that refers to the backdoored ANN to generate the target SNN will have a high risk of learning the hidden backdoor information. We express this conversion process in the formula as $\hat{\mathcal{M}}_{a}\longrightarrow \hat{\mathcal{M}}_{s}$. To verify the migration ability of backdoor information in $\hat{\mathcal{M}}_{a}\longrightarrow \hat{\mathcal{M}}_{s}$, we evaluate both learning rules ($\mathcal{LR_C}$ and $\mathcal{LR_H}$) that need to rely on the conversion step. We select CIFAR10 as the target dataset, and the experimental results are recorded in Table \ref{tab:backdoor_migration_capability}.
\begin{table}[!t]
\small
\centering
\caption{Backdoor migration capability under $\mathcal{LR_C}$ and $\mathcal{LR_H}$ with different poisoning rates.}
\begin{tabular}{c|ccc|ccc}
\hline \hline
\multirow{2}{*}{\textbf{PR}} & \multicolumn{3}{c|}{\textbf{$\mathcal{LR_C}$} (\%)}                                                                                                    & \multicolumn{3}{c}{\textbf{$\mathcal{LR_H}$} (\%)}                                                                  \\ \cline{2-7} 
                             & \textbf{$\hat{\mathcal{M}}_{a}(\hat{X})$} & \multicolumn{1}{c}{\textbf{\begin{tabular}[c]{@{}c@{}}$\hat{\mathcal{M}}_{s}(\hat{X})$\\ (1/2)\end{tabular}}} & \multicolumn{1}{c|}{\textbf{MR}} & \textbf{$\hat{\mathcal{M}}_{a}(\hat{X})$} & \textit{\textbf{\begin{tabular}[c]{@{}c@{}}$\hat{\mathcal{M}}_{s}(\hat{X})$\\ ($Com4$)\end{tabular}}} & \textbf{MR} \\ \hline
0.02                         & 96.09        & 60.67                                                                                & 63.14                          & 95.56        & 82.19                                                                  & 86.01     \\
0.04                         & 96.95        & 81.66                                                                                & 84.22                          & 96.45        & 84.01                                                                  & 87.01     \\
0.06                         & 97.9         & 91.22                                                                                & 93.17                          & 96.78        & 85.69                                                                  & 88.54    \\
0.08                         & 98.05        & 92.46                                                                                & 94.30                          & 97.44        & 86.12                                                                  & 88.38     \\
0.01                         & 98.37        & 96.26                                                                                & 97.86                          & 97.36        & 86.97                                                                  & 89.33     \\
0.12                         & 98.43        & 97.46                                                                                & 99.01                          & 97.49        & 87.03                                                                  & 89.27     \\
0.14                         & 98.45        & 96.73                                                                                & 98.25                          & 97.81        & 87.99                                                                  & 89.96     \\
0.16                         & 98.72        & 93.32                                                                                & 94.52                          & 98.04        & 89.38                                                                  & 91.17     \\
0.18                         & 98.74        & 98.6                                                                                 & 99.86                          & 97.97        & 88.1                                                                   & 89.93     \\
0.2                          & 98.64        & 96.41                                                                                & 97.73                          & 98.21        & 89.37                                                                  & 91        \\ \hline \hline
\end{tabular}
\label{tab:backdoor_migration_capability}
\end{table}

For $\mathcal{LR_C}$, the backdoor migration rate in $\hat{\mathcal{M}}_{a} \longrightarrow \hat{\mathcal{M}}_{s}$ can always be kept above 90\% (except for $PR=\{0.02, 0.04\}$) with the Spike-Norm method with a scaling of 1/2 as an example. Even the MR can reach over 99\% in some cases, e.g., $PR=\{0.12, 0.18\}$. Furthermore, we find that such a high backdoor migration rate also occurs under other conversion methods, e.g., the MR under MaxNorm can all remain above 99\% (Table \ref{tab:back_migration_sup}). Therefore, the phenomenon of backdoor migration is prevalent in $\mathcal{LR_C}$, and the backdoor information has a strong migration ability, making it sustainable in SNN.
For $\mathcal{LR_H}$, the MR is equally high and can be consistently above 85\%, even up to 91.17\% ($PR=0.16$). As a result, the relatively robust $\mathcal{LR_H}$ cannot avoid backdoor migration during the conversion process. Note that the $\mathcal{LR_H}$ section in Table \ref{tab:backdoor_migration_capability} records the backdoor migration rate in the case of a general backdoor attack, i.e., the malicious dataset is used in both the training process of the backdoored ANN and the conversion process of the SNN (i.e., $Com4$ in the Table \ref{tab:poi_com}). We find that when using a clean dataset as reference dataset for $\hat{\mathcal{M}}_{a} \longrightarrow \hat{\mathcal{M}}_{s}$ (i.e., $Com3$ in the Table \ref{tab:poi_com}), the average MR can decrease from 92.21\% to 23.6\% (Table \ref{tab:back_migration_sup}). Therefore, using the clean dataset as the reference for $\hat{\mathcal{M}}_{a} \longrightarrow \hat{\mathcal{M}}_{s}$ can reduce the migration rate of backdoors to some extent. However, it cannot completely stop the migration of backdoor information in the conversion process.

\begin{figure}[!t]
    \centering
    \includegraphics[width=0.48\textwidth]{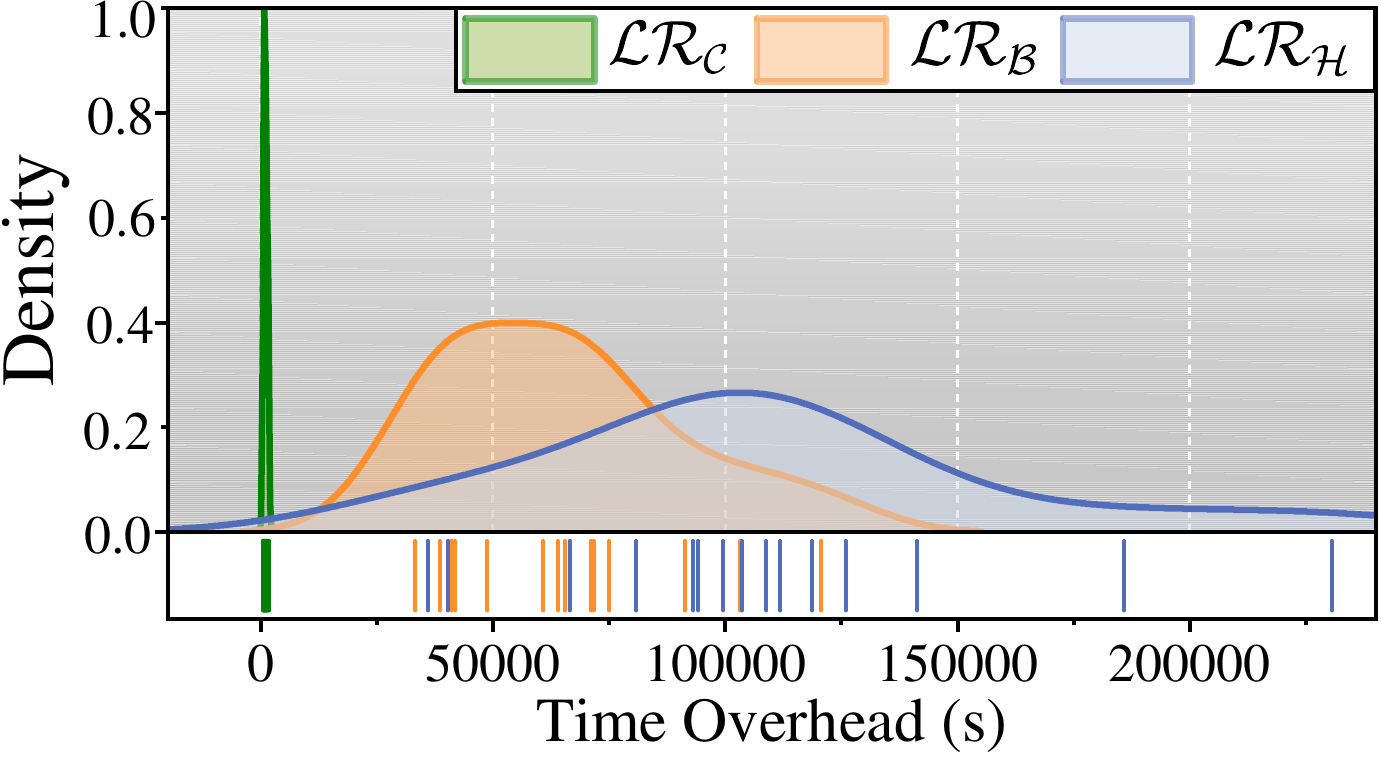}
    \caption{The distribution of computational time overhead for three supervised learning rules with different poisoning rates.}
    \label{fig: computational_time}
\end{figure}

\subsection{Time Overhead of Backdoor Injection}


This section evaluates the time overhead required to inject backdoors for different learning rules. To evaluate the time overhead of the three learning rules in the backdoor injection process more clearly, we perform time statistics on their complete backdoor injection process. This complete process refers to the process from initializing the model (ANN or SNN) to the final output of the backdoored SNN. Figure \ref{fig: computational_time} records the specific time overhead under different learning rules. Obviously, $\mathcal{LR_C}$ has the lowest time overhead, although it requires relatively larger timesteps to ensure model accuracy. Moreover, the time overhead of $\mathcal{LR_B}$ is much larger than $\mathcal{LR_C}$, which is one of the reasons why researchers are actively exploring novel training methods. $\mathcal{LR_H}$ is proposed to solve the problem of the huge time overhead of $\mathcal{LR_B}$. However, we find that $\mathcal{LR_H}$ costs the most time overhead due to its high robustness when performing backdoor injection during iterative training. Furthermore, the STDP-based weight update method of $\mathcal{LR_H}$ also increases the time overhead.

\section{Discussion}


\subsection{Trigger Position and Target Class}



\begin{table}[!t]
\small
\centering
\caption{Attack performance with different trigger positions.}
\begin{tabular}{c|cc|cc|cc}
\hline \hline
\multirow{2}{*}{\textbf{Position}} & \multicolumn{2}{c|}{\textbf{$\mathcal{LR_B}$} (\%)} & \multicolumn{2}{c|}{\textbf{$\mathcal{LR_C}$} (\%)} & \multicolumn{2}{c}{\textbf{$\mathcal{LR_H}$} (\%)} \\ \cline{2-7} 
                                                                                 & \textbf{ACC}     & \textbf{ASR}     & \textbf{ACC}     & \textbf{ASR}    & \textbf{ACC}     & \textbf{ASR}    \\ \hline
top-left                                                                                & 73.21            & 97.43            & 64.22            & 91.76            & 87.48            & 86.79           \\
top-right                                                                                & 73.87            & 95.8            & 63.17            & 91.05           & 87.18            & 83.37           \\
bottom-left                                                                                & 73.29            & 97.62            & 65.45            & 93.75           & 87.29            & 89.95           \\
bottom-right                                                                                & 72.63            & 96.51            & 66.02            & 98.12           & 88.58            & 86.97           \\
\hline \hline
\end{tabular}
\label{tab:trigger_position}
\end{table}
\begin{table}[!t]
\small
\centering
\caption{Attack performance with different target classes.}
\begin{tabular}{c|cc|cc|cc}
\hline \hline
\multirow{2}{*}{\textbf{TC}} & \multicolumn{2}{c|}{\textbf{$\mathcal{LR_B}$} (\%)} & \multicolumn{2}{c|}{\textbf{$\mathcal{LR_C}$} (\%)} & \multicolumn{2}{c}{\textbf{$\mathcal{LR_H}$} (\%)} \\ \cline{2-7} 
                                                                                 & \textbf{ACC}     & \textbf{ASR}     & \textbf{ACC}     & \textbf{ASR}    & \textbf{ACC}     & \textbf{ASR}    \\ \hline
0                                                                                & 73.55            & 96.81            & 66.27            & 90.5            & 87.71            & 87.22           \\
1                                                                                & 72.63            & 96.51            & 66.02            & 98.12           & 88.58            & 86.97           \\
2                                                                                & 74.12            & 95.99            & 62.74            & 94.35           & 87.22            & 87.42           \\
3                                                                                & 74.48            & 96.54            & 69.17            & 92.21           & 88.12            & 88.26           \\
4                                                                                & 74.09            & 95.31            & 62.48            & 91.22           & 87.32            & 86.5            \\
5                                                                                & 73.77            & 95.54            & 59.78            & 90.24           & 87.49            & 85.69           \\
6                                                                                & 73.79            & 96.59            & 61.28            & 96.33           & 87.25            & 86.92           \\
7                                                                                & 73.1             & 96.73            & 59.02            & 90.64           & 87.27            & 86.95           \\
8                                                                                & 73.59            & 97.1             & 63.51            & 93.14           & 87.46            & 86.67           \\
9                                                                                & 73.62            & 94.72            & 64.75            & 93.64           & 86.29            & 87.08           \\ \hline \hline
\end{tabular}
\label{tab:differnet_target_classes}
\end{table}

To explore the impact of trigger position and Target Class (TC) selection on backdoor attacks against SNNs, we conduct experiments on ResNet18 and CIFAR10. The specific experimental results are recorded in Tables \ref{tab:trigger_position} and \ref{tab:differnet_target_classes}. To ensure the stealthiness of the backdoor attack while preserving as many key pixels of the original image as possible, we embed the trigger pattern in the four corners of the image. In Table \ref{tab:trigger_position}, it is evident that different trigger positions affect the victim model in virtually equal measure. Furthermore, we find differences in the performance of backdoor attacks for different learning rules when facing different target classes. As can be seen from Table \ref{tab:differnet_target_classes}, for $\mathcal{LR_B}$ and $\mathcal{LR_H}$, there is almost no difference in attack performance with different target classes, as the ASRs can all be kept near the mean. However, the challenging attack class exists in $\mathcal{LR_C}$. For instance, when TC=$\{0, 5, 7\}$, the ACC and ASR of the converted SNN are significantly lower than other target classes when facing a backdoor attack. We believe this is because SNNs trained by $\mathcal{LR_C}$ rely heavily on the weight transformation of the ANN. However, the weights of ANN reflect how much it learns for each class and generally vary. Therefore, this also gives the converted SNN a certain selection tendency for different classes.

\subsection{Potential Defense Methods}

This section discusses some backdoor defense methods for SNNs, including backdoor detection and elimination. We have preliminarily verified the feasibility of these methods, which can be used as heuristic exploratory work for subsequent defense studies.

\begin{figure}[!t]
    \footnotesize
    \centering
    \subfigure[Clean Model]{\label{fig:poison-clean-voltage}
    {\includegraphics[width=0.24\textwidth]{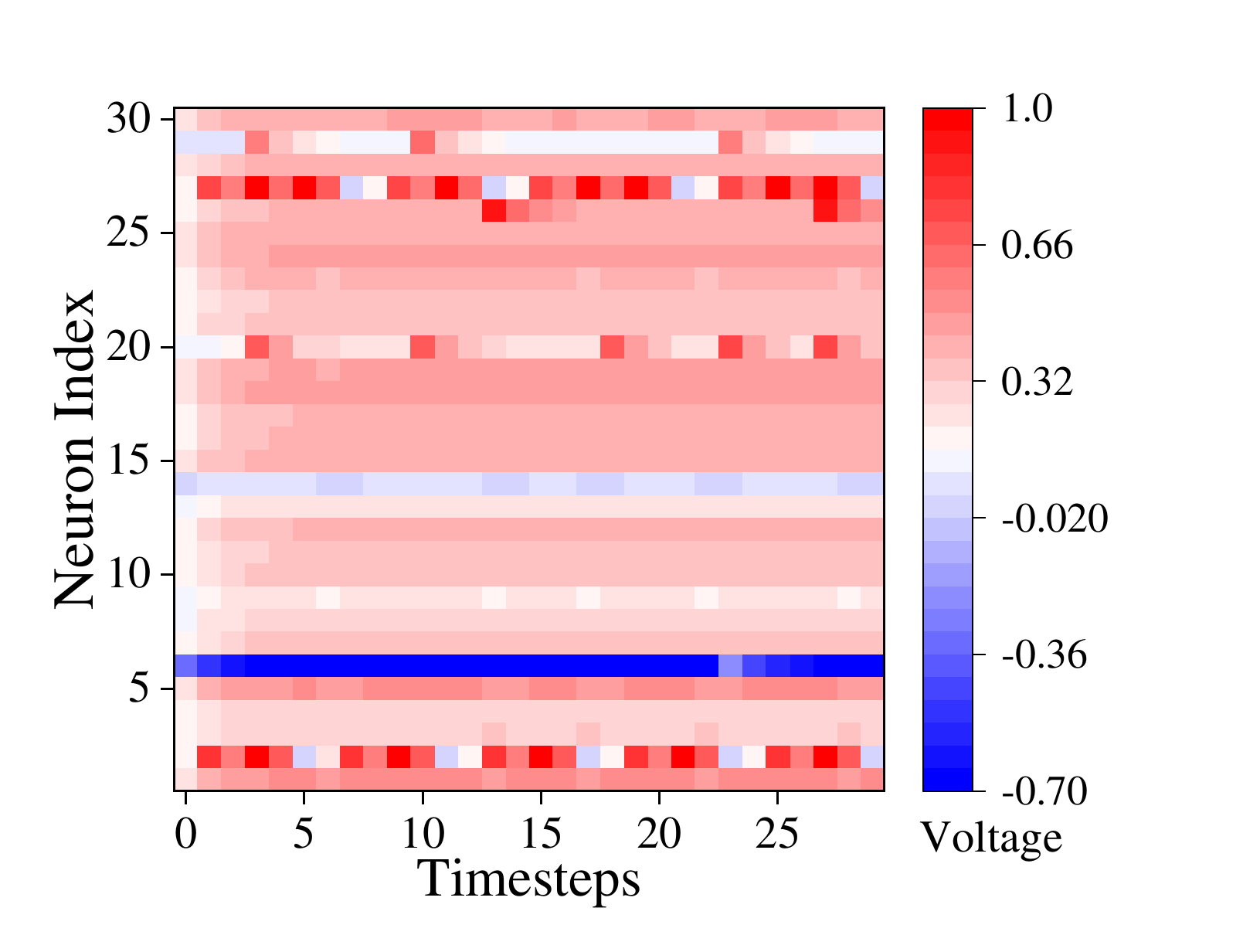}}}\hfill
    \subfigure[Backdoored Model]{\label{fig:poison-poison-voltage}
    {\includegraphics[width=0.228\textwidth]{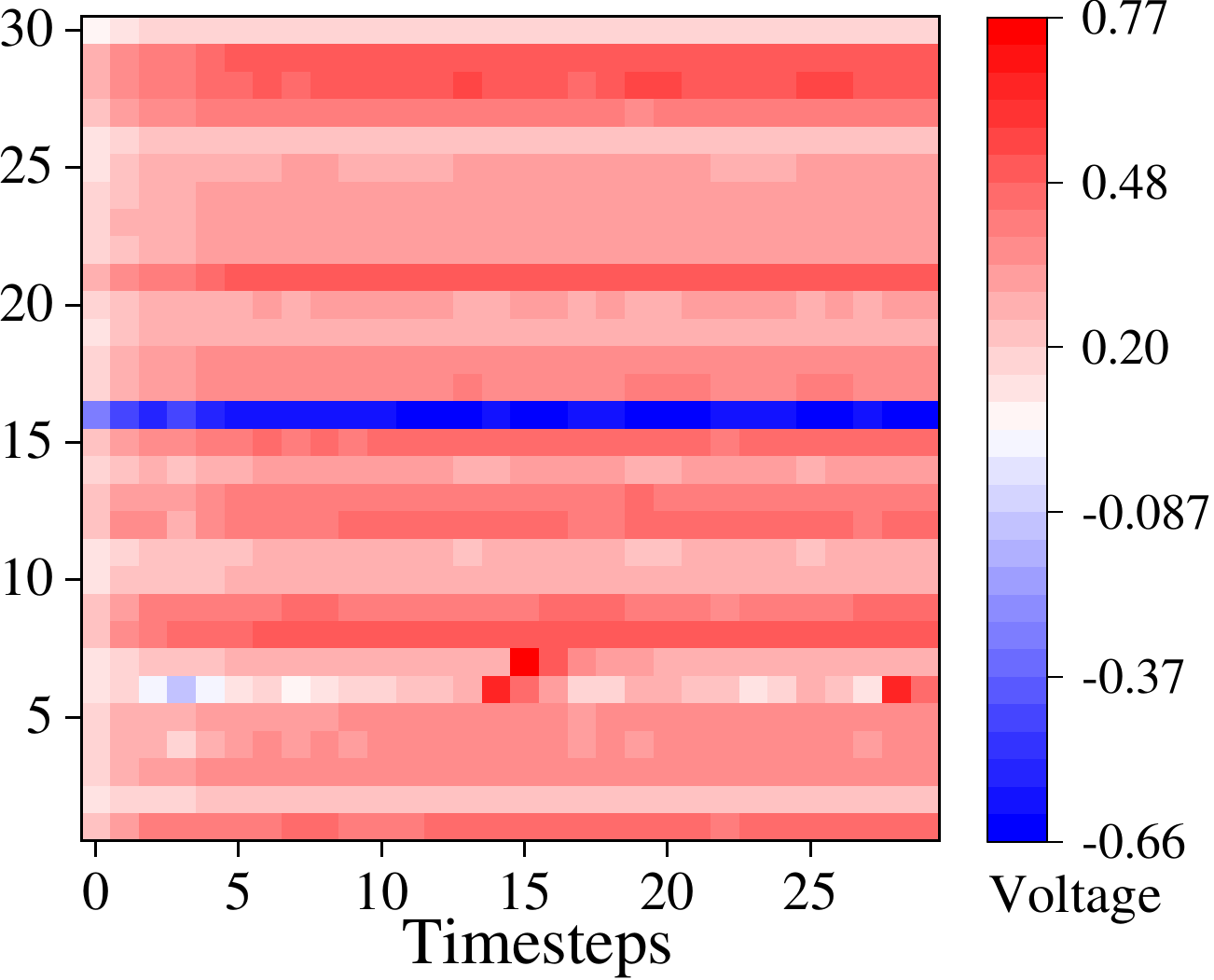}}}\hfill
    \subfigure[Clean Model]{\label{fig:poison-clean-voltage}
    {\includegraphics[width=0.236\textwidth]{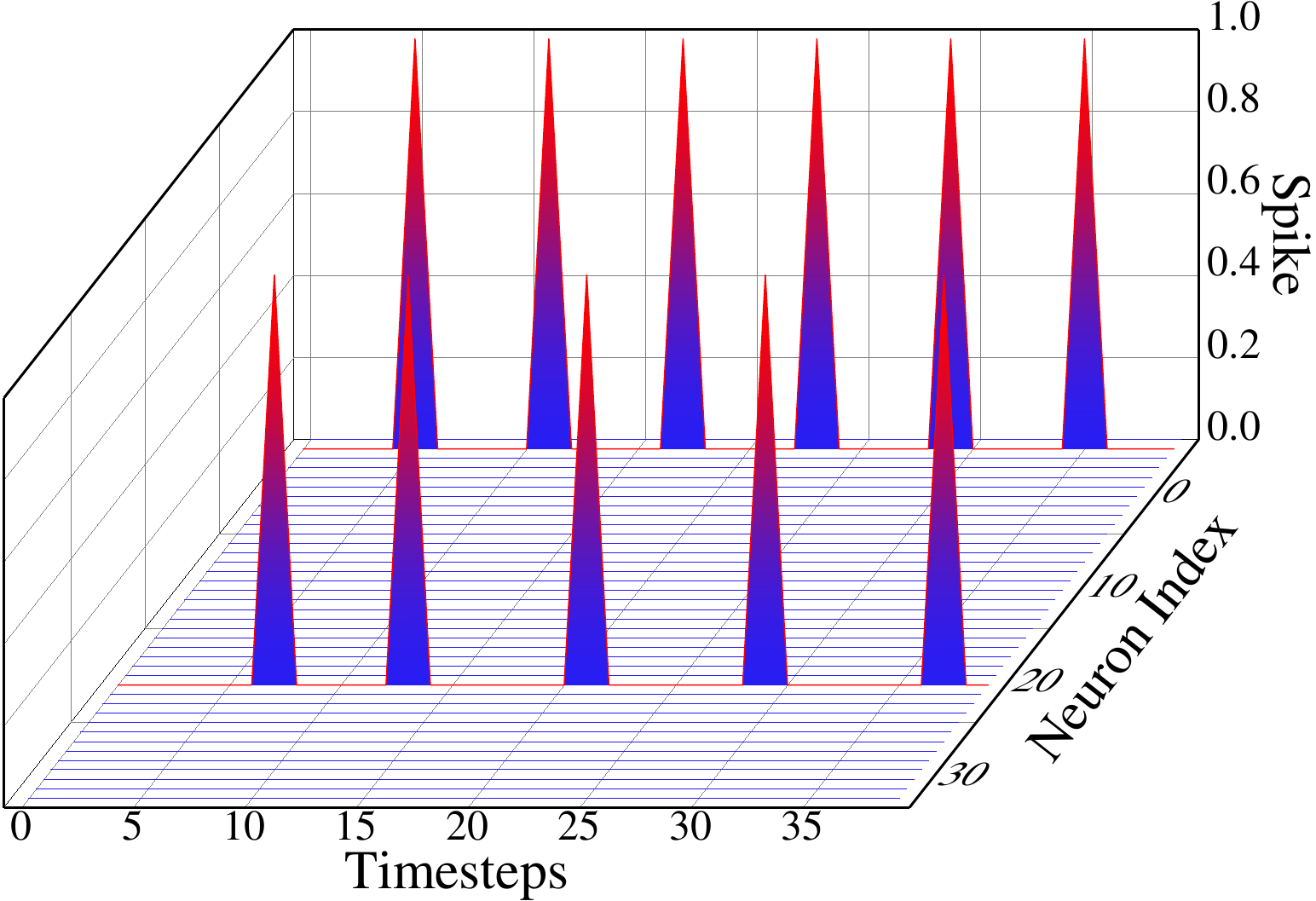}}}\hfill
    \subfigure[Backdoored Model]{\label{fig:poison-poison-voltage}
    {\includegraphics[width=0.232\textwidth]{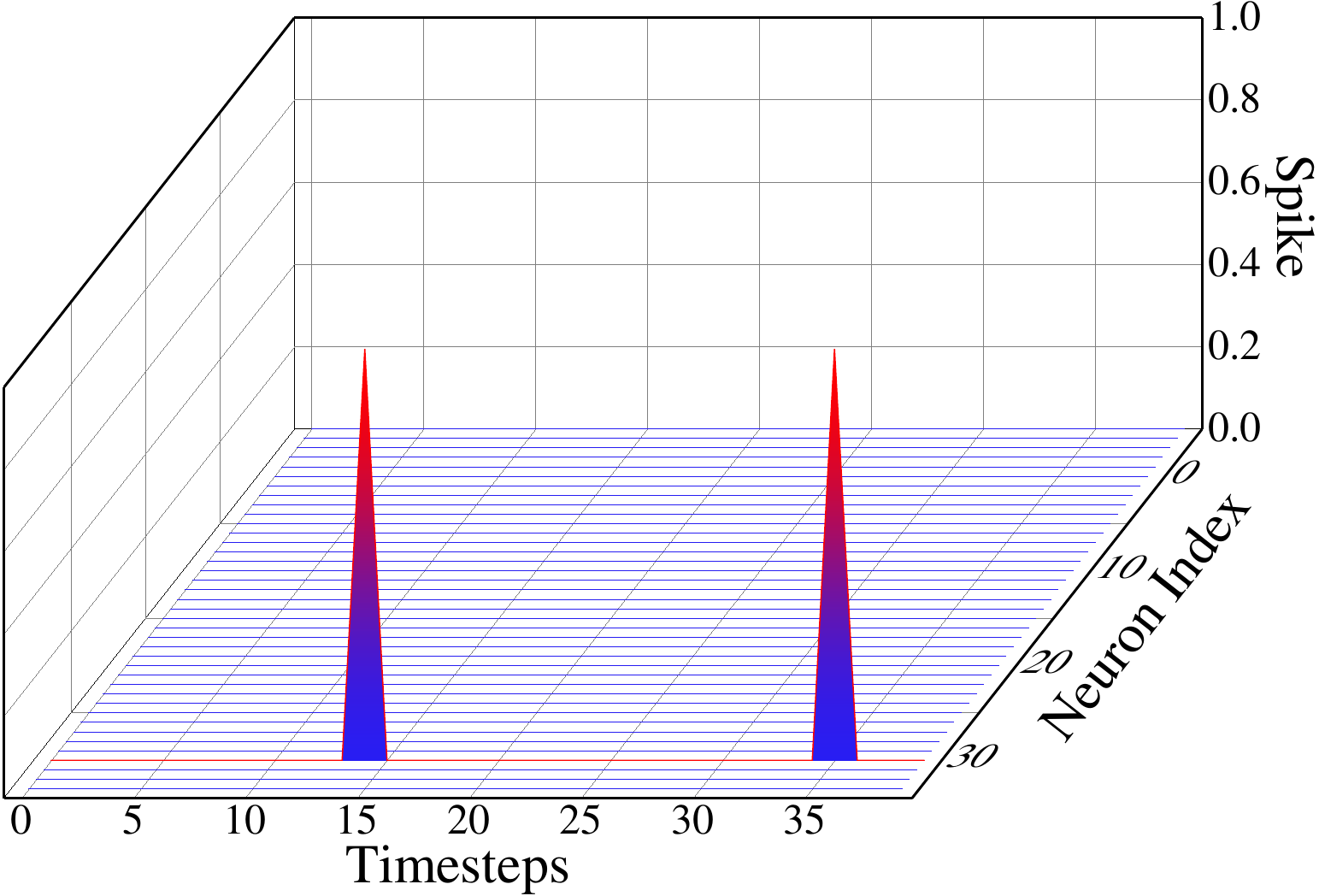}}}\hfill
    \caption{The cumulative voltage and output spike sequences of the last spike layer in spike-based ResNet18 and the differences between the clean and the backdoored model.}
    \label{fig:detection-voltage_spike}
\end{figure}

\subsubsection{Backdoor Detection}

This section discusses the feasible methods for backdoor detection. It is known from previous detection works on ANNs ~\cite{detection-AC, detection-Beatrix} that the presence of backdoor information can be determined by analyzing the activation values of the models, as the activation values of the clean and backdoored models show a significant separation. However, the layer outputs of SNNs are no longer activation values in real-valued form but rather spike sequences. Moreover, the spikes in the previous layer significantly affect the accumulative voltage of the neurons in the next layer, which further affects the output spike sequences of the next layer. Therefore, we believe that analyzing the cumulative voltage and output spike sequences in SNN can yield separations similar to that of activation values in ANNs. We conduct experiments for the clean and backdoored models under the same malicious samples. Specifically, we analyzed the cumulative voltage and output spike sequences for the same layer (the LIF neuron layer of the last BasicBlock in ResNet18) for both the clean and backdoored models. Figure \ref{fig:detection-voltage_spike} records the specific experimental results. We find that there is indeed a significant difference between the clean model and the backdoored model in cumulative voltage and output spike sequences. 
Note that this detection method can be used not only for detecting backdoored models but also for detecting poisoned samples. However, the difficulty lies in the fact that this initial detection scheme needs to ensure that immaculate models or data samples are available for reference. 
Therefore, only in the case of ensuring that we have an immaculate reference (model or dataset) can we perform a preliminary detection of the backdoor based on the voltage and output spike differences between the to-be-detected model and the reference model.

\begin{figure}[!t]
    \footnotesize
    \centering
    \subfigure[Model performance changes]{\label{fig:retrain-VGG11-tend}
    {\includegraphics[width=0.242\textwidth]{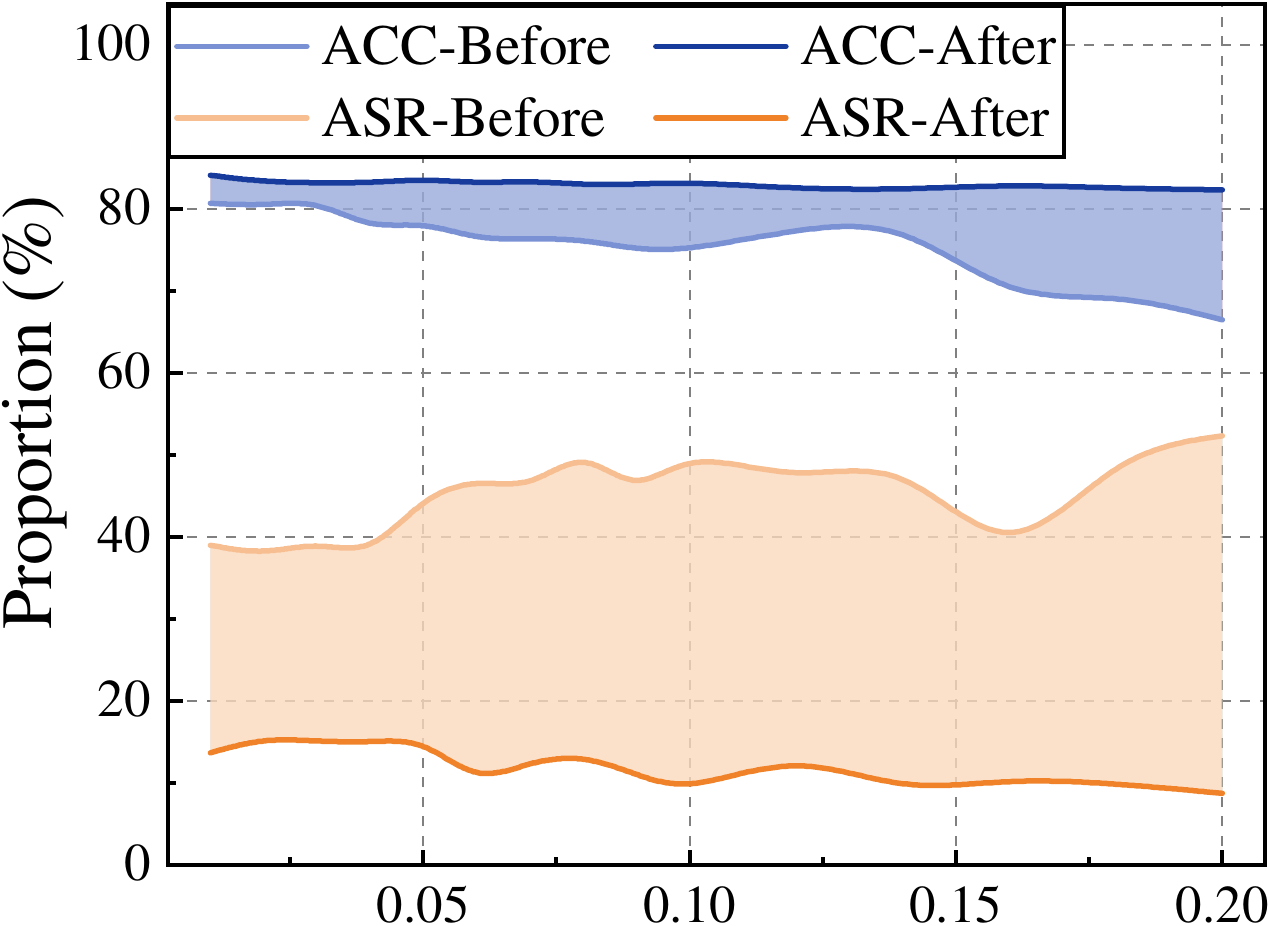}}}\hfill
    \subfigure[Average change]{\label{fig:retrain-VGG11-average}
    {\includegraphics[width=0.226\textwidth]{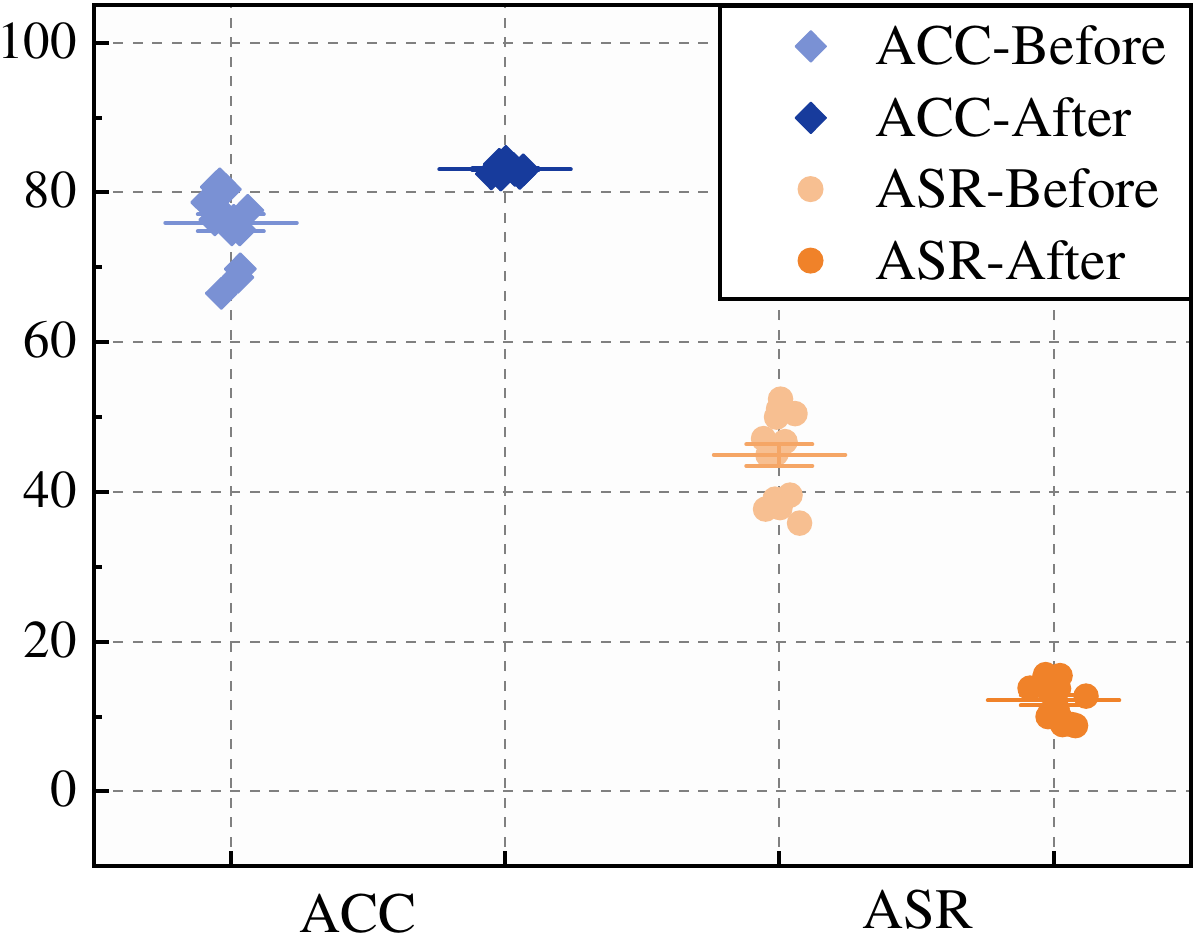}}}\hfill
    \subfigure[Model performance changes]{\label{fig:retrain-Res18-tend}
    {\includegraphics[width=0.242\textwidth]{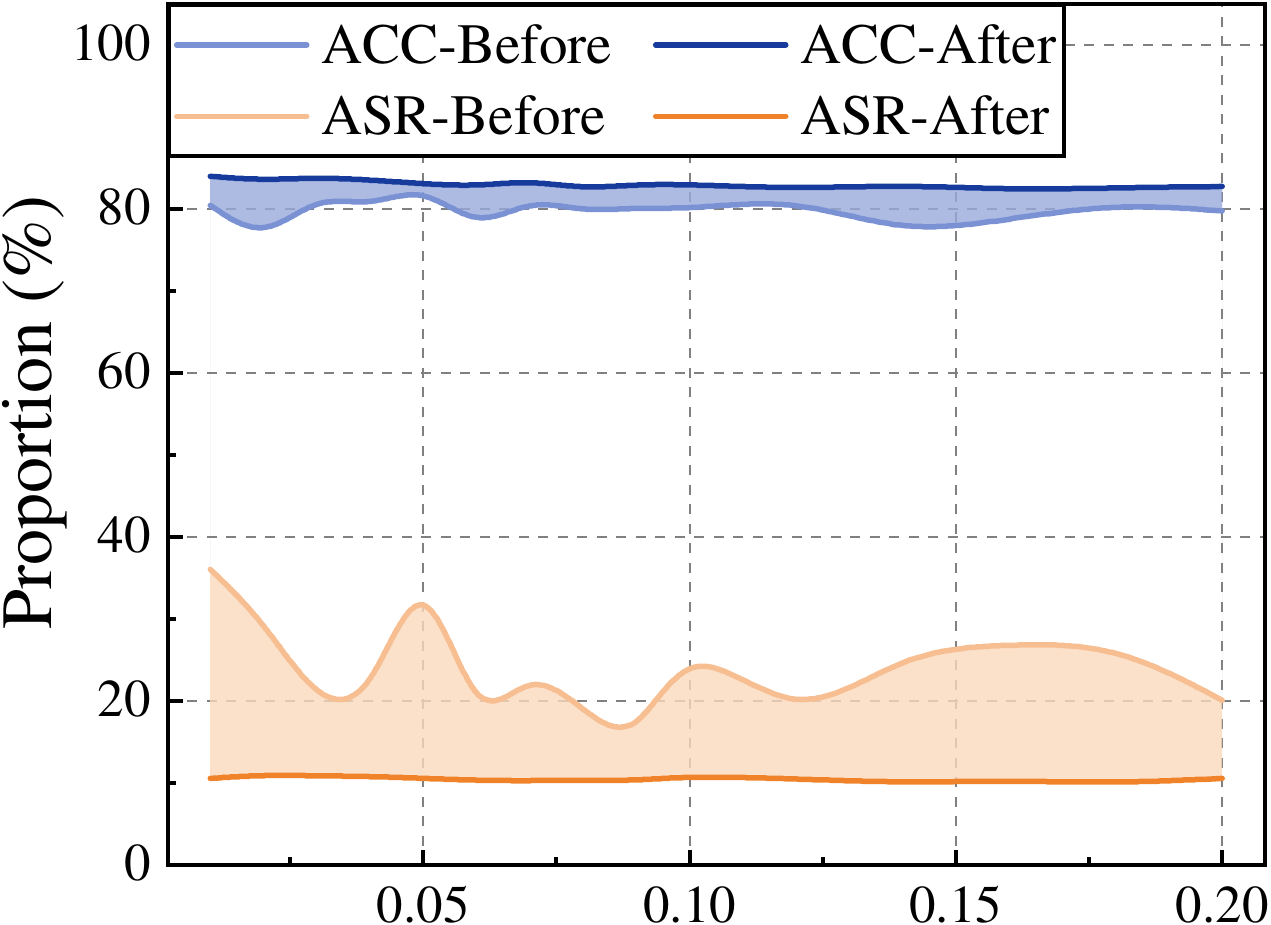}}}\hfill
    \subfigure[Average change]{\label{fig:retrain-Res18-average}
    {\includegraphics[width=0.226\textwidth]{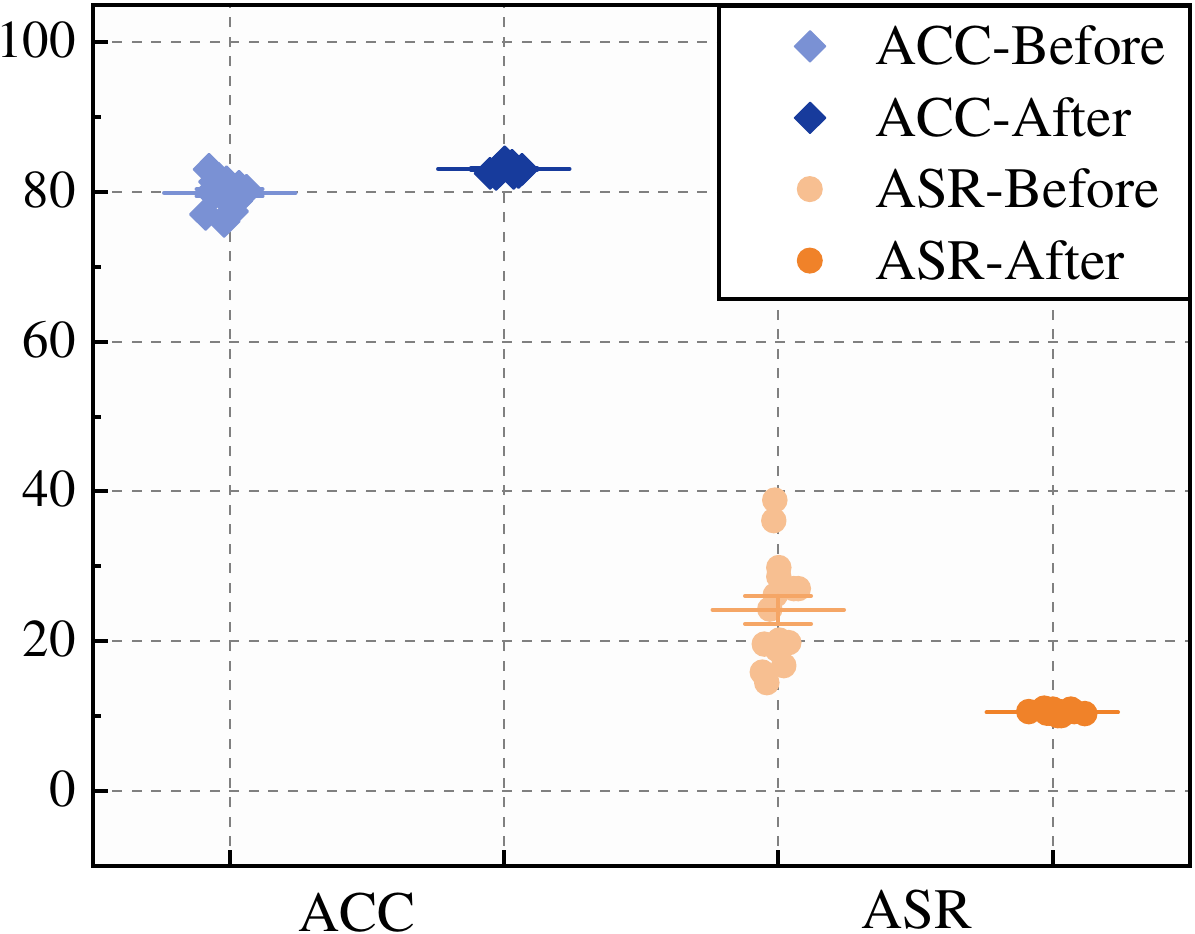}}}\hfill
    \caption{The changes in the model performance after the fine-tuning-based elimination method for VGG11 (a)-(b) and ResNet18 (c)-(d) on CIFAR10.}
    \label{fig:defense-retrain}
\end{figure}

\subsubsection{Backdoor Elimination}

This section discusses methods for backdoor elimination, provided that the presence of backdoor information in the victim model has been identified through detection methods. Based on previous summarized research work on backdoor elimination methods for ANNs ~\cite{Backdoor_Survey}, it is known that backdoor elimination consists of two main strategies: fine-tuning and pruning. Moreover, according to the experimental findings in Sec. \ref{Sec:LR_H}, it can be seen that $\mathcal{LR_H}$ ($Com3$), which uses the clean dataset as the reference for model conversion, can mitigate the impact caused by the backdoor lurking in the model, i.e., effectively reduce the ASR. Therefore, concerning the above conclusions, we believe that using a completely clean dataset for fine-tuning the backdoored model in the SNN step and thus consciously erasing the backdoor information is a feasible means to remove the backdoor. To verify the feasibility of this approach, we conduct a validation experiment on $\mathcal{LR_H}$ ($Com3$) for this inspiration. Specifically, we use a completely clean dataset as the reference dataset for both the conversion and the training process in the SNN step of $\mathcal{LR_H}$, intending to utilize the inherent self-repairing ability of $\mathcal{LR_H}$ to adjust the model weights and determine the thresholds dynamically, thus reducing the impact of backdoor information. Figure \ref{fig:defense-retrain} records the specific experimental results.

The overall trends before and after fine-tuning of VGG11 (Figure \ref{fig:retrain-VGG11-tend}) and ResNet18 (Figure \ref{fig:retrain-Res18-tend}) show that the ASR of the model after fine-tuning at different poisoning rates is significantly reduced, accompanied by an increase in ACC. According to Figure \ref{fig:retrain-VGG11-average}, the average ASR drop of VGG11 can reach about 32.75\%, while the average ACC improvement is about 7.19\%. The largest ASR decreases occur at $PR=\{0.1, 0.14, 0.18, 0.2\}$, where the ASR decreases are over 40\%. Their ASR after fine-tuning can even drop to around 9\%, which is even lower than the ASR of the clean model. For ResNet18 (Figure \ref{fig:retrain-Res18-average}), the average ASR decreases by 13.67\%, and the average ACC improves by about 3.2\%. The ASR of the ResNet18 model with different poisoning rates using fine-tuning stays around 11\%. In contrast, the fine-tuning method achieves better results for backdoor elimination on the VGG11 model, including decreased ASR and increased ACC. Note that this method is temporarily unable to pinpoint and overwrite malicious weights during the fine-tuning process but can only achieve backdoor removal-like results through global fine-tuning.


\section{Related Work}

\subsection{SNNs}
The SNN model structure was formally proposed by ~\cite{SNN_proposation_1, SNN_proposation_2} and is known as the third generation of neural networks after ANN. SNN is more biologically plausible than ANN. 
Initially, SNNs are primarily trained using variants of the Hebbian rule-based unsupervised learning method ~\cite{Hebb_rule, learning_window, STDP, tempotron, ReSuMe, STDP-Learn}. 
Furthermore, reward-based learning rules ~\cite{Neuromodulated_STDP, reward-modulated_STDP, R-STDP_robot} are also very effective unsupervised training methods. 
Subsequently, some supervised learning rules are proposed and contribute to the rapid development of SNNs, including backpropagation-based ~\cite{Spikeprop, SuperSpike, STBP}, conversion-based ~\cite{MaxNorm, RobustNorm, Spike-Norm}, and hybrid learning rules ~\cite{STDB}. 
Backpropagation-based learning rules have long faced the dilemma that SNNs are non-differentiable until ~\cite{STBP} proposed the surrogate gradient. The surrogate gradients allow SNNs to be trained using backpropagation-based methods and have grown rapidly since then. 
The conversion-based learning rules are based on the maturity of ANN training, which transforms a well-performing ANN directly into a well-performing SNN by weight normalization strategies. The initial conversion framework and constraints are first proposed by ~\cite{conv_framework}. Subsequent studies further improve the performance and minimize the structural constraints for ANN to SNN conversion ~\cite{MaxNorm, RobustNorm, Spike-Norm}. 
However, the backpropagation-based learning rules require a high computational time overhead, and the conversion-based methods rely on giant time steps (sometimes larger than 3000) to ensure well-maintained performance, leading to significant time latency. Therefore, ~\cite{STDB} first proposes a hybrid learning rule to avoid the above problems. 
At present, SNNs are developing rapidly, and several advanced spike-based model structures have been proposed, such as Spiking Graph Convolutional Networks (SGCN) ~\cite{SGCN}, SpikeGPT ~\cite{SpikeGPT}, Spike-driven Transformer ~\cite{Spikeformer,spike-driven-transformer}, SpikeBERT ~\cite{SpikeBERT}, and Spiking Denoising Diffusion Probabilistic Models (SDDPMs) ~\cite{SDDPM}.

\subsection{Data Poisoning-based Backdoor Attack}

The Backdoor attack is strongly purposeful and highly stealthy. 
Generally, the backdoored model behaves normally for clean samples and only outputs the targeted result preset by the attacker when it encounters malicious samples with a specific trigger pattern. 
The concept of backdoor attack was first proposed by ~\cite{data-poison-init}, which mainly accomplishes backdoor injection by embedding triggers into the training dataset. 
Some works have proven that a backdoor attack can fool the application systems in a practical environment ~\cite{data-poison-init, BA-physicalworld}. 
Ref. ~\cite{BadNets} injects backdoors by accessing the training supply chain of the model, which further improves the applicable scenarios of backdoor attacks. 
There are also some research attempts to further enhance the stealthiness of backdoor attacks from the perspective of trigger optimization ~\cite{invisible-BA, imperceptible-BA, BA-physicalworld}. 
Moreover, backdoor attacks can also be realized by weighted poisoning ~\cite{TANN:NDSS-2018, ModelReuse} or bit-flipping ~\cite{TBT:CVPR-2020, ProFlip:ICCV-2021, ETBT, TBD:USENIX-2019}. 
Additionally, backdoor attacks can not only target traditional CNN models but also migrate to self-attention-based transformer structures, such as NLP models (e.g. BERT ~\cite{BERT}
and LSTM ~\cite{LSTM}) ~\cite{Piccolo-BA-NLP} and vision transformers ~\cite{BA-ViT-init, DBIA, TrojViT}. 
Currently, few studies have focused on the security threats of SNNs when targeting backdoor attacks. Only ~\cite{Poster:BA, Sneaky_Spikes} preliminarily verified the feasibility of backdoor attacks based on data poisoning on the backpropagation-based learning rules and neuromorphic datasets of SNNs, thus exposing the security issues in SNN training.

\section{Conclusion}

This work represents a solid step towards exploring the vulnerability of SNN learning rules through data poisoning-based Backdoor attacks. We show both empirically and analytically that (i) backdoor attacks can easily pose a security threat to the training process of different supervised learning rules, (ii) there are some differences in robustness between different learning rules, and SNNs no longer have the advantage of high robustness over ANNs when facing backdoor attacks, (iii) backdoor information can be hidden in the final SNNs with a very high migration rate.

This paper also provides reference ideas for some future work. First, learning rules relying on conversion have the most severe security vulnerabilities. Hence, the identification and elimination of backdoor information for the whole conversion process is an urgent problem that needs to be solved. Moreover, the so-called efficient hybrid learning rules are not an excellent solution to the problem of time overhead, and further exploration of efficient and secure learning rules is needed. Finally, we must focus on exploring defenses against backdoor attacks, especially designing methods that accurately and timely detect, locate, and eliminate backdoors.

{\footnotesize \bibliographystyle{acm}
\bibliography{ref}}

\begin{thebibliography}{10}

\bibitem{Spikeprop}
Error-backpropagation in temporally encoded networks of spiking neurons.
\newblock {\em Neurocomputing 48}, 1 (2002), 17--37.

\bibitem{Poster:BA}
{\sc Abad, G., Ersoy, O., Picek, S., Ram{\'{\i}}rez{-}Dur{\'{a}}n, V.~J., and Urbieta, A.}
\newblock Poster: Backdoor attacks on spiking nns and neuromorphic datasets.
\newblock In {\em {CCS}\/} (2022), {ACM}, pp.~3315--3317.

\bibitem{Sneaky_Spikes}
{\sc Abad, G., Ersoy, O., Picek, S., and Urbieta, A.}
\newblock Sneaky spikes: Uncovering stealthy backdoor attacks in spiking neural networks with neuromorphic data.
\newblock In {\em NDSS\/} (2024).

\bibitem{DVS128Gesture}
{\sc Amir, A., Taba, B., Berg, D., Melano, T., McKinstry, J., Di~Nolfo, C., Nayak, T., Andreopoulos, A., Garreau, G., Mendoza, M., Kusnitz, J., Debole, M., Esser, S., Delbruck, T., Flickner, M., and Modha, D.}
\newblock A low power, fully event-based gesture recognition system.
\newblock In {\em 2017 IEEE Conference on Computer Vision and Pattern Recognition (CVPR)\/} (2017), pp.~7388--7397.

\bibitem{Hebb_rule}
{\sc Attneave, F., B., M., and Hebb, D.~O.}
\newblock The organization of behavior: A neuropsychological theory.

\bibitem{first-data-poisoning}
{\sc Barreno, M., Nelson, B., Sears, R., Joseph, A.~D., and Tygar, J.~D.}
\newblock Can machine learning be secure?
\newblock In {\em AsiaCCS\/} (2006), {ACM}, pp.~16--25.

\bibitem{adver_examples_4}
{\sc Bu, T., Ding, J., Hao, Z., and Yu, Z.}
\newblock Rate gradient approximation attack threats deep spiking neural networks.
\newblock In {\em {CVPR}\/} (2023), {IEEE}, pp.~7896--7906.

\bibitem{SDDPM}
{\sc Cao, J., Wang, Z., Guo, H., Cheng, H., Zhang, Q., and Xu, R.}
\newblock Spiking denoising diffusion probabilistic models.
\newblock In {\em 2024 IEEE/CVF Winter Conference on Applications of Computer Vision (WACV)\/} (2024), pp.~4900--4909.

\bibitem{conv_framework}
{\sc Cao, Y., Chen, Y., and Khosla, D.}
\newblock Spiking deep convolutional neural networks for energy-efficient object recognition.
\newblock {\em International Journal of Computer Vision 113\/} (05 2015), 54--66.

\bibitem{detection-AC}
{\sc Chen, B., Carvalho, W., Baracaldo, N., Ludwig, H., Edwards, B., Lee, T., Molloy, I.~M., and Srivastava, B.}
\newblock Detecting backdoor attacks on deep neural networks by activation clustering.
\newblock In {\em SafeAI@AAAI\/} (2019), vol.~2301 of {\em {CEUR} Workshop Proceedings}, CEUR-WS.org.

\bibitem{ProFlip:ICCV-2021}
{\sc Chen, H., Fu, C., Zhao, J., and Koushanfar, F.}
\newblock Proflip: Targeted trojan attack with progressive bit flips.
\newblock In {\em {ICCV}\/} (2021), pp.~7698--7707.

\bibitem{data-poison-init}
{\sc Chen, X., Liu, C., Li, B., Lu, K., and Song, D.}
\newblock Targeted backdoor attacks on deep learning systems using data poisoning.
\newblock {\em CoRR abs/1712.05526\/} (2017).

\bibitem{DVS-1}
{\sc Delbruck, T., et~al.}
\newblock Frame-free dynamic digital vision.
\newblock In {\em Proceedings of Intl. Symp. on Secure-Life Electronics, Advanced Electronics for Quality Life and Society\/} (2008), vol.~1, Citeseer, pp.~21--26.

\bibitem{SNN_vs_ANN}
{\sc Deng, L., Wu, Y., Hu, X., Liang, L., Ding, Y., Li, G., Zhao, G., Li, P., and Xie, Y.}
\newblock Rethinking the performance comparison between {SNNS} and {ANNS}.
\newblock {\em Neural Networks 121\/} (2020), 294--307.

\bibitem{BERT}
{\sc Devlin, J., Chang, M., Lee, K., and Toutanova, K.}
\newblock {BERT:} pre-training of deep bidirectional transformers for language understanding.
\newblock In {\em {NAACL-HLT} {(1)}\/} (2019), Association for Computational Linguistics, pp.~4171--4186.

\bibitem{STDP-Learn}
{\sc Diehl, P., and Cook, M.}
\newblock Unsupervised learning of digit recognition using spike-timing-dependent plasticity.
\newblock {\em Frontiers in Computational Neuroscience 9\/} (2015).

\bibitem{MaxNorm}
{\sc Diehl, P.~U., Neil, D., Binas, J., Cook, M., Liu, S., and Pfeiffer, M.}
\newblock Fast-classifying, high-accuracy spiking deep networks through weight and threshold balancing.
\newblock In {\em {IJCNN}\/} (2015), {IEEE}, pp.~1--8.

\bibitem{dead_neuron_problem}
{\sc Eshraghian, J.~K., Ward, M., Neftci, E.~O., Wang, X., Lenz, G., Dwivedi, G., Bennamoun, M., Jeong, D.~S., and Lu, W.~D.}
\newblock Training spiking neural networks using lessons from deep learning.
\newblock {\em Proceedings of the IEEE 111}, 9 (2023), 1016--1054.

\bibitem{Spikingjelly}
{\sc Fang, W., Yanqi, C., Ding, J., Yu, Z., Masquelier, T., Chen, D., Huang, L., Zhou, H., Li, G., and Tian, Y.}
\newblock Spikingjelly: An open-source machine learning infrastructure platform for spike-based intelligence.
\newblock {\em Science Advances 9}, 40 (2023), eadi1480.

\bibitem{Neuromodulated_STDP}
{\sc Frémaux, N., and Gerstner, W.}
\newblock Neuromodulated spike-timing-dependent plasticity, and theory of three-factor learning rules.
\newblock {\em Frontiers in Neural Circuits 9\/} (2016).

\bibitem{BadNets}
{\sc Gu, T., Liu, K., Dolan-Gavitt, B., and Garg, S.}
\newblock Badnets: Evaluating backdooring attacks on deep neural networks.
\newblock {\em IEEE Access 7\/} (2019), 47230--47244.

\bibitem{tempotron}
{\sc Gütig, R., and Sompolinsky, H.}
\newblock The tempotron: a neuron that learns spike-timing based decisions.
\newblock {\em Reviews in the neurosciences 16\/} (01 2005), S27--S27.

\bibitem{ResNet}
{\sc He, K., Zhang, X., Ren, S., and Sun, J.}
\newblock Deep residual learning for image recognition.
\newblock In {\em {CVPR}\/} (2016), {IEEE} Computer Society, pp.~770--778.

\bibitem{LSTM}
{\sc Hochreiter, S., and Schmidhuber, J.}
\newblock Long short-term memory.
\newblock {\em Neural Computation 9}, 8 (1997), 1735--1780.

\bibitem{TBD:USENIX-2019}
{\sc Hong, S., Frigo, P., Kaya, Y., Giuffrida, C., and Dumitras, T.}
\newblock Terminal brain damage: Exposing the graceless degradation in deep neural networks under hardware fault attacks.
\newblock In {\em {USENIX} Security Symposium\/} (2019), pp.~497--514.

\bibitem{ModelReuse}
{\sc Ji, Y., Zhang, X., Ji, S., Luo, X., and Wang, T.}
\newblock Model-reuse attacks on deep learning systems.
\newblock In {\em Proceedings of the 2018 ACM SIGSAC Conference on Computer and Communications Security\/} (New York, NY, USA, 2018), CCS '18, Association for Computing Machinery, p.~349–363.

\bibitem{ETBT}
{\sc Jin, L., Jiang, W., Zhan, J., and Wen, X.}
\newblock Highly evasive targeted bit-trojan on deep neural networks.
\newblock {\em IEEE Transactions on Computers 73}, 9 (2024), 2350--2363.

\bibitem{Backdoor_Survey}
{\sc Jin, L., Wen, X., Jiang, W., and Zhan, J.}
\newblock A survey of trojan attacks and defenses to deep neural networks, 2024.

\bibitem{R-STDP_robot}
{\sc Juarez-Lora, A., Ponce-Ponce, V.~H., Sossa, H., and Rubio-Espino, E.}
\newblock R-stdp spiking neural network architecture for motion control on a changing friction joint robotic arm.
\newblock {\em Frontiers in Neurorobotics 16\/} (2022).

\bibitem{Modelzoo}
{\sc Koh, J.~Y.}
\newblock Model zoo.

\bibitem{CIFAR10}
{\sc Krizhevsky, A.}
\newblock Learning multiple layers of features from tiny images.

\bibitem{Spike-Thrift}
{\sc Kundu, S., Datta, G., Pedram, M., and Beerel, P.~A.}
\newblock Spike-thrift: Towards energy-efficient deep spiking neural networks by limiting spiking activity via attention-guided compression.
\newblock In {\em {WACV}\/} (2021), {IEEE}, pp.~3952--3961.

\bibitem{MNIST}
{\sc LeCun, Y., Bottou, L., Bengio, Y., and Haffner, P.}
\newblock Gradient-based learning applied to document recognition.
\newblock {\em Proc. {IEEE} 86}, 11 (1998), 2278--2324.

\bibitem{CIFAR10-DVS}
{\sc Li, H., Liu, H., Ji, X., Li, G., and Shi, L.}
\newblock Cifar10-dvs: An event-stream dataset for object classification.
\newblock {\em Frontiers in Neuroscience 11\/} (2017).

\bibitem{invisible-BA}
{\sc Li, S., Xue, M., Zhao, B. Z.~H., Zhu, H., and Zhang, X.}
\newblock Invisible backdoor attacks on deep neural networks via steganography and regularization.
\newblock {\em IEEE Trans. Dependable Secur. Comput. 18}, 5 (sep 2021), 2088–2105.

\bibitem{adver_examples_2}
{\sc Liang, L., Hu, X., Deng, L., Wu, Y., Li, G., Ding, Y., Li, P., and Xie, Y.}
\newblock Exploring adversarial attack in spiking neural networks with spike-compatible gradient.
\newblock {\em {IEEE} Trans. Neural Networks Learn. Syst. 34}, 5 (2023), 2569--2583.

\bibitem{DVS-2}
{\sc Lichtsteiner, P., Posch, C., and Delbruck, T.}
\newblock A 128$\times$ 128 120 db 15 $\mu$s latency asynchronous temporal contrast vision sensor.
\newblock {\em IEEE Journal of Solid-State Circuits 43}, 2 (2008), 566--576.

\bibitem{adver_examples_3}
{\sc Lin, X., Dong, C., Liu, X., and Zhang, Y.}
\newblock Spa: An efficient adversarial attack on spiking neural networks using spike probabilistic.
\newblock In {\em 2022 22nd IEEE International Symposium on Cluster, Cloud and Internet Computing (CCGrid)\/} (2022), pp.~366--375.

\bibitem{TANN:NDSS-2018}
{\sc Liu, Y., Ma, S., Aafer, Y., Lee, W., Zhai, J., Wang, W., and Zhang, X.}
\newblock Trojaning attack on neural networks.
\newblock In {\em {NDSS}\/} (2018).

\bibitem{Piccolo-BA-NLP}
{\sc Liu, Y., Shen, G., Tao, G., An, S., Ma, S., and Zhang, X.}
\newblock Piccolo: Exposing complex backdoors in {NLP} transformer models.
\newblock In {\em {SP}\/} (2022), {IEEE}, pp.~2025--2042.

\bibitem{SpikeBERT}
{\sc Lv, C., Li, T., Xu, J., Gu, C., Ling, Z., Zhang, C., Zheng, X., and Huang, X.}
\newblock Spikebert: {A} language spikformer trained with two-stage knowledge distillation from {BERT}.
\newblock {\em CoRR abs/2308.15122\/} (2023).

\bibitem{DBIA}
{\sc Lv, P., Ma, H., Zhou, J., Liang, R., Chen, K., Zhang, S., and Yang, Y.}
\newblock {DBIA:} data-free backdoor attack against transformer networks.
\newblock In {\em {ICME}\/} (2023), {IEEE}, pp.~2819--2824.

\bibitem{detection-Beatrix}
{\sc Ma, W., Wang, D., Sun, R., Xue, M., Wen, S., and Xiang, Y.}
\newblock The "beatrix" resurrections: Robust backdoor detection via gram matrices.
\newblock In {\em {NDSS}\/} (2023), The Internet Society.

\bibitem{SNN_proposal}
{\sc Maass, W.}
\newblock On the computational complexity of networks of spiking neurons.
\newblock In {\em {NIPS}\/} (1994), {MIT} Press, pp.~183--190.

\bibitem{SNN_proposation_1}
{\sc Maass, W.}
\newblock On the computational complexity of networks of spiking neurons.
\newblock In {\em Proceedings of the 7th International Conference on Neural Information Processing Systems\/} (Cambridge, MA, USA, 1994), NIPS'94, MIT Press, p.~183–190.

\bibitem{SNN_proposation_2}
{\sc Maass, W.}
\newblock Lower bounds for the computational power of networks of spiking neurons.
\newblock {\em Neural Computation 8}, 1 (1996), 1--40.

\bibitem{Third_NN}
{\sc Maass, W.}
\newblock Networks of spiking neurons: The third generation of neural network models.
\newblock {\em Neural Networks 10}, 9 (1997), 1659--1671.

\bibitem{reward-modulated_STDP}
{\sc Mozafari, M., Ganjtabesh, M., Nowzari-Dalini, A., Thorpe, S.~J., and Masquelier, T.}
\newblock Bio-inspired digit recognition using reward-modulated spike-timing-dependent plasticity in deep convolutional networks.
\newblock {\em Pattern Recognition 94\/} (2019), 87--95.

\bibitem{N-MNIST}
{\sc Orchard, G., Jayawant, A., Cohen, G.~K., and Thakor, N.}
\newblock Converting static image datasets to spiking neuromorphic datasets using saccades.
\newblock {\em Frontiers in Neuroscience 9\/} (2015).

\bibitem{Overview_of_SNN}
{\sc Pietrzak, P., Szczęsny, S., Huderek, D., and Przyborowski, L.}
\newblock Overview of spiking neural network learning approaches and their computational complexities.
\newblock {\em Sensors 23}, 6 (2023).

\bibitem{ReSuMe}
{\sc Ponulak, F., and Kasinski, A.~J.}
\newblock Supervised learning in spiking neural networks with resume: Sequence learning, classification, and spike shifting.
\newblock {\em Neural Comput. 22}, 2 (2010), 467--510.

\bibitem{STDP}
{\sc qiang Bi, G., and ming Poo, M.}
\newblock Synaptic modifications in cultured hippocampal neurons: Dependence on spike timing, synaptic strength, and postsynaptic cell type.
\newblock {\em Journal of Neuroscience 18}, 24 (1998), 10464--10472.

\bibitem{learning_window}
{\sc qiang Bi, G., and ming Poo, M.}
\newblock Synaptic modifications in cultured hippocampal neurons: Dependence on spike timing, synaptic strength, and postsynaptic cell type.
\newblock {\em Journal of Neuroscience 18}, 24 (1998), 10464--10472.

\bibitem{TBT:CVPR-2020}
{\sc Rakin, A.~S., He, Z., and Fan, D.}
\newblock {TBT:} targeted neural network attack with bit trojan.
\newblock In {\em {CVPR}\/} (2020), pp.~13195--13204.

\bibitem{STDB}
{\sc Rathi, N., Srinivasan, G., Panda, P., and Roy, K.}
\newblock Enabling deep spiking neural networks with hybrid conversion and spike timing dependent backpropagation.
\newblock In {\em {ICLR}\/} (2020), OpenReview.net.

\bibitem{RobustNorm}
{\sc Rueckauer, B., Lungu, I.-A., Hu, Y., Pfeiffer, M., and Liu, S.-C.}
\newblock Conversion of continuous-valued deep networks to efficient event-driven networks for image classification.
\newblock {\em Frontiers in Neuroscience 11\/} (2017).

\bibitem{Spike-Norm}
{\sc Sengupta, A., Ye, Y., Wang, R., Liu, C., and Roy, K.}
\newblock Going deeper in spiking neural networks: Vgg and residual architectures.
\newblock {\em Frontiers in Neuroscience 13\/} (2019).

\bibitem{DVS-4}
{\sc Serrano-Gotarredona, T., and Linares-Barranco, B.}
\newblock A 128$\,\times$ 128 1.5
\newblock {\em IEEE Journal of Solid-State Circuits 48}, 3 (2013), 827--838.

\bibitem{adver_examples_1}
{\sc Sharmin, S., Panda, P., Sarwar, S.~S., Lee, C., Ponghiran, W., and Roy, K.}
\newblock A comprehensive analysis on adversarial robustness of spiking neural networks.
\newblock In {\em {IJCNN}\/} (2019), {IEEE}, pp.~1--8.

\bibitem{inherent_robustness}
{\sc Sharmin, S., Rathi, N., Panda, P., and Roy, K.}
\newblock Inherent adversarial robustness of deep spiking neural networks: Effects of discrete input encoding and non-linear activations.
\newblock In {\em {ECCV} {(29)}\/} (2020), vol.~12374 of {\em Lecture Notes in Computer Science}, Springer, pp.~399--414.

\bibitem{VGG}
{\sc Simonyan, K., and Zisserman, A.}
\newblock Very deep convolutional networks for large-scale image recognition.
\newblock In {\em {ICLR}\/} (2015).

\bibitem{BA-ViT-init}
{\sc Subramanya, A., Saha, A., Koohpayegani, S.~A., Tejankar, A., and Pirsiavash, H.}
\newblock Backdoor attacks on vision transformers.
\newblock {\em CoRR abs/2206.08477\/} (2022).

\bibitem{BA-physicalworld}
{\sc Wenger, E., Passananti, J., Bhagoji, A.~N., Yao, Y., Zheng, H., and Zhao, B.~Y.}
\newblock Backdoor attacks against deep learning systems in the physical world.
\newblock In {\em 2021 IEEE/CVF Conference on Computer Vision and Pattern Recognition (CVPR)\/} (2021), pp.~6202--6211.

\bibitem{BPTT}
{\sc Werbos, P.~J.}
\newblock Backpropagation through time: what it does and how to do it.
\newblock {\em Proc. {IEEE} 78}, 10 (1990), 1550--1560.

\bibitem{STBP}
{\sc Wu, Y., Deng, L., Li, G., Zhu, J., and Shi, L.}
\newblock Spatio-temporal backpropagation for training high-performance spiking neural networks.
\newblock {\em Frontiers in Neuroscience 12\/} (2018).

\bibitem{DVS-3}
{\sc Yang, M., Liu, S.-C., and Delbruck, T.}
\newblock A dynamic vision sensor with 1
\newblock {\em IEEE Journal of Solid-State Circuits 50}, 9 (2015), 2149--2160.

\bibitem{spike-driven-transformer}
{\sc Yao, M., Hu, J., Zhou, Z., Yuan, L., Tian, Y., Xu, B., and Li, G.}
\newblock Spike-driven transformer.
\newblock In {\em NeurIPS\/} (2023).

\bibitem{Attention_SNN}
{\sc Yao, M., Zhao, G., Zhang, H., Hu, Y., Deng, L., Tian, Y., Xu, B., and Li, G.}
\newblock Attention spiking neural networks.
\newblock {\em {IEEE} Trans. Pattern Anal. Mach. Intell. 45}, 8 (2023), 9393--9410.

\bibitem{SuperSpike}
{\sc Zenke, F., and Ganguli, S.}
\newblock Superspike: Supervised learning in multilayer spiking neural networks.
\newblock {\em Neural Computation 30}, 6 (2018), 1514--1541.

\bibitem{TrojViT}
{\sc Zheng, M., Lou, Q., and Jiang, L.}
\newblock Trojvit: Trojan insertion in vision transformers.
\newblock In {\em {CVPR}\/} (2023), {IEEE}, pp.~4025--4034.

\bibitem{imperceptible-BA}
{\sc Zhong, N., Qian, Z., and Zhang, X.}
\newblock Imperceptible backdoor attack: From input space to feature representation.
\newblock In {\em Proceedings of the Thirty-First International Joint Conference on Artificial Intelligence, {IJCAI-22}\/} (7 2022), L.~D. Raedt, Ed., International Joint Conferences on Artificial Intelligence Organization, pp.~1736--1742.
\newblock Main Track.

\bibitem{SCNN}
{\sc Zhou, S., Li, X., Chen, Y., Chandrasekaran, S.~T., and Sanyal, A.}
\newblock Temporal-coded deep spiking neural network with easy training and robust performance.
\newblock In {\em {AAAI}\/} (2021), {AAAI} Press, pp.~11143--11151.

\bibitem{Spikeformer}
{\sc Zhou, Z., Zhu, Y., He, C., Wang, Y., Yan, S., Tian, Y., and Yuan, L.}
\newblock Spikformer: When spiking neural network meets transformer.
\newblock In {\em {ICLR}\/} (2023), OpenReview.net.

\bibitem{SpikeGPT}
{\sc Zhu, R., Zhao, Q., and Eshraghian, J.~K.}
\newblock Spikegpt: Generative pre-trained language model with spiking neural networks.
\newblock {\em CoRR abs/2302.13939\/} (2023).

\bibitem{SGCN}
{\sc Zhu, Z., Peng, J., Li, J., Chen, L., Yu, Q., and Luo, S.}
\newblock Spiking graph convolutional networks.
\newblock In {\em {IJCAI}\/} (2022), ijcai.org, pp.~2434--2440.

\end{thebibliography}

\theendnotes

\appendix

\section{Model Structures}
\label{sec:model_structures}
We use VGG11 and ResNet18 as the primary model structure in our experiments. The specific spike-based model structure is referred to ~\cite{STDB, Spikingjelly} and shown in Tables \ref{tab:spike-based res18}, \ref{tab:spike-based VGG11}, \ref{tab:block_in_vgg11}.

\begin{table}[ht]
\caption{Model structure of spike-based ResNet18.}
\centering
\small
\begin{tabular}{cc|cc}
\hline \hline
\multicolumn{2}{c|}{\textbf{Spiek-based ResNet18}}    & \multicolumn{2}{c}{\textbf{Spike-based BasicBlock}}   \\ \hline
\textbf{Layer Type} & \textbf{Layer Size} & \textbf{Layer Type} & \textbf{Layer Size} \\ \hline
Conv                & 3*3*1               & Conv                & 3*3*1               \\
BactchNorm          & -                   & BatchNorm           & -                   \\
\textcolor{red}{Spike Layer}         & -                   & \textcolor{red}{Spike Layer}         & -                   \\
MaxPool             & 2*2*2               & Conv                & 3*3*1               \\
(BasicBlock*2)*4    &                     & BatchNorm           & -                   \\
AvgPool             & -                   & \textcolor{red}{Spike Layer}         & -                   \\
FC                  & -                   &                     &                     \\ \hline \hline
\end{tabular}
\label{tab:spike-based res18}
\end{table}

\begin{table}[ht]
\centering
\caption{Model Structure of spike-based VGG11.}
\begin{tabular}{ccc}
\hline \hline
\multicolumn{3}{c}{\textbf{Spike-based VGG11}}                                                                   \\ \hline
\multicolumn{2}{c}{\textbf{Layer Type}}                                                    & \textbf{Layer Size} \\ \hline
\multirow{2}{*}{\begin{tabular}[c]{@{}c@{}}Feature\\ Extractor\end{tabular}} & Block1 * 2  &                     \\
                                                                             & Block2 * 3  &                     \\
                                                                             & AvgPool     & -                   \\
Classifier                                                                   & FC          & -                   \\
                                                                             & \textcolor{red}{Spike Layer} & -                   \\
                                                                             & Dropout     & 0.5                 \\
Classifier                                                                   & FC          & -                   \\
                                                                             & \textcolor{red}{Spike Layer} & -                   \\
                                                                             & Dropout     & 0.5                 \\
                                                                             & FC          & -                   \\ \hline \hline
\end{tabular}
\label{tab:spike-based VGG11}
\end{table}

\begin{table}[ht]
\caption{The structure of the block in spike-based VGG11.}
\centering
\begin{tabular}{cc|cc}
\hline \hline
\multicolumn{2}{c|}{\textbf{Spike-based Block1}}     & \multicolumn{2}{c}{\textbf{Spike-based Block2}}      \\ \hline
\textbf{Layer Type} & \textbf{Layer Size} & \textbf{Layer Type} & \textbf{Layer Size} \\ \hline
Conv                & 3*3*1               & Conv                & 3*3*1               \\
\textcolor{red}{Spike Layer}         & -                   & \textcolor{red}{Spike Layer}         & -                   \\
MaxPool             & 2*2*2               & Conv                & 3*3*1               \\
                    &                     & \textcolor{red}{Spike Layer}         & -                   \\
                    &                     & MaxPool             & 2*2*2               \\ \hline \hline
\end{tabular}
\label{tab:block_in_vgg11}
\end{table}

\section{Hyper-parameter Settings}


To ensure that the best-performing models can be obtained under different learning rules, we select different time steps for the three learning rules. Specifically, $T=100$ is set for $\mathcal{LR_B}$ and $\mathcal{LR_H}$, while $T=400$ is set for $\mathcal{LR_C}$. Note that typically, $\mathcal{LR_C}$ needs to utilize a large timestep ($T>=3000$) to ensure better performance during the conversion process. However, we find no significant difference in performance after model conversion at $T=400$ and $T=3000$. Therefore, to ensure the model has stable performance while saving computational time, we set $T=400$ for $\mathcal{LR_H}$. Moreover, we set $\alpha=0.5$, $\beta = 0.5$, $\lambda=0.5$, and $\mu = 0.5$ for optimization process. Additionally, we always keep $\tau = 2.0$ for LIF neurons.
The specific hyper-parameters during training for different learning rules are as follows:
\begin{itemize}
    \item $\mathcal{LR_B}$: We train the spike-based VGG11 and ResNet18 from scratch, using: Adam, 200 epochs, 0.001 lr, 0.9 momentum, 64 batches. 
    \item $\mathcal{LR_C}$: We train the VGG11 and ResNet18 from scratch, using: Adam, 200 epochs, 0.001 lr, 0.9 momentum, 64 batches.
    \item $\mathcal{LR_H}$: First, We train the VGG11 and ResNet18 from scratch, using: SGD, 200 epochs, 0.01 lr, 0.9 momentum, 64 batches. Then, we train the converted VGG11 and ResNet18 after the conversion step, using Adam, 200 epochs, 0.0001 lr, 0.9 momentum, 64 batches.
\end{itemize}


\begin{figure}[ht]
    \footnotesize
    \centering
    \subfigure[Dead neuron problem]{\label{fig:dirac-function}
    {\includegraphics[width=0.235\textwidth]{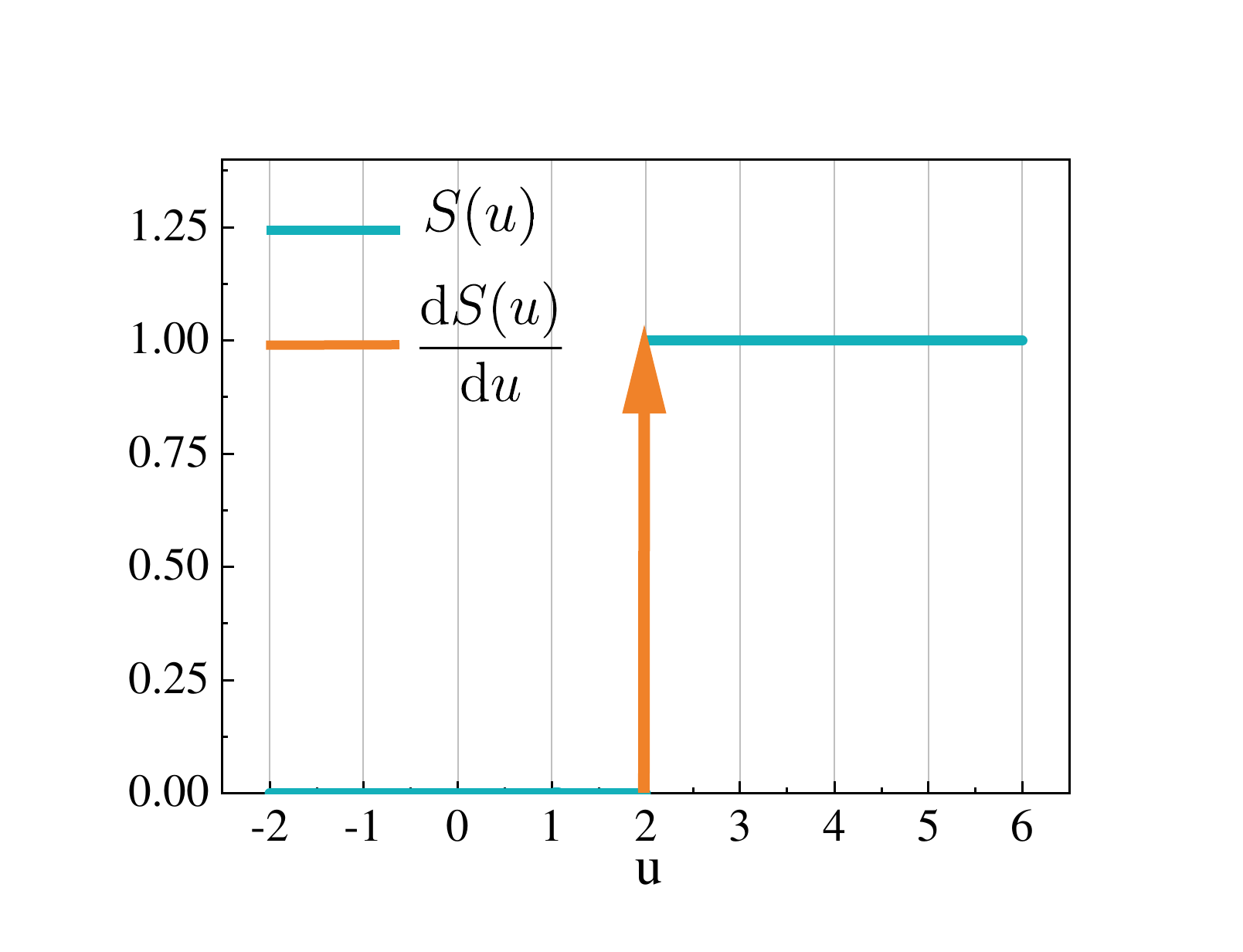}}}\hfill
    \subfigure[Surrogate gradient]{\label{fig:surrogate-gradient}
    {\includegraphics[width=0.232\textwidth]{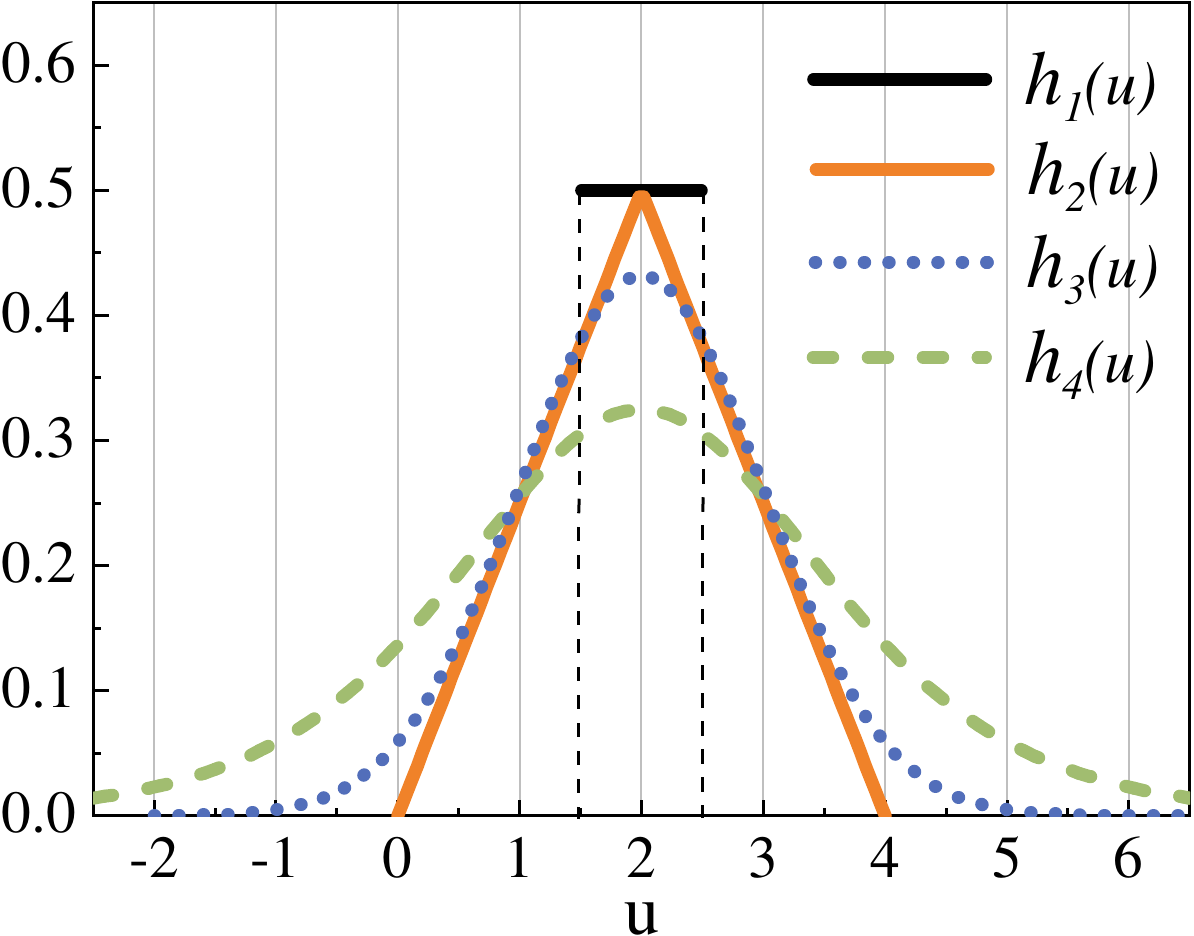}}}\hfill
    \caption{Dead neuron problem and surrogate gradients.}
    \label{fig:dirac-function-surrogate-gradient}
\end{figure}

\begin{figure}[ht]
    \footnotesize
    \centering
    \subfigure[Firing rate of IF neuron]{\label{fig:firing-rate}
    {\includegraphics[width=0.233\textwidth]{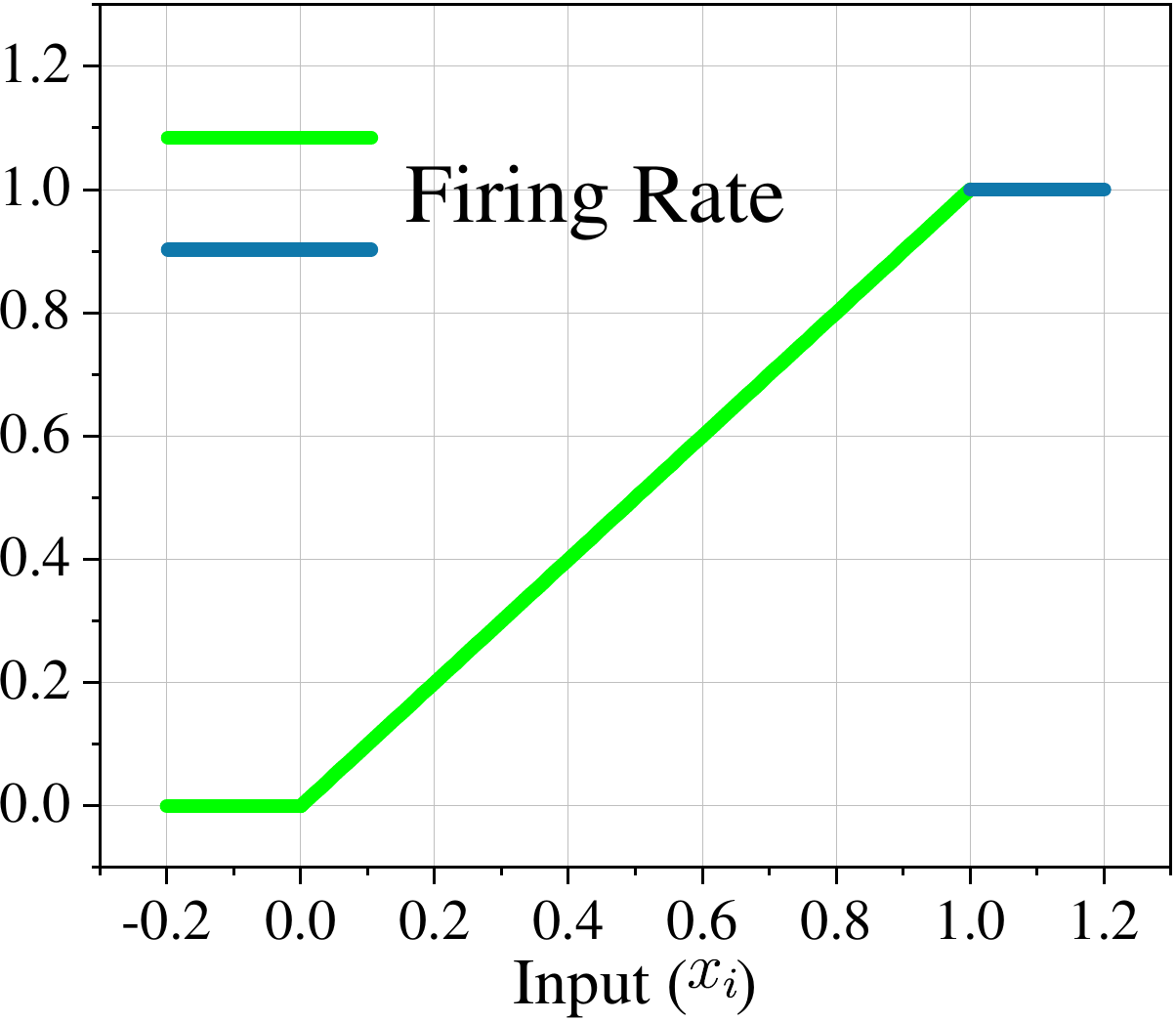}}}\hfill
    \subfigure[Activation value of ReLU]{\label{fig:ReLU-activation}
    {\includegraphics[width=0.233\textwidth]{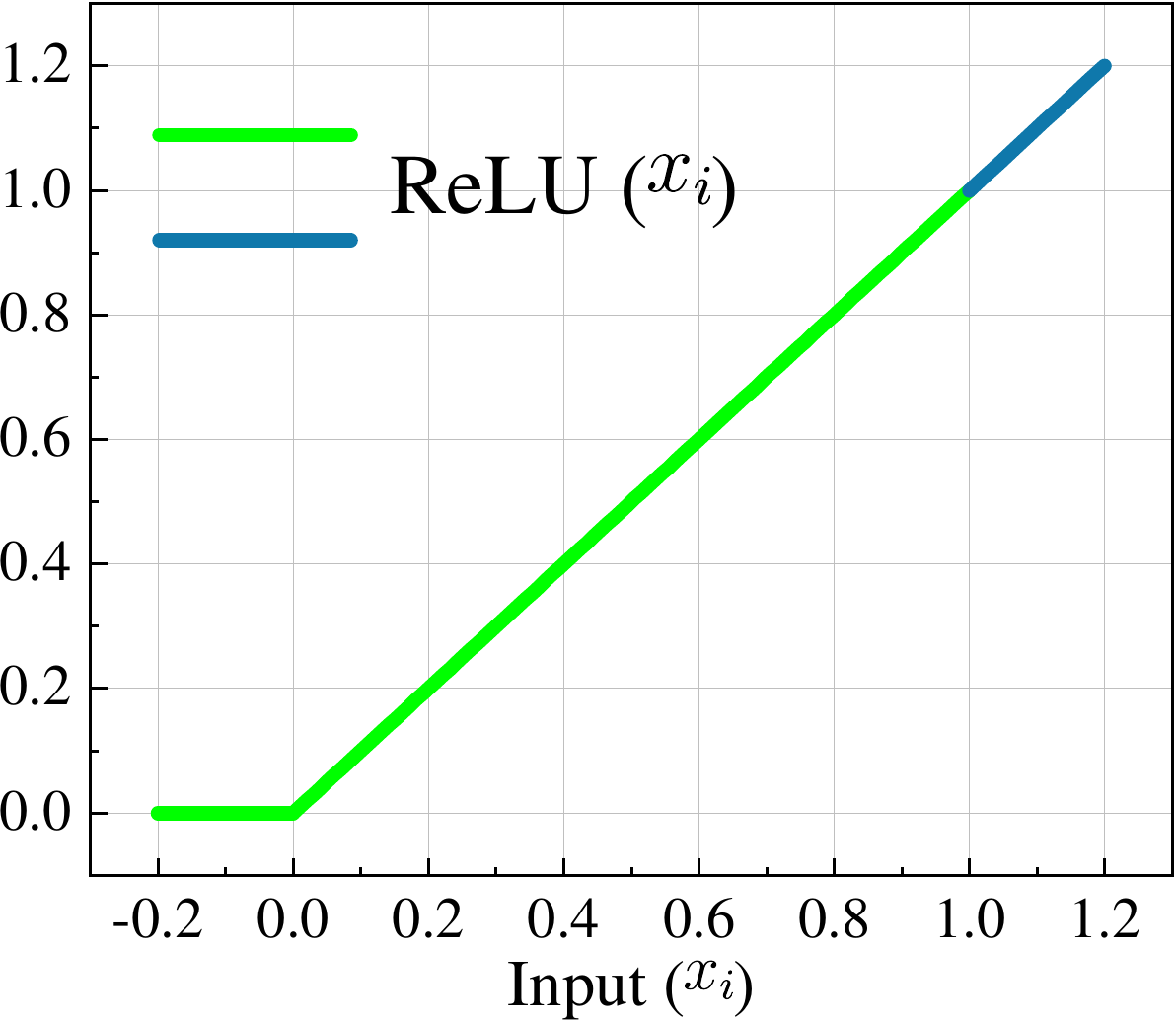}}}\hfill
    \caption{Comparison between IF neuron and ReLU.}
    \label{fig:IF-ReLU}
\end{figure}

\begin{figure}[ht]
    \footnotesize
    \centering
    \subfigure[White]{\label{fig:white-trigger}
    {\includegraphics[width=0.232\textwidth]{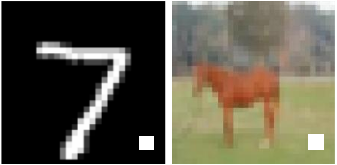}}}\hfill
    \subfigure[Random]{\label{fig:random-trigger}
    {\includegraphics[width=0.232\textwidth]{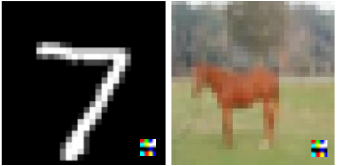}}}\hfill
    \caption{Two types of trigger patterns for traditional image datasets (e.g. MNIST ~\cite{MNIST} and CIFAR10 ~\cite{CIFAR10}).}
    \label{fig:trigger_pattern_tra}
\end{figure}

\begin{figure}[h]
    \footnotesize
    \centering
    \subfigure[Polarity=0]{\label{fig:polarity-0}
    {\includegraphics[width=0.15\textwidth]{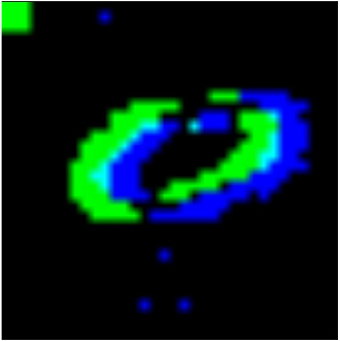}}}\hfill
    \subfigure[Polarity=1]{\label{fig:polarity-1}
    {\includegraphics[width=0.15\textwidth]{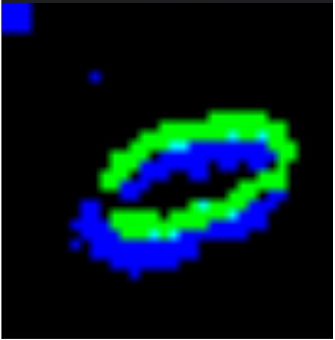}}}\hfill
    \subfigure[Polarity=2]{\label{fig:polarity-2}
    {\includegraphics[width=0.15\textwidth]{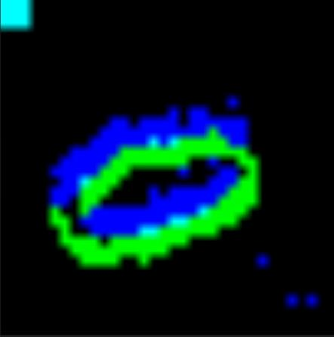}}}\hfill
    \caption{Three types of polarity-based trigger patterns for neuromorphic datasets (e.g. N-MNIST ~\cite{N-MNIST}, CIFAR10-DVS ~\cite{CIFAR10-DVS}, DVS128Gesture ~\cite{DVS128Gesture}). Note that the neuromorphic dataset is in the data format of frame animation, and here we only intercepted one of the image frames as an example.}
    \label{fig:trigger_pattern_neuro}
\end{figure}

\begin{table}[t]
\caption{Different combinations to implement backdoor injection for hybrid learning rules ($\mathcal{LR_H}$).}
\centering
\resizebox{0.49\textwidth}{13mm}{
\begin{tabular}{cccc}
\hline \hline
\multirow{2}{*}{\textbf{Combination}} & \multicolumn{2}{c}{\textbf{Poisoning Step}}              & \multirow{2}{*}{\textbf{Output Model}} \\ \cline{2-3}
                                        & \textbf{ANN step}          & \textbf{SNN step}          &                                        \\ \hline
\textbf{$Com1$}                                       & \ding{56} & \ding{56} & Clean Model                            \\ \hline
\textbf{$Com2$}                                       & \ding{56} & \ding{52} & Backdoored Model                         \\ \hline
\textbf{$Com3$}                              & \ding{52} & \ding{56} & Backdoored Model                         \\ \hline
\textbf{$Com4$}                              & \ding{52} & \ding{52} & Backdoored Model                         \\ \hline \hline
\end{tabular}
}
\label{tab:poi_com}
\end{table}

\begin{figure}[!h]
\footnotesize
\centering
\subfigure[ACC of MNIST]{\label{fig:ACC-MR}
{\includegraphics[width=0.234\textwidth]{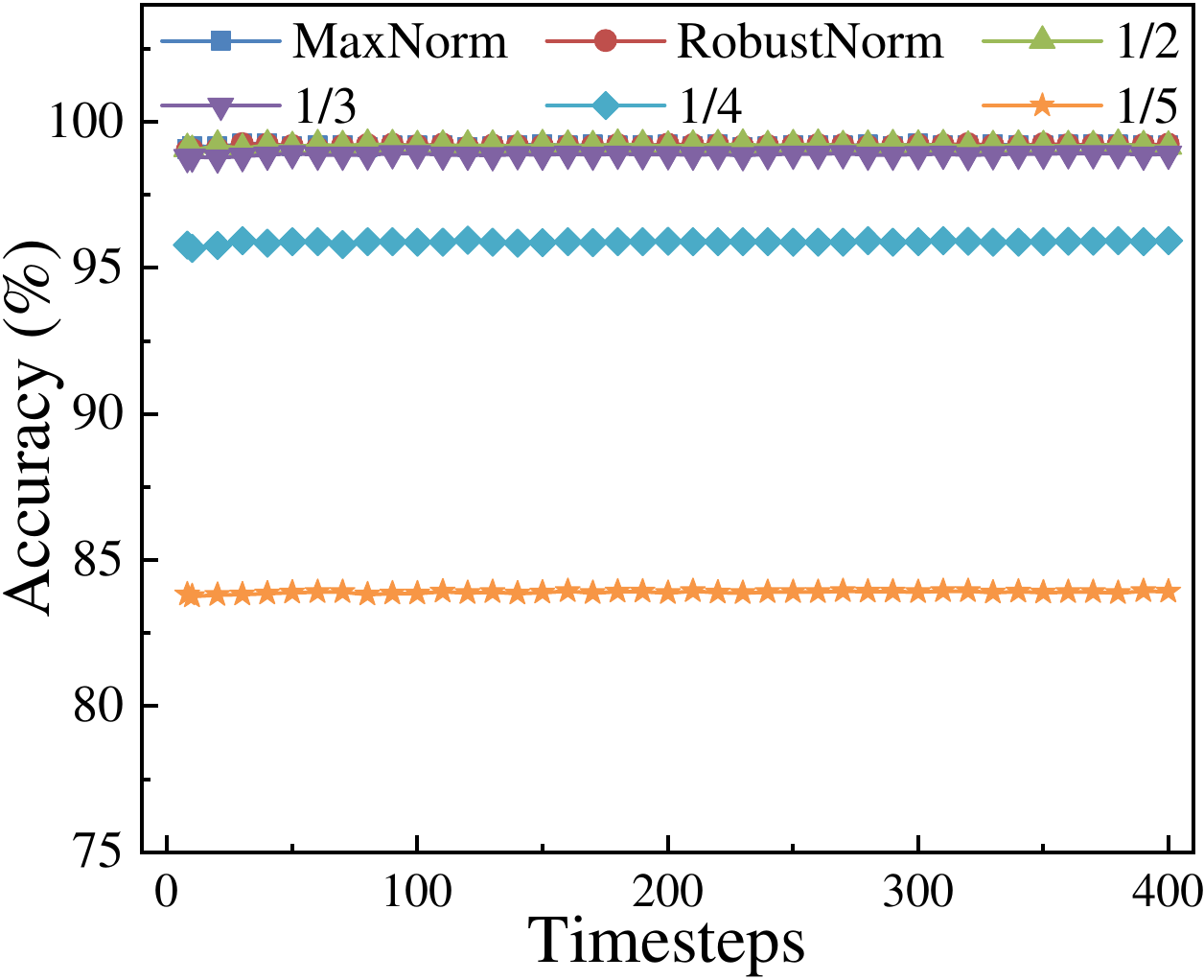}}}\hfill
\subfigure[ASR of MNIST]{\label{fig:ASR-MR}
{\includegraphics[width=0.234\textwidth]{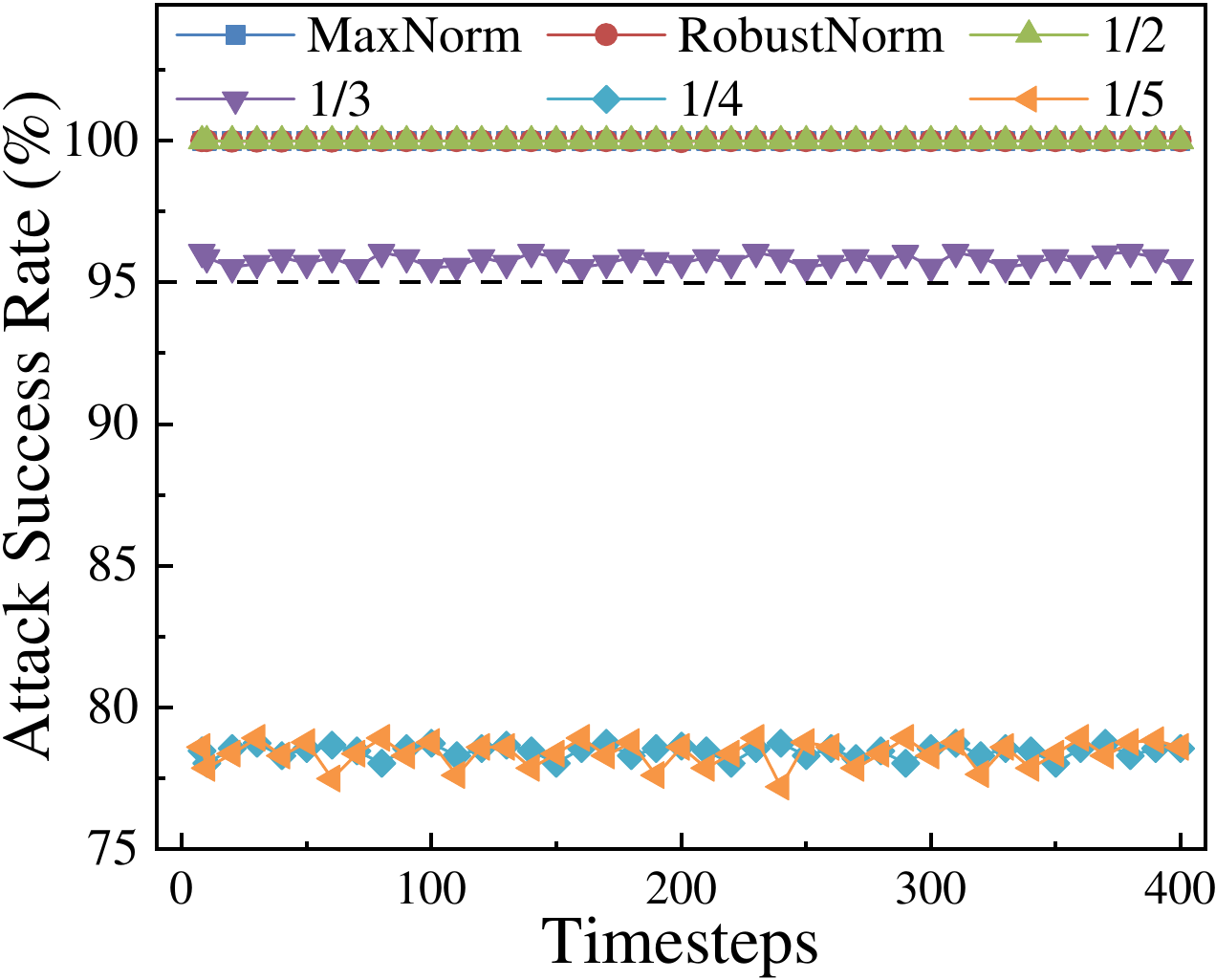}}}\hfill
\subfigure[ACC of CIFAR10]{\label{fig:ACC-CR}
{\includegraphics[width=0.234\textwidth]{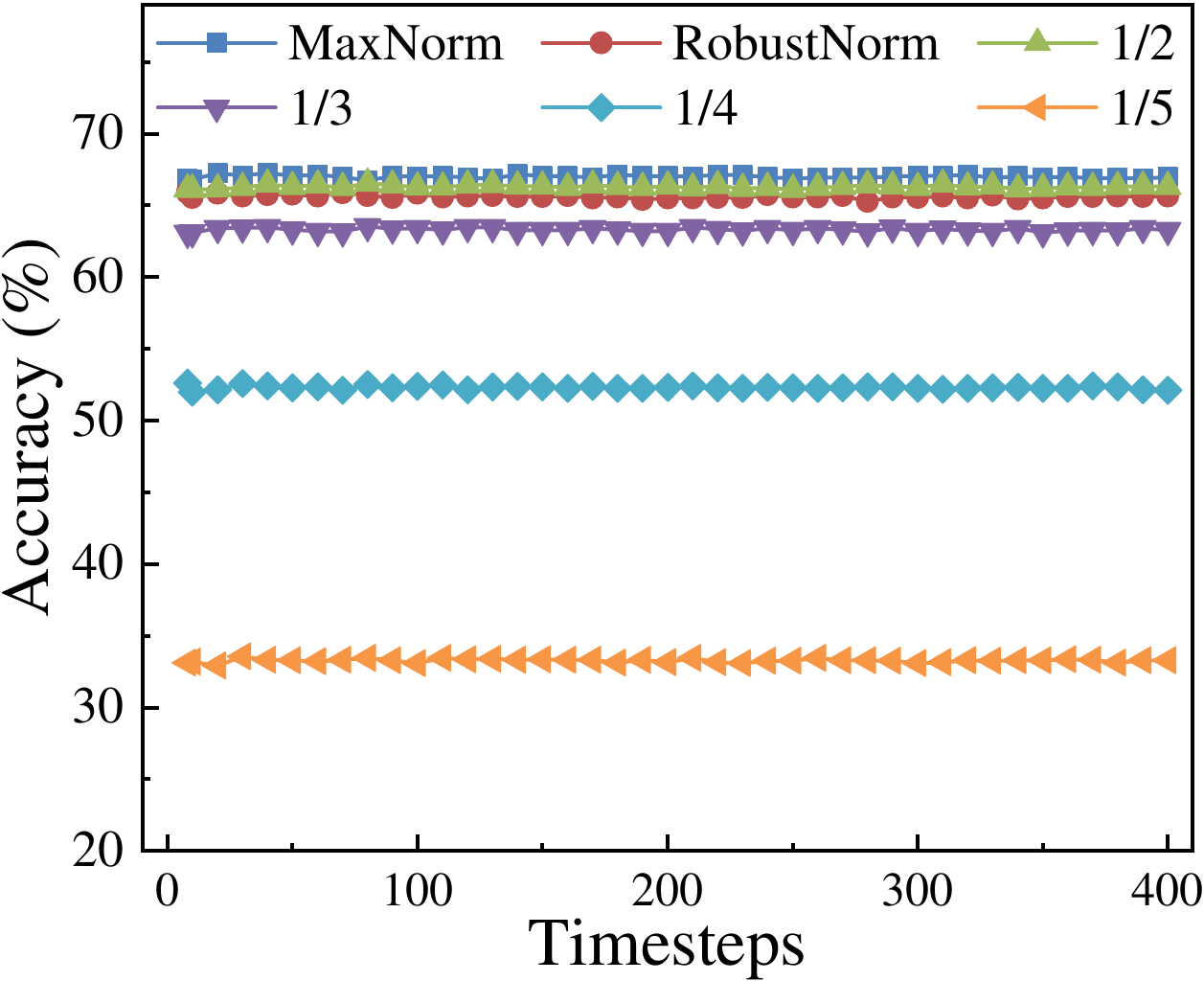}}}\hfill
\subfigure[ASR of CIFAR10]{\label{fig:ASR-CR}
{\includegraphics[width=0.234\textwidth]{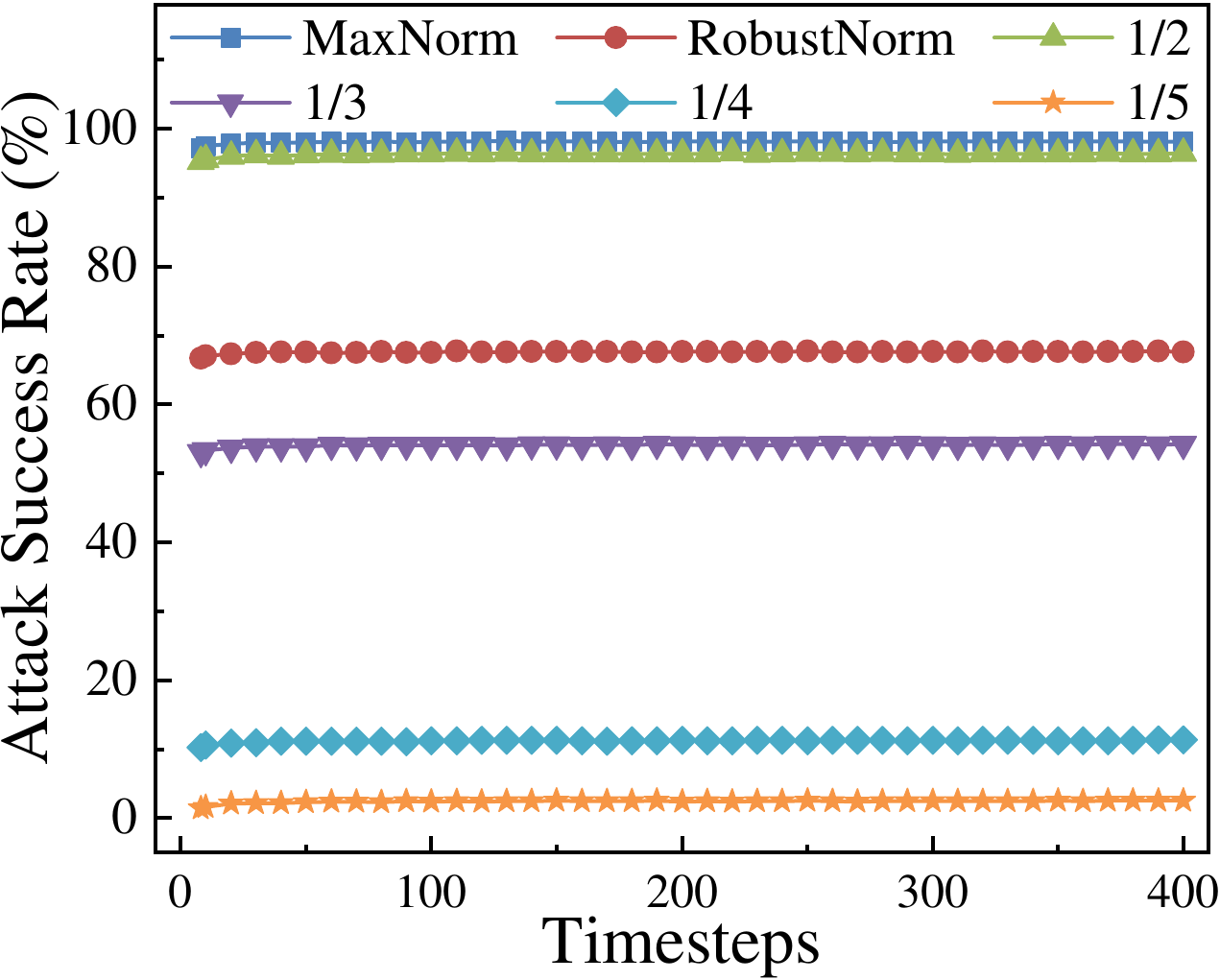}}}\hfill
\caption{The performance of spike-based ResNet18 trained by conversion-based learning rules ($\mathcal{LR_C}$) with different conversion strategies and increasing timesteps under backdoor attacks.}
\label{fig:conversion}
\end{figure}

\begin{figure}[!t]
\footnotesize
\centering
\subfigure[VGG11-MNIST]{\label{fig:ANN2SNN-VGG11-MNIST}
{\includegraphics[width=0.233\textwidth]{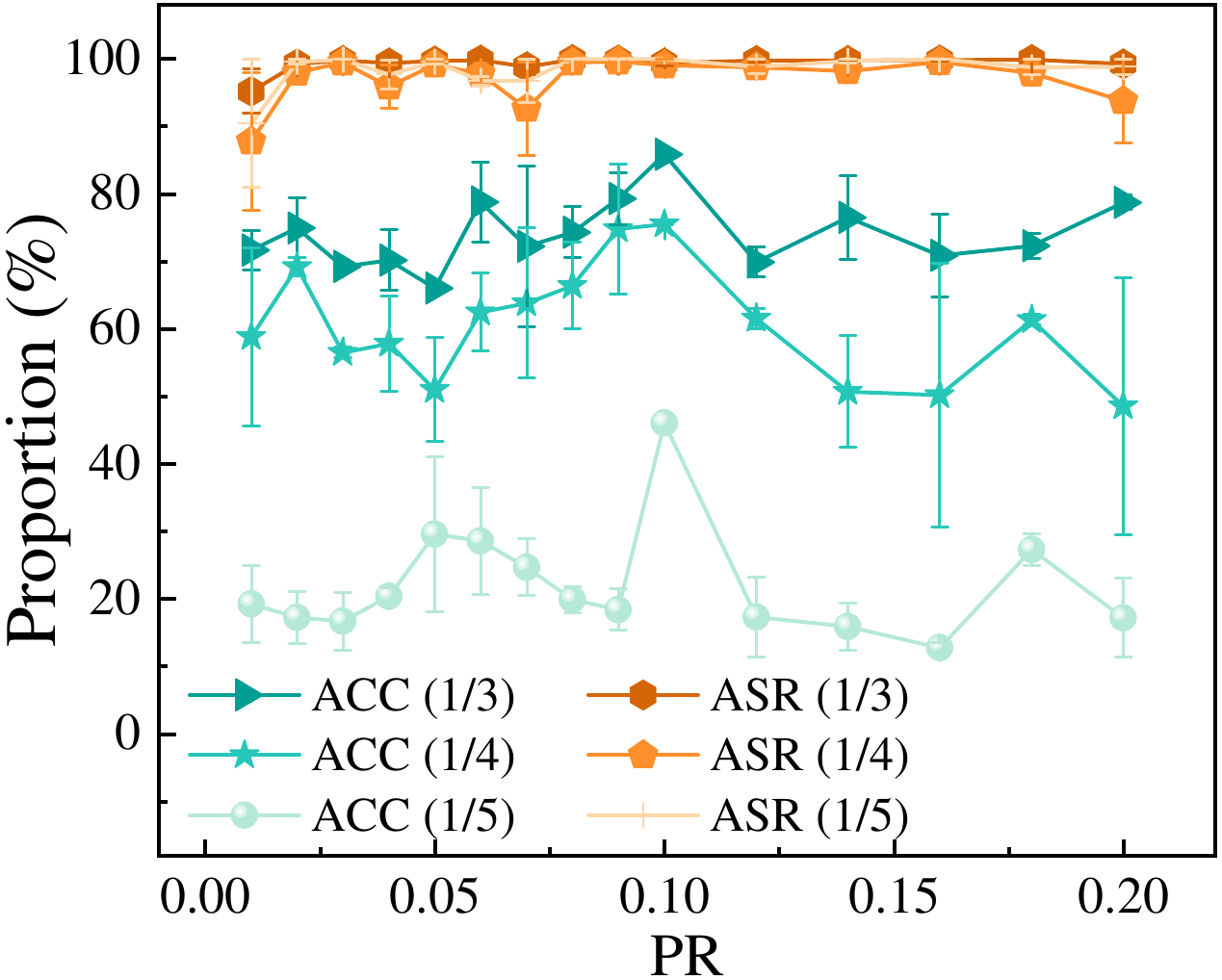}}}\hfill
\subfigure[VGG11-CIFAR10]{\label{fig:ANN2SNN-VGG11-CIFAR10}
{\includegraphics[width=0.233\textwidth]{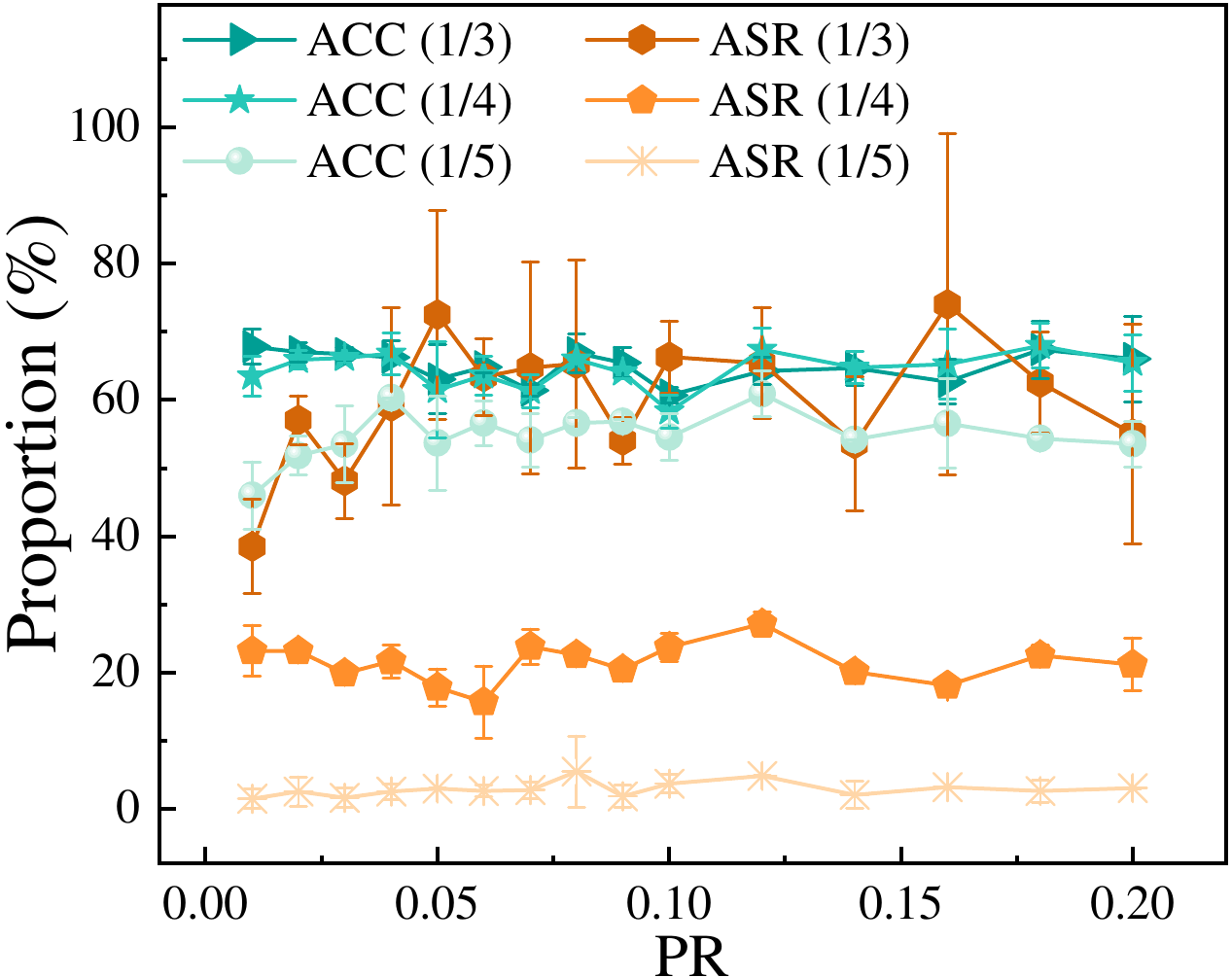}}}\hfill
\subfigure[ResNet18-MNIST]{\label{fig:ANN-RES18-MNIST}
{\includegraphics[width=0.233\textwidth]{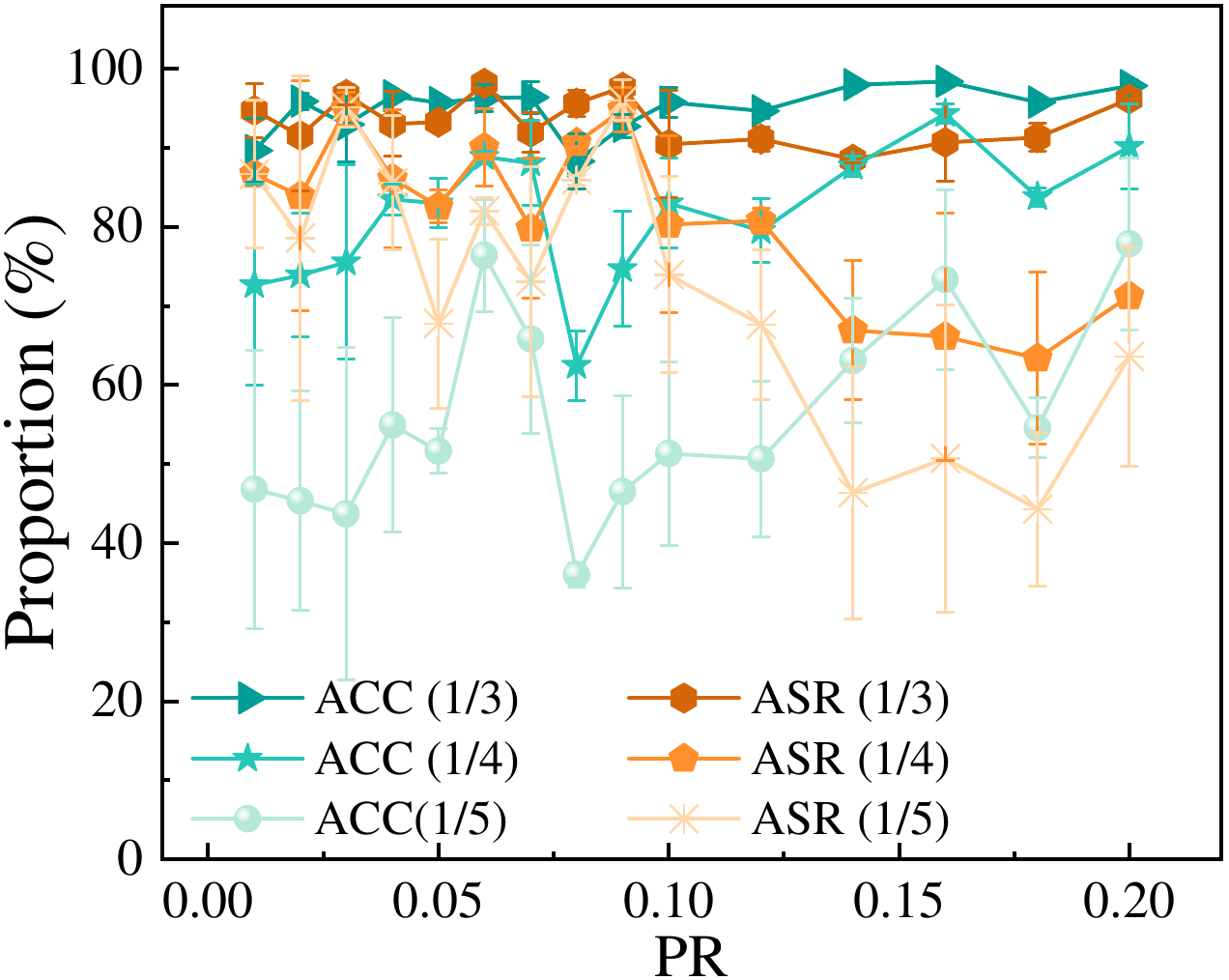}}}\hfill
\subfigure[ResNet18-CIFAR10]{\label{fig:ANN2SNN-RES18-CIFAR10}
{\includegraphics[width=0.233\textwidth]{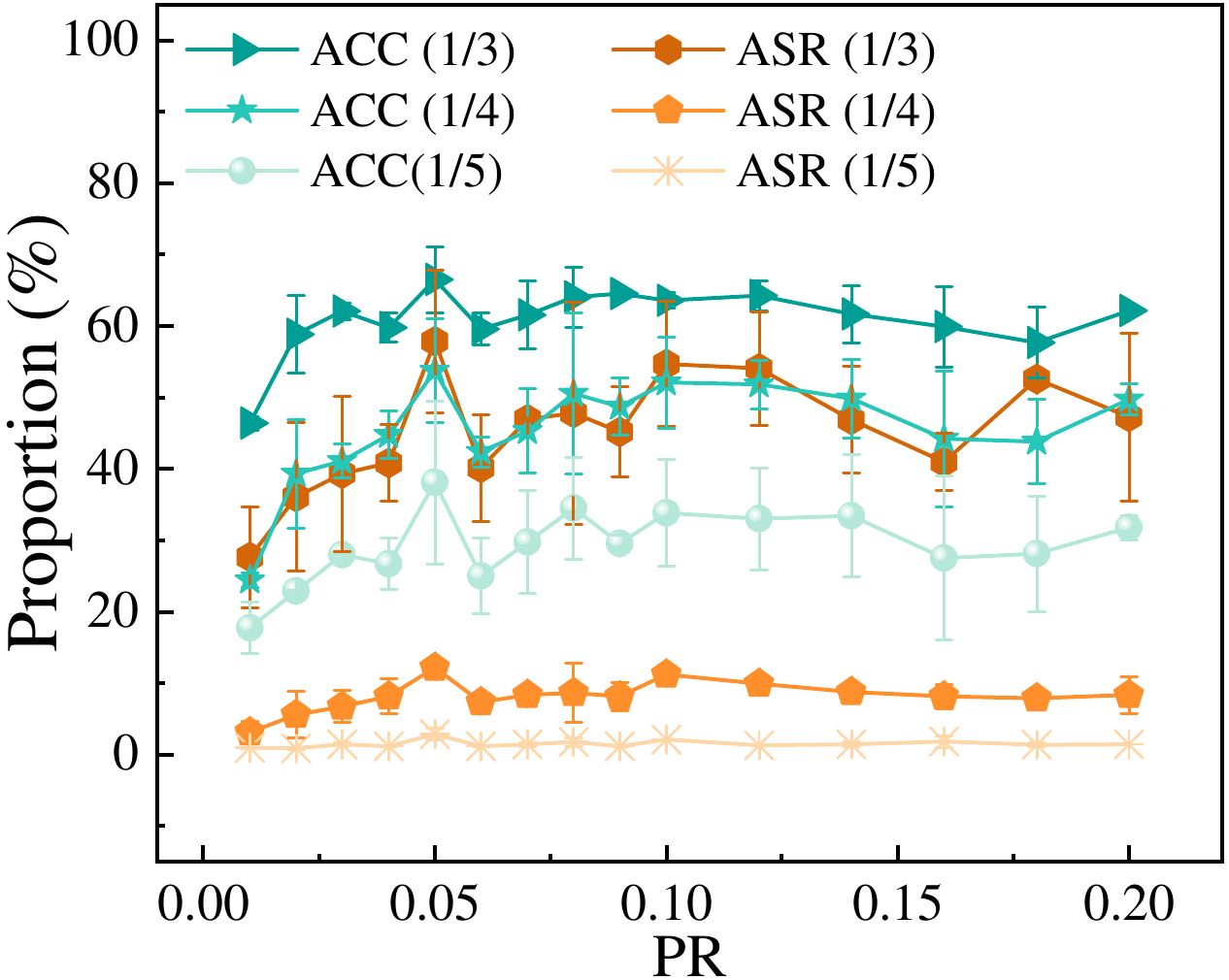}}}\hfill
\caption{The performance of converted SNN models trained by conversion-based learning rules ($\mathcal{LR_C}$) with different conversion strategies under backdoor attacks with different poisoning rates.}
\label{fig:conversion-sup}
\end{figure}

\begin{table}[ht]
\centering
\small
\caption{Backdoor migration capability under $\mathcal{LR_C}$ and $\mathcal{LR_H}$ with different poisoning rates.}
\begin{tabular}{c|ccc|ccc}
\hline \hline
\multirow{2}{*}{\textbf{PR}} & \multicolumn{3}{c|}{\textbf{$\mathcal{LR_C}$} (\%)}                                                                 & \multicolumn{3}{c}{\textbf{$\mathcal{LR_H}$} (\%)}                                                               \\ \cline{2-7}
                             & \textbf{$\hat{\mathcal{M}}_{a}(\cdot)$} & \textbf{\begin{tabular}[c]{@{}c@{}}$\hat{\mathcal{M}}_{s}(\cdot)$\\ (Max)\end{tabular}} & \textbf{MR} & \textbf{$\hat{\mathcal{M}}_{a}(\cdot)$} & \textbf{\begin{tabular}[c]{@{}c@{}}$\hat{\mathcal{M}}_{s}(\cdot)$\\ ($Com3$)\end{tabular}} & \textbf{MR} \\ \hline
0.02                         & 96.09           & 95.72                                                               & 99.61     & 95.56           & 29.82                                                            & 31.21     \\
0.04                         & 96.95           & 96.74                                                               & 99.78     & 96.45           & 19.57                                                            & 20.29     \\
0.06                         & 97.9            & 97.67                                                               & 99.77     & 96.78           & 15.82                                                            & 16.35     \\
0.08                         & 98.05           & 97.84                                                               & 99.79     & 97.44           & 18.88                                                            & 19.38     \\
0.01                         & 98.37           & 98.23                                                               & 99.86     & 97.36           & 28.62                                                            & 29.40     \\
0.12                         & 98.43           & 98.1                                                                & 99.66     & 97.49           & 16.69                                                            & 17.12     \\
0.14                         & 98.45           & 98.26                                                               & 99.81     & 97.81           & 26.16                                                            & 26.75     \\
0.16                         & 98.72           & 98.22                                                               & 99.49     & 98.04           & 26.97                                                            & 27.51     \\
0.18                         & 98.74           & 98.35                                                               & 99.61     & 97.97           & 27.01                                                            & 27.57     \\
0.2                          & 98.64           & 98.51                                                               & 99.87     & 98.21           & 20.12                                                            & 20.49     \\ \hline \hline
\end{tabular}
\label{tab:back_migration_sup}
\end{table}

\end{document}